\documentclass[letterpaper,11pt]{article}
\pdfoutput=1

\usepackage[utf8]{inputenc}

\usepackage{jheppub}
\usepackage{multirow, amsthm}
\usepackage{braket}
\usepackage{mathrsfs}
\usepackage{tikz}
\usepackage[caption=false]{subfig}
\usepackage{xspace}
\usepackage{xcolor}
\usepackage{float,color}
\usepackage{siunitx}
\usepackage[ruled,norelsize]{algorithm2e}
\usepackage[countmax]{subfloat}
\usepackage [autostyle, english = american]{csquotes}
\MakeOuterQuote{"}
\graphicspath{ {./figs/} }
\usepackage{placeins}
\usepackage{cleveref}

\usepackage{accents}
\newlength{\dhatheight}

\providecommand{\href}[2]{#2}
\usepackage{comment}

\definecolor{darkred}{rgb}{0.5,0.0,0.0}
\definecolor{darkblue}{rgb}{0.0,0.0,0.9}
\definecolor{darkerblue}{rgb}{0.0,0.0,0.5}
\definecolor{darkgreen}{rgb}{0.0,0.5,0.0}
\definecolor{black}{rgb}{0.0,0.0,0.0}
\definecolor{brown}{rgb}{0.6,0.4,0.2}

\newcommand{\bpn}[1]{\marginpar{\raggedright\scriptsize\textbf{\textcolor{red}{bpn}}}  \textbf{\textcolor{red}{(#1 --bpn)}}}

\title{\boldmath Investigating the Topology Dependence of \\ Quark and Gluon Jets}
\author[1,2]{Samuel Bright-Thonney}
\author[1]{and Benjamin Nachman}
\affiliation[1]{\normalsize\it Physics Division, Lawrence Berkeley National Laboratory, Berkeley, CA 94720, USA}
\affiliation[2]{\normalsize\it Physics Department, Cornell University, 109 Clark Hall, Ithaca, New York 14853, USA}

\emailAdd{sbright@berkeley.edu}
\emailAdd{bpnachman@lbl.gov}

\abstract{
As most target final states for searches and measurements at the Large Hadron Collider have a particular quark/gluon composition, tools for distinguishing quark- from gluon-initiated jets can be very powerful.  In addition to the difficulty of the classification task, quark-versus-gluon jet tagging is challenging to calibrate.  The difficulty arises from the topology dependence of quark-versus-gluon jet tagging: since quarks and gluons have net quantum chromodynamic color charge while only colorless hadrons are measured, the radiation pattern inside a jet of a particular type depends on the rest of its environment.  Given a definition of a quark or gluon jet, this paper studies the topology dependence of such jets in simulation.  A set of phase space regions and jet substructure observables are identified for further comparative studies between generators and eventually in data.}

\begin{document} 
\maketitle
\flushbottom

\section{Introduction}
\label{sec:intro}

Classifying jets as initiated from a quark or a gluon can be useful for improving Standard Model (SM) measurements~\cite{Aaboud:2016cns,ATLAS-CONF-2016-063,Aad:2016oit,ATLAS-CONF-2015-002,Sirunyan:2017jej,Sirunyan:2018ygk,Khachatryan:2015bnx,Khachatryan:2014dea} and searches for physics beyond the SM~\cite{Aaboud:2017eta,Sirunyan:2017pjw,Sirunyan:2017lzl,Sirunyan:2017wif} at the Large Hadron Collider (LHC).   As gluons are in the adjoint representation of the Quantum Chromodynamic (QCD) SU(3) gauge group while quarks are in the fundamental representation, gluons carry both color and anti-color quantum numbers while quarks have only a single color.  Therefore, gluon jets tend to have more constituents and a broader radiation pattern than quark jets\footnote{More precisely, the Altarelli-Parisi splitting functions~\cite{Altarelli:1977zs} contain a factor of $C_A=3$ for gluon radiation from a gluon and a factor of $C_F=4/3$ for gluon radiation from a quark.  The multiplicity and width of gluon jets are therefore approximately $9/4$ bigger than for quarks, on average.}.  Recent developments in quark versus gluon jet tagging have resulted from advances in the theoretical~\cite{Larkoski:2014pca,Gras:2017jty,Frye:2017yrw}, phenomenological~\cite{Gallicchio:2011xq,Gallicchio:2012ez}, and experimental~\cite{Aad:2014gea,ATLAS-CONF-2016-034,ATL-PHYS-PUB-2017-009,CMS-PAS-JME-13-002,CMS-DP-2016-070,ATL-PHYS-PUB-2017-017,CMS-DP-2017-027} understanding of quark-versus-gluon jet tagging as well as the development of powerful machine learning techniques that can utilize the entire jet internal radiation pattern~\cite{ATL-PHYS-PUB-2017-017,CMS-DP-2017-027,Komiske:2016rsd,Dery:2017fap,Metodiev:2017vrx,Luo:2017ncs,Komiske:2018oaa,Cheng:2017rdo}.  

The goal of this paper is to study one of the key challenges with quark-versus-gluon jet tagging: the topology dependence\footnote{"Topology dependence" is used interchangeably with "process dependence"; the latter is more precise, but the former is used more often in practice.}.  Since quarks and gluons have color charge but only color neutral hadrons are observed, the energy flow of jets formed from quarks and gluons depends on the rest of the event.  Traditionally, quark-versus-gluon jet tagging algorithms have been calibrated by comparing the substructure of jets from two different event samples with different gluon fractions.  However, this method will not close exactly when the gluon or quark jets from one sample do not have the same statistical properties as those from the second sample.  Evidence for such non-closures was present in the Run 1 studies from the ATLAS collaboration~\cite{Aad:2014gea,ATLAS-CONF-2016-034}, though this interpretation may be obscured due to detector effects (not unfolded).  One clear example of topology dependence is the study of colorflow in $t\bar{t}$ events, using color singlet $W$ boson decays into quarks~\cite{Gallicchio:2010sw,Aad:2015lxa,Aaboud:2018ibj,Abazov:2011vh}.  The radiation pattern inside one of the jets resulting from the $W$ decay significantly (though subtly) depends on the orientation of its companion jet.  Color singlet hadronic decays highlight another case in which the jet $p_\text{T}$ and parton type are insufficient for describing the full radiation pattern.  In such decays, half of the boson mass is the relevant scale for jet fragmentation even though the jet $p_\text{T}$ can be arbitrarily small.  For inclusive jets in $pp$ collisions, the jet $p_\text{T}$ accounts for most of the variation in the fragmentation, but some variation may be captured by $p_z$, albeit with large variance from the longitudinal boost of the center of mass frame.  Subtler differences in the process dependence of the soft radiation around quark and gluon jets has been studied both numerically and analytically~\cite{stewart2015dissecting}.   These, and potentially other effects, are investigated systematically across observables and processes.

This paper is organized as follows.  Section~\ref{sec:qgt} briefly reviews the jet substructure observables considered for the comparative study.  The various topologies are introduced in Sec.~\ref{sec:base} as well as a set of baseline results.  Variations that include the $p_\text{T}$, simulator, and quark- or gluon-jet labeling scheme appear in Sec.~\ref{ptdep}--\ref{sec:her7}.  The paper concludes with conclusions and future outlook in Sec.~\ref{sec:conclusions}.

\section{Jet tagging \& observables}
\label{sec:qgt}

\subsection{Quark/gluon jet tagging}
Before presenting the results of our analysis, it is necessary to state precisely what is meant by a "quark jet" and "gluon jet". A number of definitions have been proposed, each of which suffer from varying degrees of ambiguity, as detailed in Ref. \cite{Gras:2017jty}. In the context of a Monte Carlo (MC) study, quark/gluon jets would ideally refer to "quark-enriched" or "gluon-enriched" regions of phase space which make no reference to individual quark or gluon partons. Given the focus of this study, the goal is to extract both quark and gluon jets from a particular channel, making it impossible to define jet flavor in this manner. Therefore, jets are classified by scanning the MC event record for the highest-energy parton whose rapidity-azimuth distance $\Delta R = \sqrt{(\Delta y)^2 + (\Delta \phi)^2}$ from the jet axis is less than the jet's radius $R$, and assigning the jet the same flavor as this parton\footnote{This method is used in the latest ATLAS~\cite{ATL-PHYS-PUB-2017-017} and CMS~\cite{CMS-DP-2016-070} quark versus gluon jet studies and is common in phenomenological studies as well.}.   Quarks and gluons from color singlet decays provide a laboratory for studying jets that are color isolated from the rest of the event.  For these topologies (more detail in Sec.~\ref{sec:base}), only those partons from the singlet decay are used for the parton labeling.  An alternative labeling scheme is investigated in Sec.~\ref{sec:qcdaware}.  Other general definitions based on ideas like jet topics~\cite{Metodiev:2018ftz,Komiske:2018vkc} are left for future studies.

\subsection{Generalized angularities}
\begin{figure}
\centering
\subfloat[]{\label{fig:sampa}\includegraphics[width=0.33\textwidth]{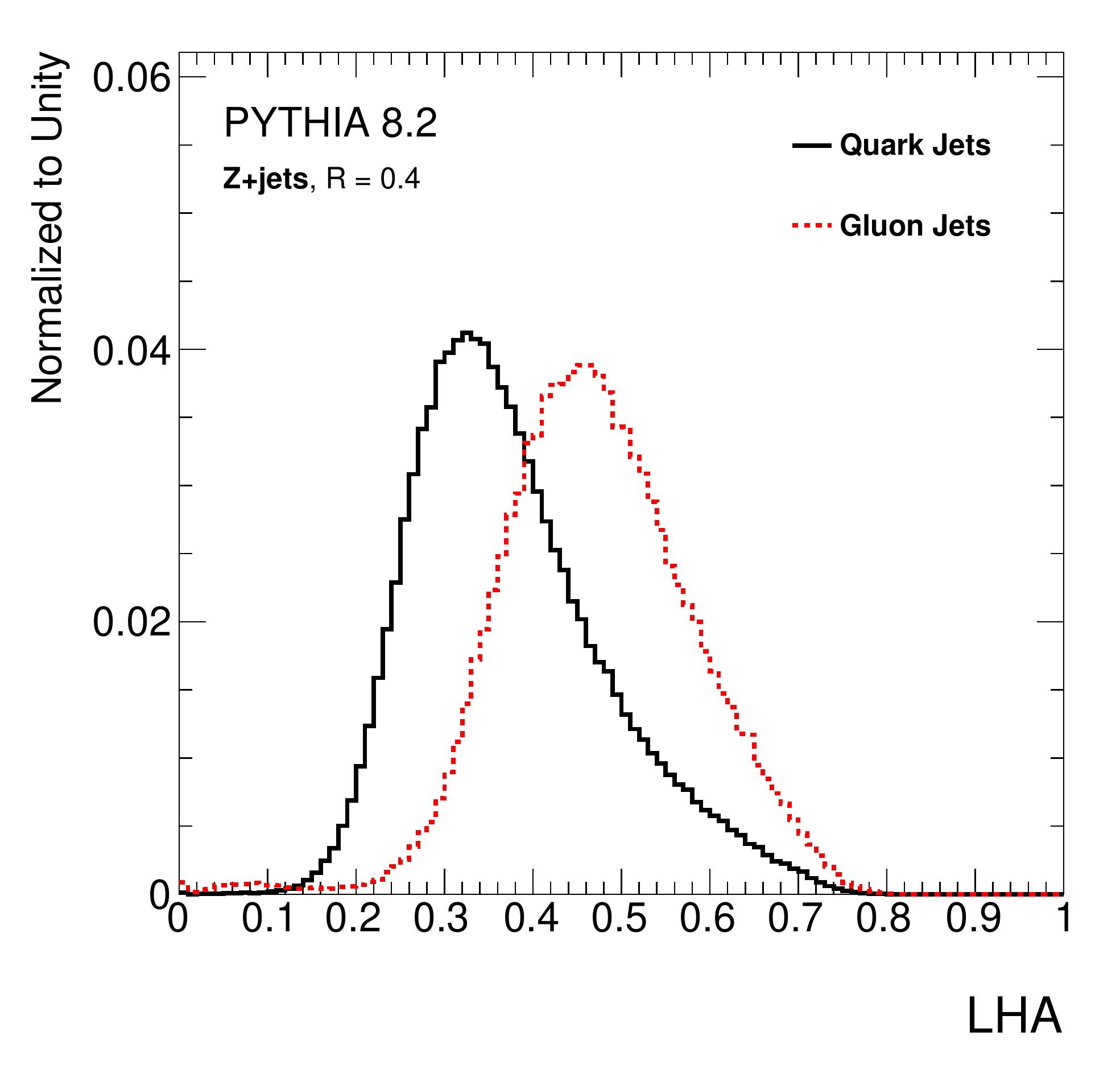}}
\subfloat[]{\label{fig:sampb}\includegraphics[width=0.33\textwidth]{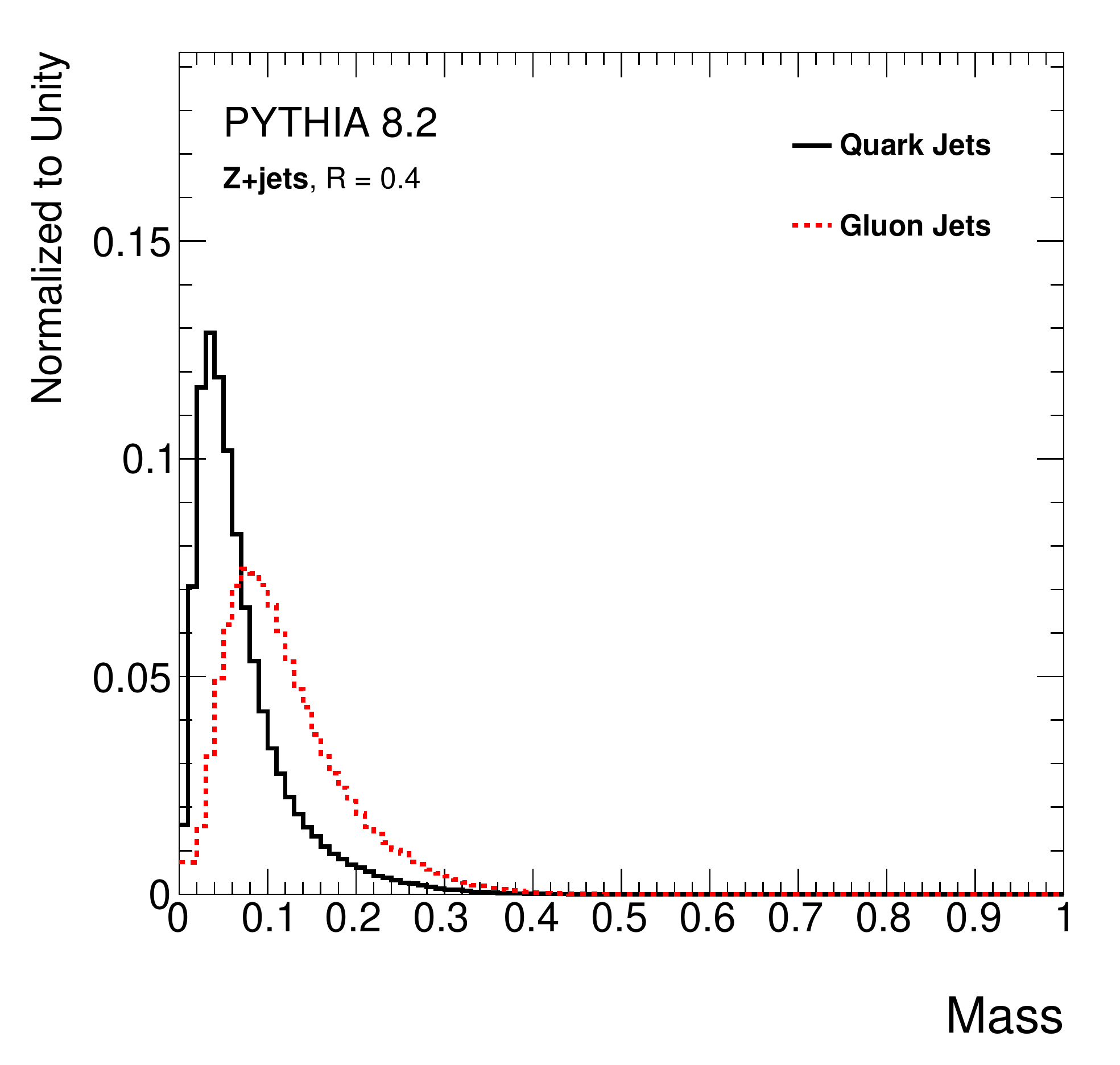}}
\subfloat[]{\label{fig:sampc}\includegraphics[width=0.33\textwidth]{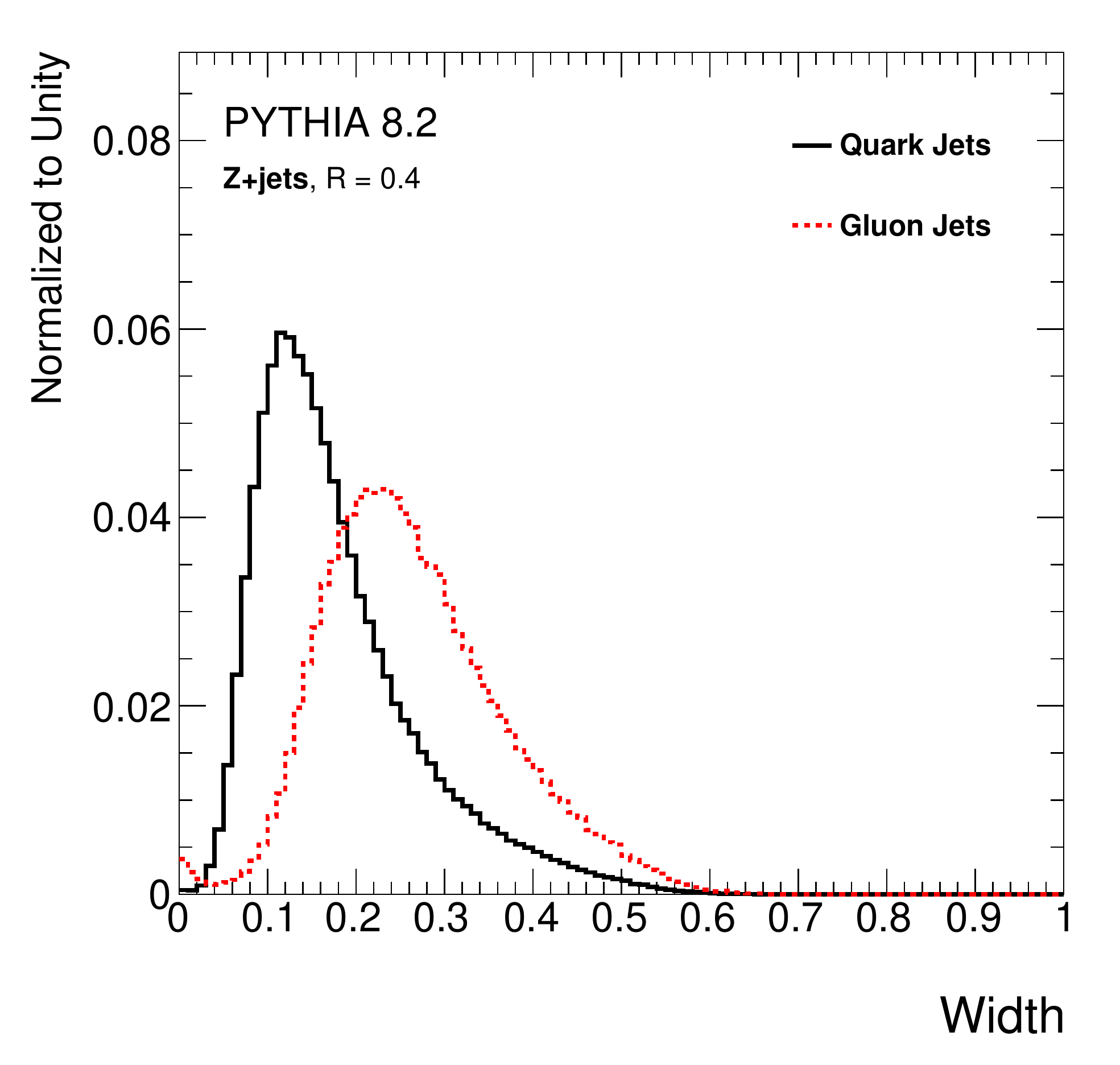}}
\\
\subfloat[]{\label{fig:sampd}\includegraphics[width=0.49\textwidth]{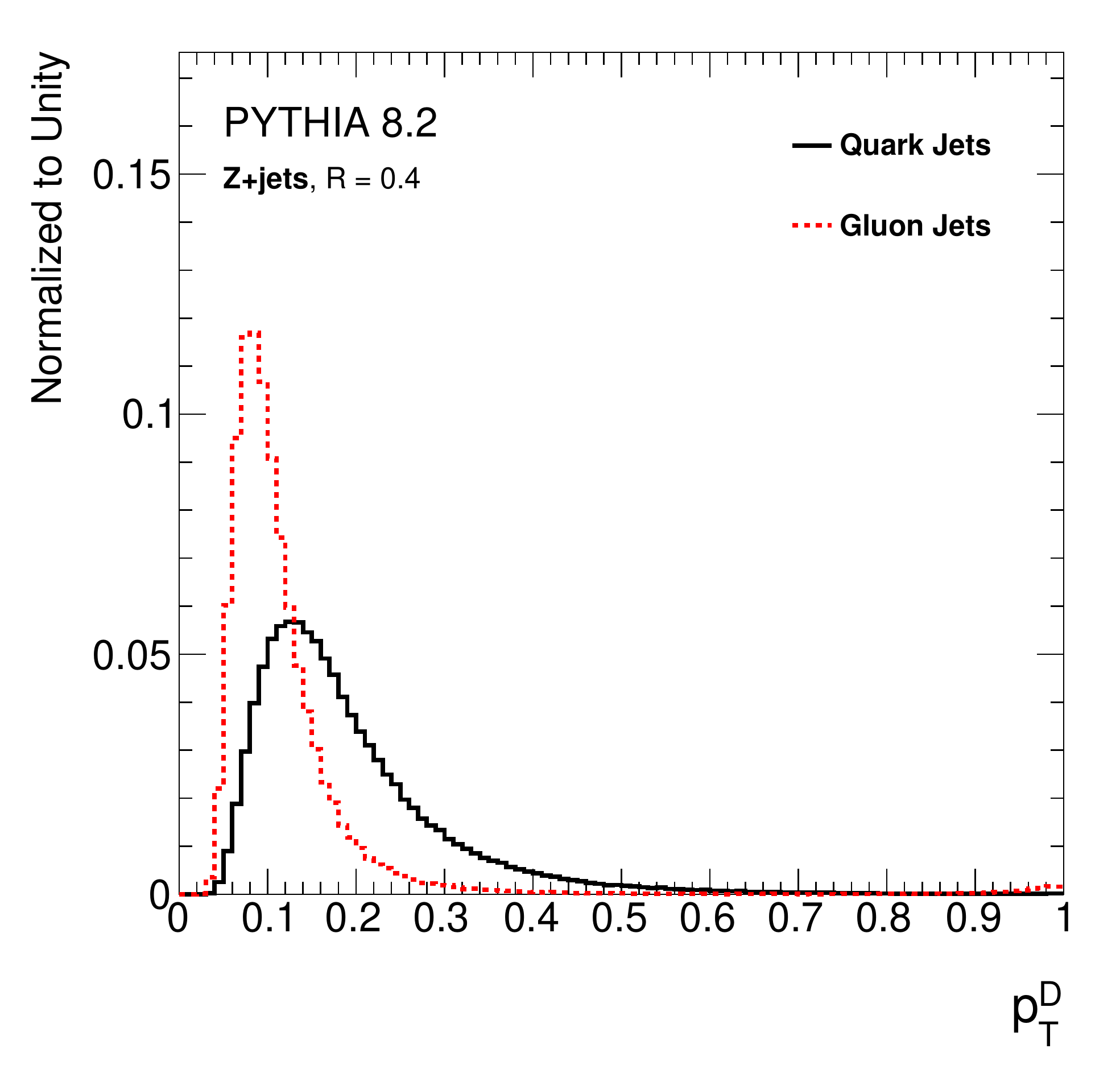}}
\subfloat[]{\label{fig:sampe}\includegraphics[width=0.49\textwidth]{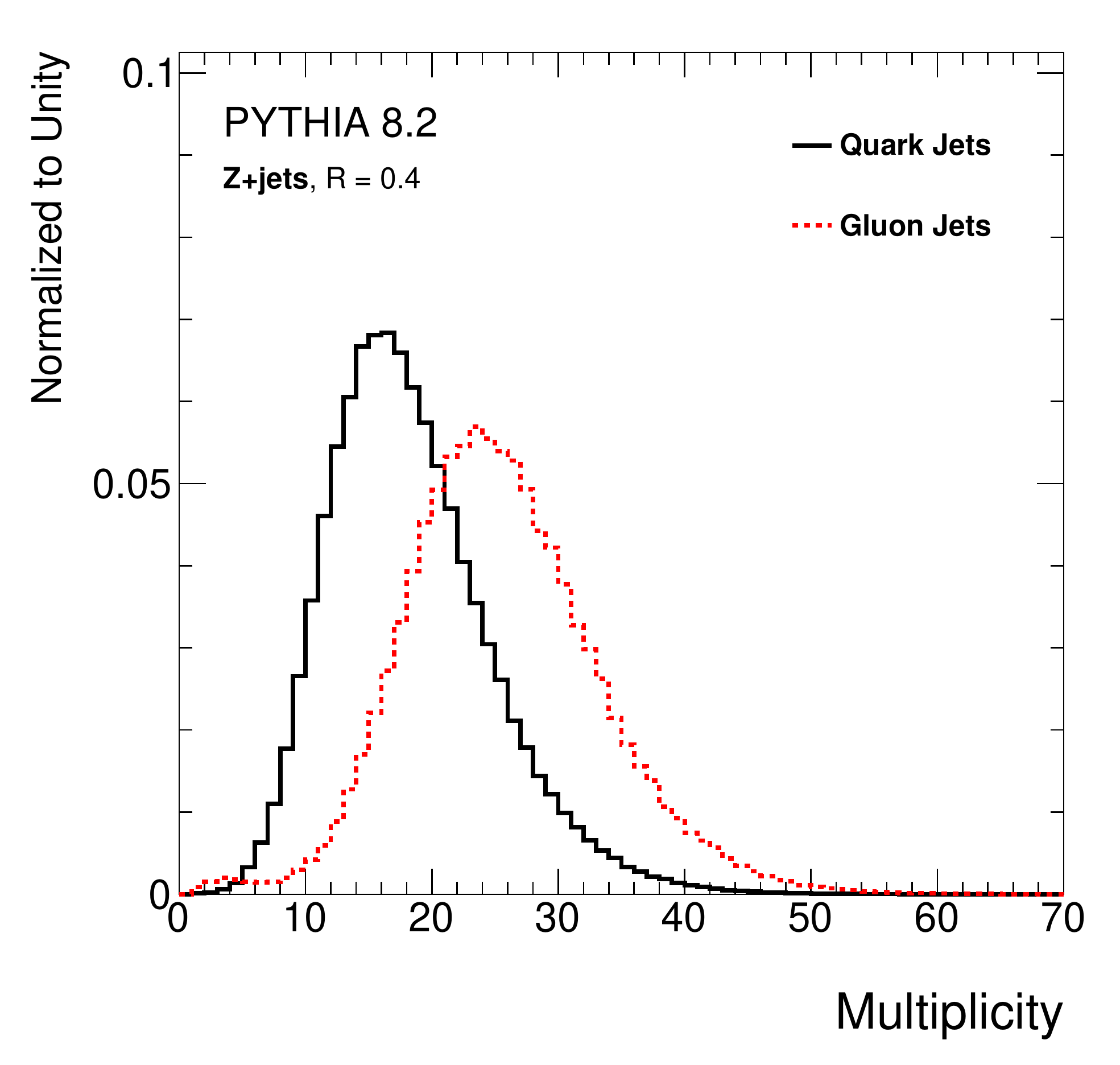}}
\caption{\label{fig:samp} Plots showing normalized distributions of the five generalized angularities in the quark and gluon jet channels from the Z+jets topology}
\end{figure}

There are a wide variety of substructure variables that have been tested in quark/gluon jet discrimination studies (see e.g. Ref.~\cite{Larkoski:2017jix,Asquith:2018igt} for a recent review). This analysis uses a particular class of generalized angularities~\cite{Larkoski:2014pca} that have been found to be effective discriminants~\cite{Gras:2017jty}, and examine how they vary amongst jets of the same flavor that originate from different toplogies. The angularities depend on two parameters ($\kappa$, $\beta$), and are defined as
\begin{equation}
\lambda_\beta^\kappa = \sum_{i\in jet}{z_i^\kappa\theta_i^\beta},
\end{equation}
where $z_i$ is the momentum fraction of jet constituent $i$, and $\theta_i$ is the normalized rapidity-azimuth angle to the jet axis. Jets are clustered using the anti-$k_t$ algorithm with $E$-scheme recombination, and 
\begin{equation}
z_i \equiv \frac{p_{Ti}}{\sum_{i \in jet}{p_{Ti}}}, \qquad \theta_i \equiv \frac{\Delta R_i}{R},
\end{equation}
where $R$ is the jet radius and $\Delta R_i$ is the rapidity-azimuth distance from constituent $i$ to the jet axis\footnote{We use the standard $E$-scheme combination axis instead of the winner-takes-all axis, as in Ref.~\cite{Gras:2017jty}.}.  Five angularities are used, each denoted by its $(\kappa,\beta)$ values~\cite{Gras:2017jty}:
\begin{equation}
\begin{split}
(0,0) &\Rightarrow \text{hadron multiplicity} \\
(2,0) &\Rightarrow (p_\text{T}^D)^2\, \text{\cite{Chatrchyan:2012sn}} \\
(1,0.5) &\Rightarrow \text{Les Houches Angularity (LHA) \cite{Gras:2017jty}} \\
(1,1) &\Rightarrow \text{width \cite{catani1992jet,rakow1981transverse,Ellis:1986ig}} \\
(1,2) &\Rightarrow \text{mass \cite{Farhi:1977sg}}.
\end{split}
\end{equation}
Angularities with $\kappa = 1$ are collinear safe and those with $\kappa > 0$ are infrared safe.  Observables that are both infrared and collinear (IRC) safe are calculable in perturbative QCD.  Some non-IRC safe observables are also under analytic control, though with non-standard perturbative expansions~\cite{Larkoski:2015lea}. Figure~\ref{fig:samp} shows sample angularity distributions from the quark and gluon jet channels in $Z$+jets events.

\clearpage

\subsection{Quantifying separation power}
Since several variables are studied across multiple topologies, it is most efficient to quantify separation power using a single number. As in Ref.~\cite{Gras:2017jty}, the classifier separation provides a quantitative summary statistic~\cite{Harrison:1998yr,Hocker:2007ht}\footnote{In the language of information theory, this is closely related to the $\chi^2$ divergence; both are $f$-divergences~\cite{csiszar63,citeulike:11934748,doi:10.1143/JPSJ.18.328} with $f(u)=(u-1)^2/(u+1)$ for the classifier separation and $f(u)=(u-1)^2$ for the $\chi^2$ divergence~\cite{Nachman:2016qyc,8507032}.  We are grateful to Ben Elder, who pointed out to us that this quantity has also been referred to as the triangular discriminator in the information theory literature~\cite{850703}.},
\begin{equation}
\label{sep}
\Delta(\lambda) = \frac{1}{2}\int d\lambda\frac{(p_1(\lambda) - p_2(\lambda))^2}{p_1(\lambda) + p_2(\lambda)}\, ,
\end{equation}
where $p_{1/2}(\lambda)$ is the probability distribution for a jet of some flavor (quark or gluon) as a function of the classifier $\lambda$ (in this case, $\lambda$ is a generalized angularity). The separation $\Delta$ ranges from 0 (no separation) to 1 (full separation).  The distributions $p_1(\lambda)$ and $p_2(\lambda)$ are equal if and only if $\Delta(\lambda)=0$.  As $0\leq \Delta\leq 1$, the classifier separation will often be referred to as a percentage (i.e. $\Delta=0$ is equivalent to $0\%$ separated). 

\section{Baseline analysis}
\label{sec:base}
In the baseline study, the behavior of angularities in quark and gluon jets from six different topologies in $pp$ collisions is investigated:
\begin{enumerate}
\item Dijets
\item $Z$+jets
\item $gg \rightarrow Hg$
\item $q\bar{q} \rightarrow Zg$
\item $H \rightarrow gg$
\item $H \rightarrow q\bar{q}$.
\end{enumerate}
Samples of one million events are generated for each topology using \texttt{PYTHIA 8.226}~\cite{Sjostrand:2007gs} with the Monash 2013 tune~\cite{skands2014tuning}, a center-of-mass energy of 13 TeV, and a $\hat{p}_T$ range of $45 \leq \hat{p}_T \leq 200$ GeV. Jets are clustered using \texttt{FASTJET 3.2.1}~\cite{Cacciari:2011ma} with the anti-$k_t$ algorithm~\cite{Cacciari:2008gp} using $E$-scheme recombination. Quark and gluon jets are identified using the procedure described in Sec.~\ref{sec:qgt}.  In order to avoid sculpting from the $\hat{p}_T$ requirement, jets are only considered if $50 < p_{T} < 150$~GeV; to emulate the acceptance of typical tracking detectors, jets must be within\footnote{The jets are clustered using $y$ instead of $\eta$, but since there is no natural mass scale, $y$ and $\eta$ are very similar and the LHC experiments currently use $\eta$ exclusively to define event selections.} $|\eta| < 2.0$.  In order to study the affect of jet radius on separation power, samples are generated for each topology with jet radii in the range $0.2 \leq R \leq 1.5$ in steps of $0.1$. In samples 2--4, the Higgs and $Z$ bosons are forced to decay into neutrinos, preventing any hadronic or leptonic decay products from interfering with other jets in the event.  The masses of the bosons are also set equal ($m_H = m_Z = 200$~GeV) in order to help control the jet $p_T$ spectrum.

\par Further selection criteria (partially inspired by Ref.~\cite{Gras:2017jty}) are applied to the events in order to ensure that differences in the radiation profile are dominated by topology effects and not from trivial kinematic differences.  In particular, in dijet and $H \rightarrow q\bar{q}/gg$ events, both the leading and subleading jets that pass the kinematic selection, regardless of flavor, are considered.  If only one jet passes the selection, then it is used.  After picking these jets, they are sorted by parton type.  In contrast, only the leading jets that pass the kinematic selection are used from $Z$+jets, $gg \rightarrow Hg$, and $q\bar{q} \rightarrow Zg$ events.  The reason for using different jets in the two sets of topologies is that either the $Z/H$ boson or the leading jet in $Z$+jets, $gg \rightarrow Hg$, and $q\bar{q} \rightarrow Zg$ events might be the hardest $p_T$ object.  It is likely that when the jet is subleading in $p_T$ to the boson in such events, the originating parton radiated more than in cases when it is leading.  Therefore, if only the leading jet in dijet events were considered, there would be a systematic difference.  This is addressed by taking two jets in topologies defined by two hard jets and one jet in topologies defined by one hard jet and a boson.  

\par
\begin{figure}
\centering
\subfloat[]{\label{fig:0a}\includegraphics[width=0.49\textwidth]{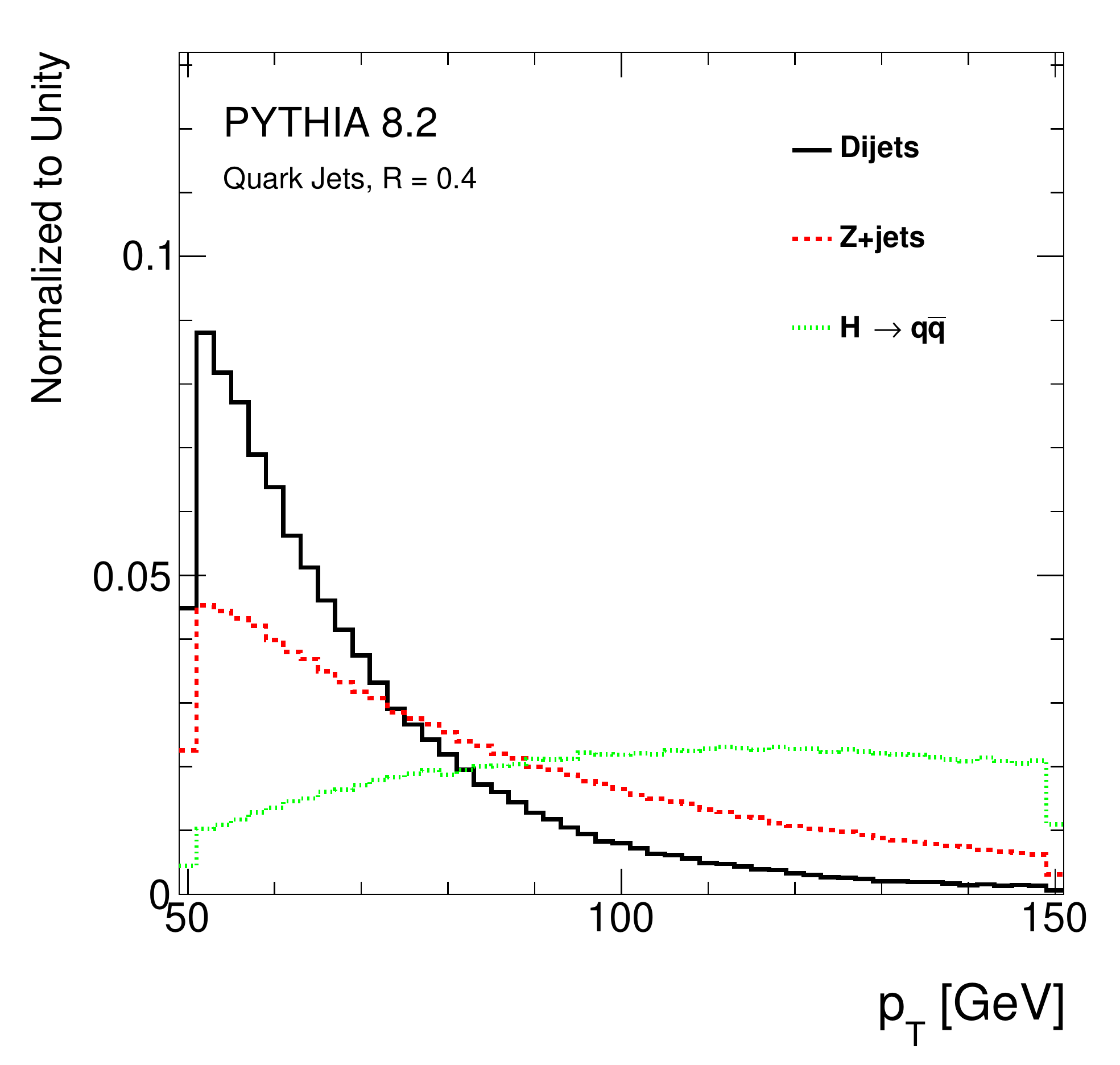}}
\hfill
\subfloat[]{\label{fig:0b}\includegraphics[width=0.49\textwidth]{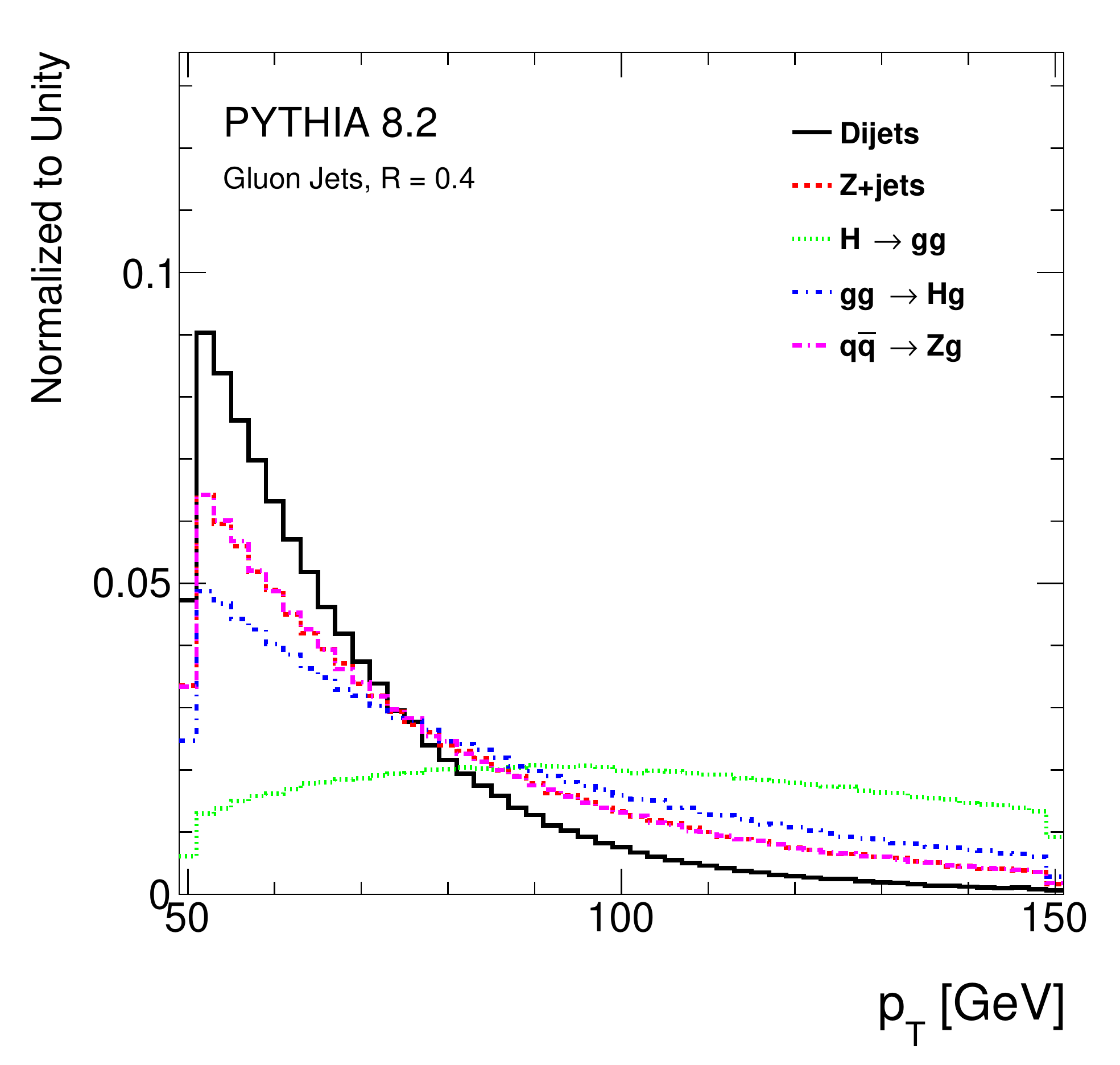}}
\\
\subfloat[]{\label{fig:0c}\includegraphics[width=0.49\textwidth]{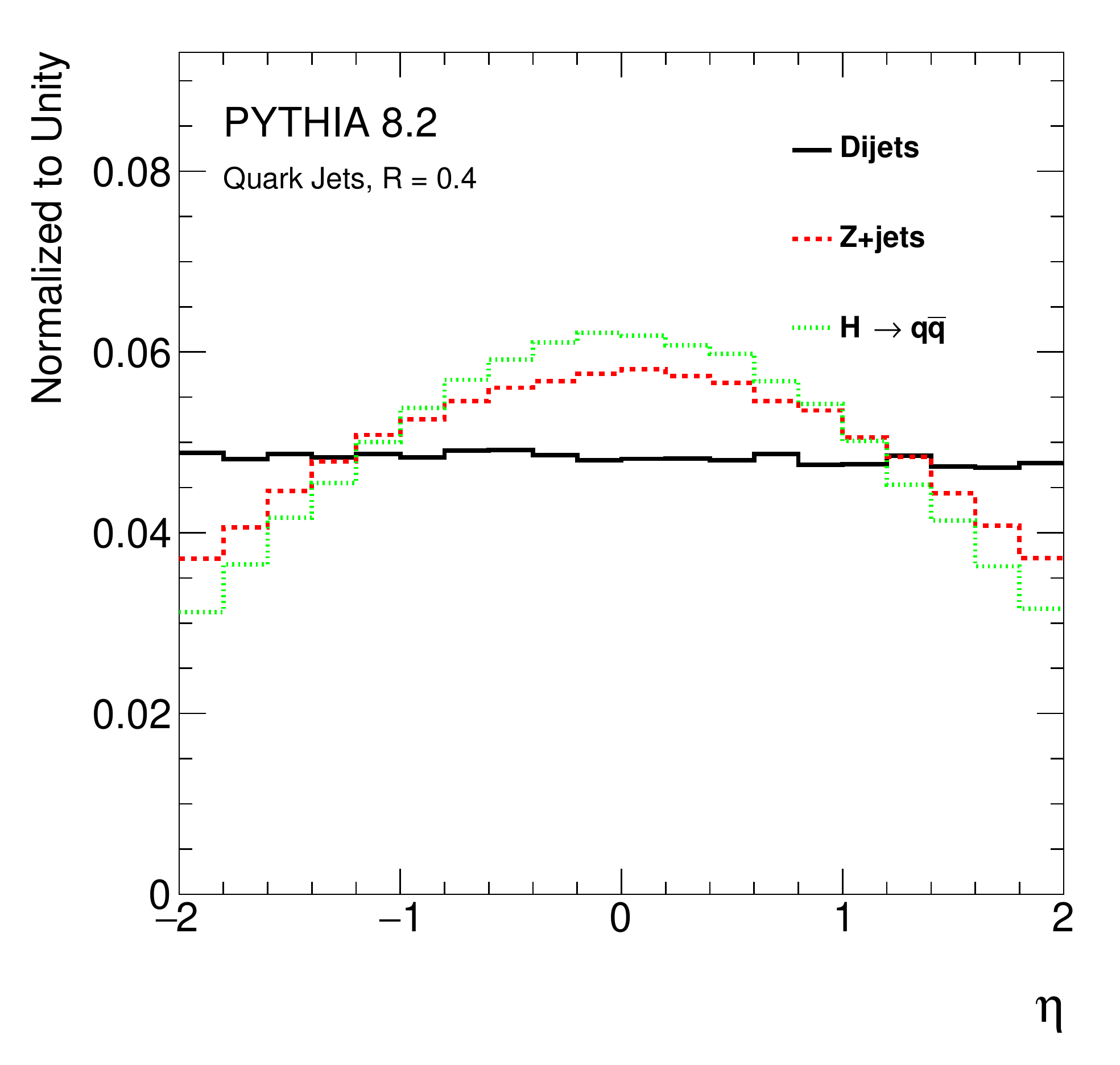}}
\hfill
\subfloat[]{\label{fig:0d}\includegraphics[width=0.49\textwidth]{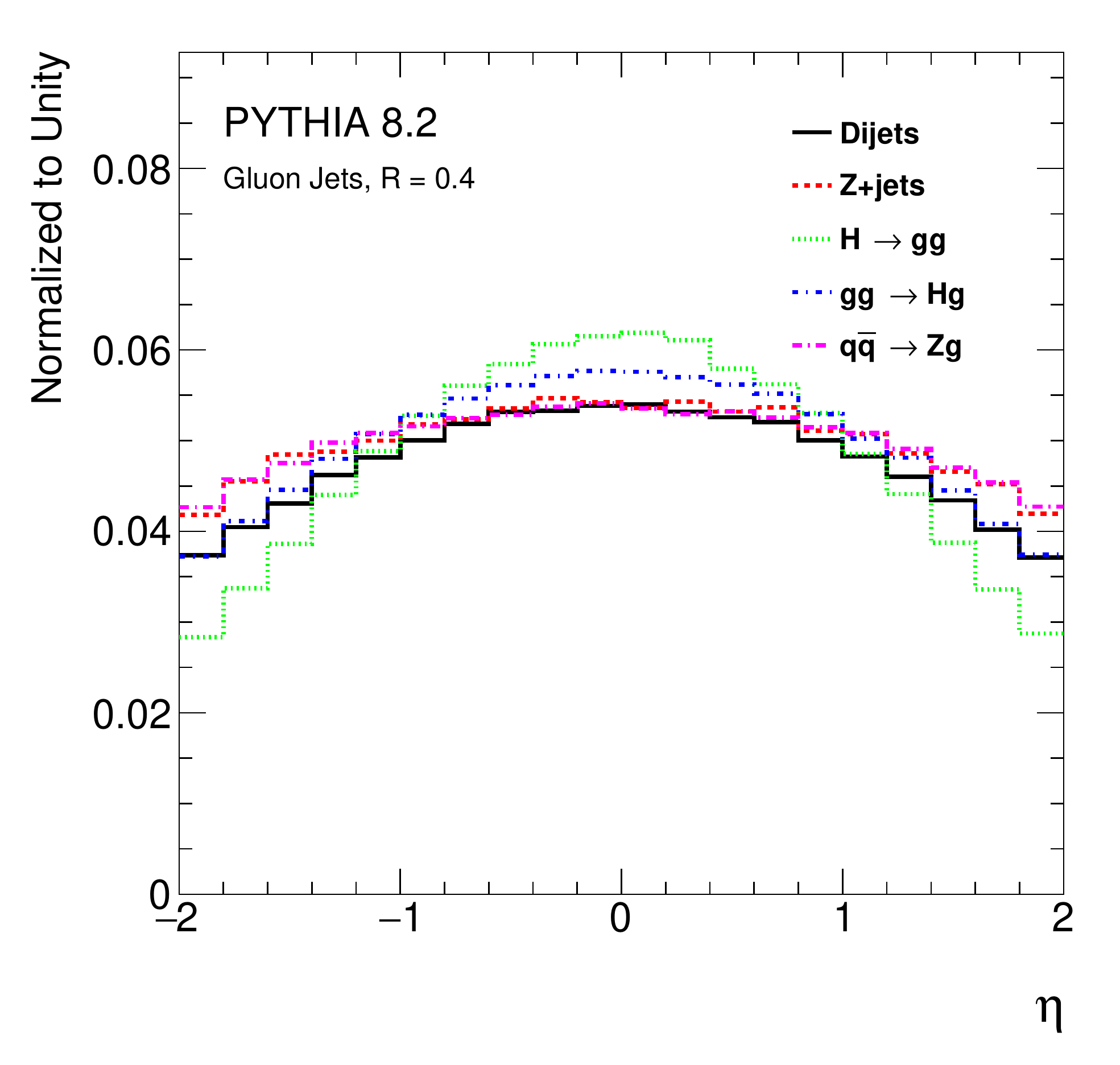}}
\caption{\label{fig:0} Plots showing normalized $p_{T}$ (top) and $\eta$ (bottom) distributions from different samples in the quark jet (left) and gluon jet (right) channels. These plots illustrate the shape differences that are rectified by the re-weighting procedure.} 

\end{figure}

 The radiation pattern inside jets depends on the jet $p_T$ and to a lesser extent on the jet $\eta$.  Differences in the $p_T$ and $\eta$ spectra between topologies therefore are also a source of trivial differences.  To remove this difference, the jet $p_T$ and $\eta$ spectra for each topology are re-weighted to match the (arbitrarily chosen) quark jet spectrum in the $Z$+jets sample.  When jet grooming is applied (Sec.~\ref{groomed}), jets are selected based on their un-groomed properties and the re-weighting is also performed with the un-groomed kinematic quantities.  Figure~\ref{fig:0} shows $p_T$ and $\eta$ distributions from each sample (normalized to unity) in the quark/gluon jet channels prior to re-weighting. The curves drawn in red in the quark jet channel (left-hand column of Fig.~\ref{fig:0}) correspond to the $Z$+jets samples, and are the distributions to which the other distributions are re-weighted for all subsequent studies.

\clearpage

\subsection{Results}

\begin{figure}[h!]
\centering
\subfloat[]{\label{fig:1a}\includegraphics[width=0.49\textwidth]{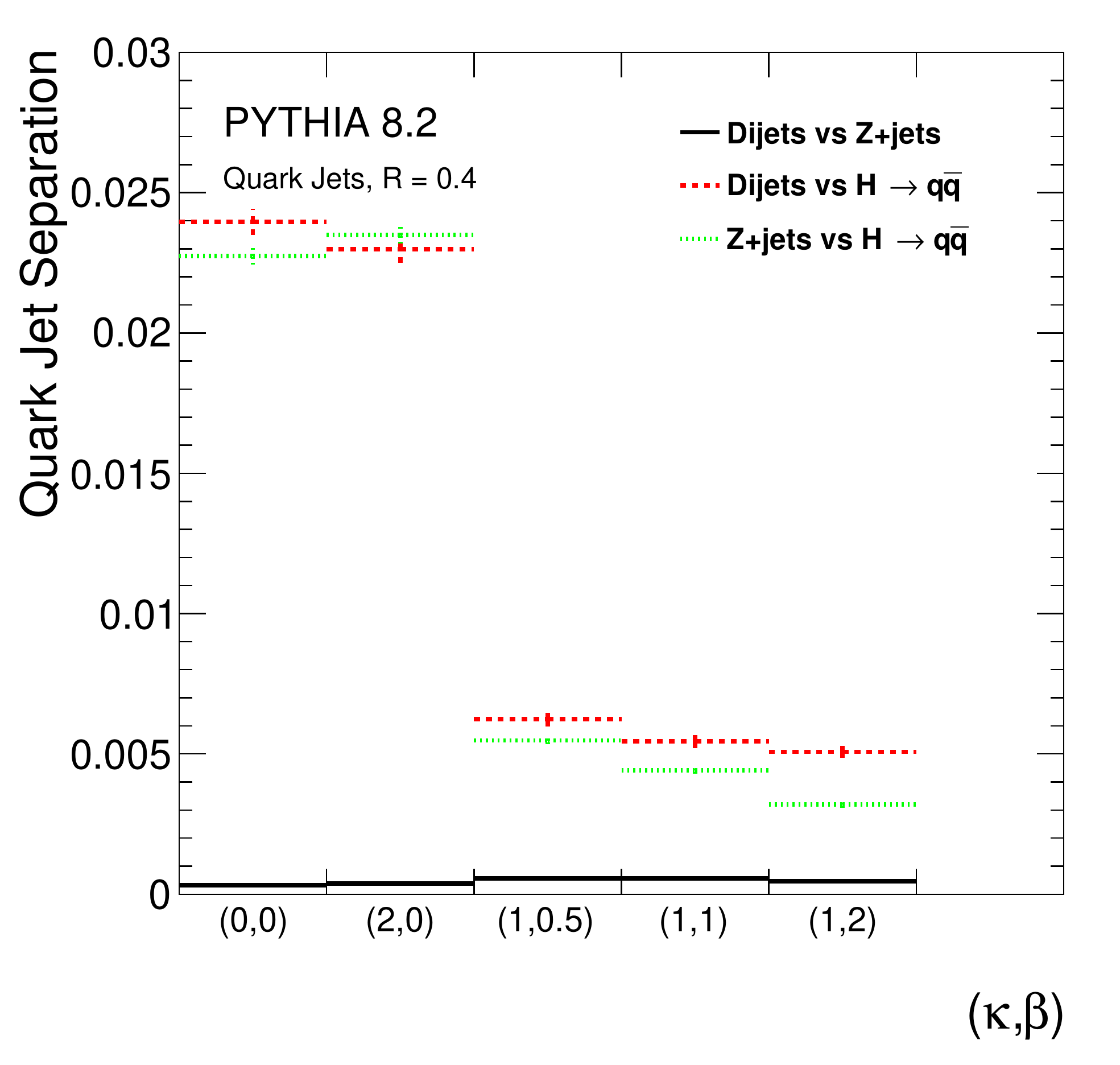}}
\hfill
\subfloat[]{\label{fig:1b}\includegraphics[width=0.49\textwidth]{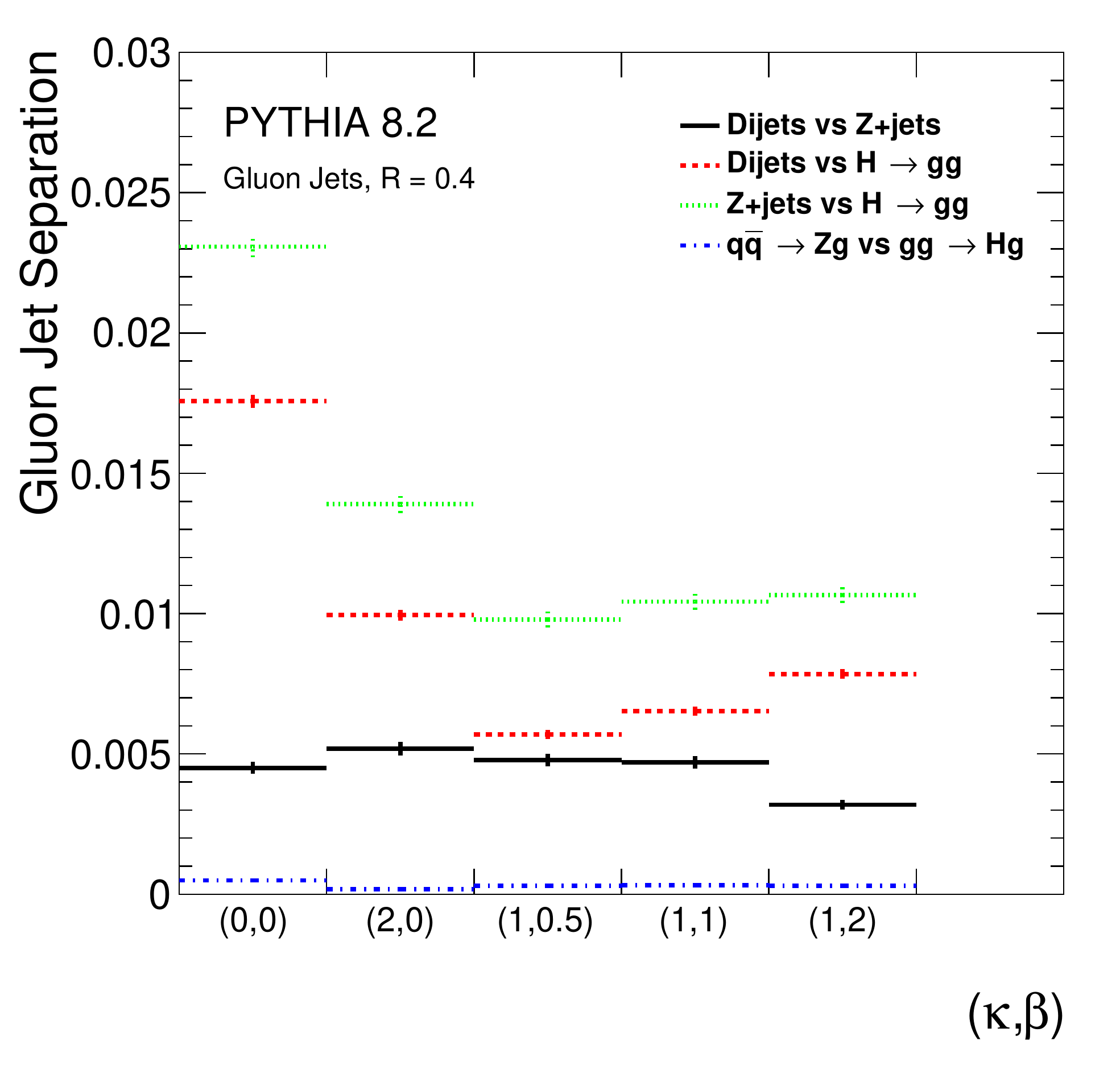}}
\\
\subfloat[]{\label{fig:1c}\includegraphics[width=0.49\textwidth]{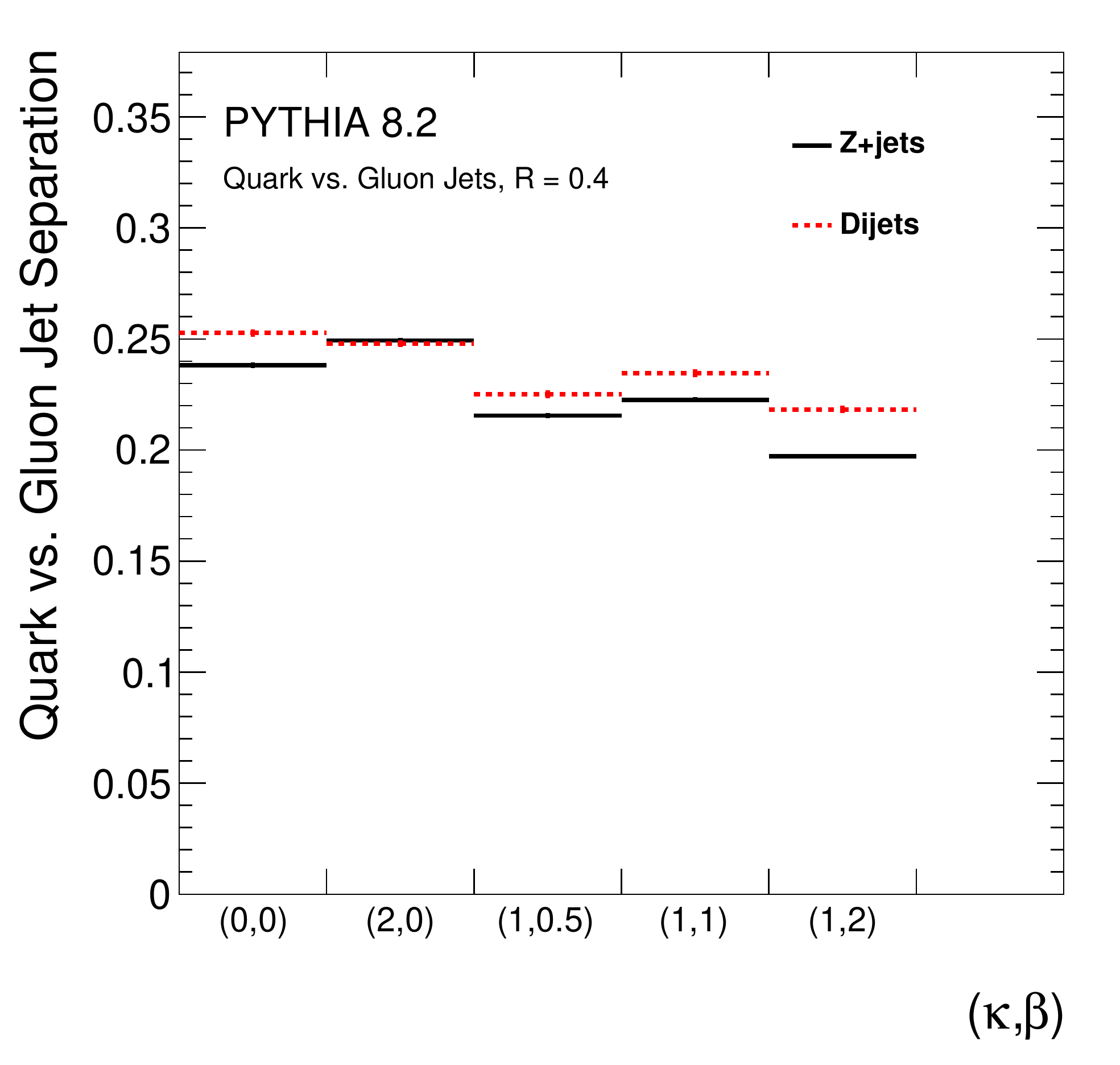}}
\caption{\label{fig:1} Classifier separation power $\Delta$ of the five different generalized angularities for (a) quark jets in different topologies, (b) gluon jets in different topologies, and (c) quark vs. gluon jets within a single topology. The plot in (c) provides benchmark values of $\Delta$ in a scenario where separation is expected, and the results in (a) and (b) can be compared against it. Error bars represent statistical uncertainty.}
\end{figure}

Figure \ref{fig:1} shows the classifier separation ($\Delta$) of the five generalized angularities for jets with radius $R = 0.4$. Following the style of Ref.~\cite{Gras:2017jty}, IRC unsafe angularities (multiplicity and $p_\text{T}^D$) are shown in the first two columns, and the IRC safe ones are shown in the last three columns. Figures~\ref{fig:1a} and~\ref{fig:1b} show same-flavor comparisons between quark jets and gluon jets, respectively, in different topologies.  In order to set the scale for $\Delta$, Fig.~\ref{fig:1c} shows the separation power for quarks versus gluons from the same topology and is similar to results presented in Ref.~\cite{Gras:2017jty}.  Compared with the quark versus gluon separation, the $\Delta$ for quark versus quark and gluon versus gluon is much smaller, for all topologies.  For example, the IRC safe angularities are separated at or below the 1\% level -- a factor of 10 or more below that of quark versus gluon jet tagging.  This means that for the purpose of quark versus gluon separation, the notions of quark and gluon jets are well-approximated as universal up to 10\% corrections.  For most searches at the LHC, variations on the order of 1\% in inter-topology separation are unlikely to have significant effects on quark versus gluon jet tagging performed at the 20--30\% separation level.  Precision measurements may consider this to be a significant effect that needs to be accounted for in the analysis.

\par  Even though $\Delta_{\text{$q$ vs. $q$}}$ and $\Delta_{\text{$g$ vs. $g$}}$ is much smaller than $\Delta_{\text{$q$ vs. $g$}}$, there is considerable variation for different angularities between pairs of topologies for $\Delta_{\text{$q$ vs. $q$}}$ and $\Delta_{\text{$g$ vs. $g$}}$.  For the quarks presented in Fig.~\ref{fig:1a}, dijets are much more similar to $Z$+jets (0.1\%) than to $H\rightarrow q\bar{q}$ (0.5-2\%).  The separation between $Z$+jets/dijets and $H\rightarrow q\bar{q}$ is larger for the IRC unsafe angularities (2\%) than for the IRC safe ones (0.5\%).  Similar trends are observed for gluons in Fig.~\ref{fig:1b}, though there are larger differences (0.5\%) between dijets and $Z$+jets and the jets are less separated for $p_T^D$ than for multiplicity.  The larger differences for IR unsafe observables suggests that soft radiation is driving the small, but clear differences between topologies.

\par
\begin{figure}[h!]
\centering
\subfloat[]{\label{fig:2a}\includegraphics[width=0.49\textwidth]{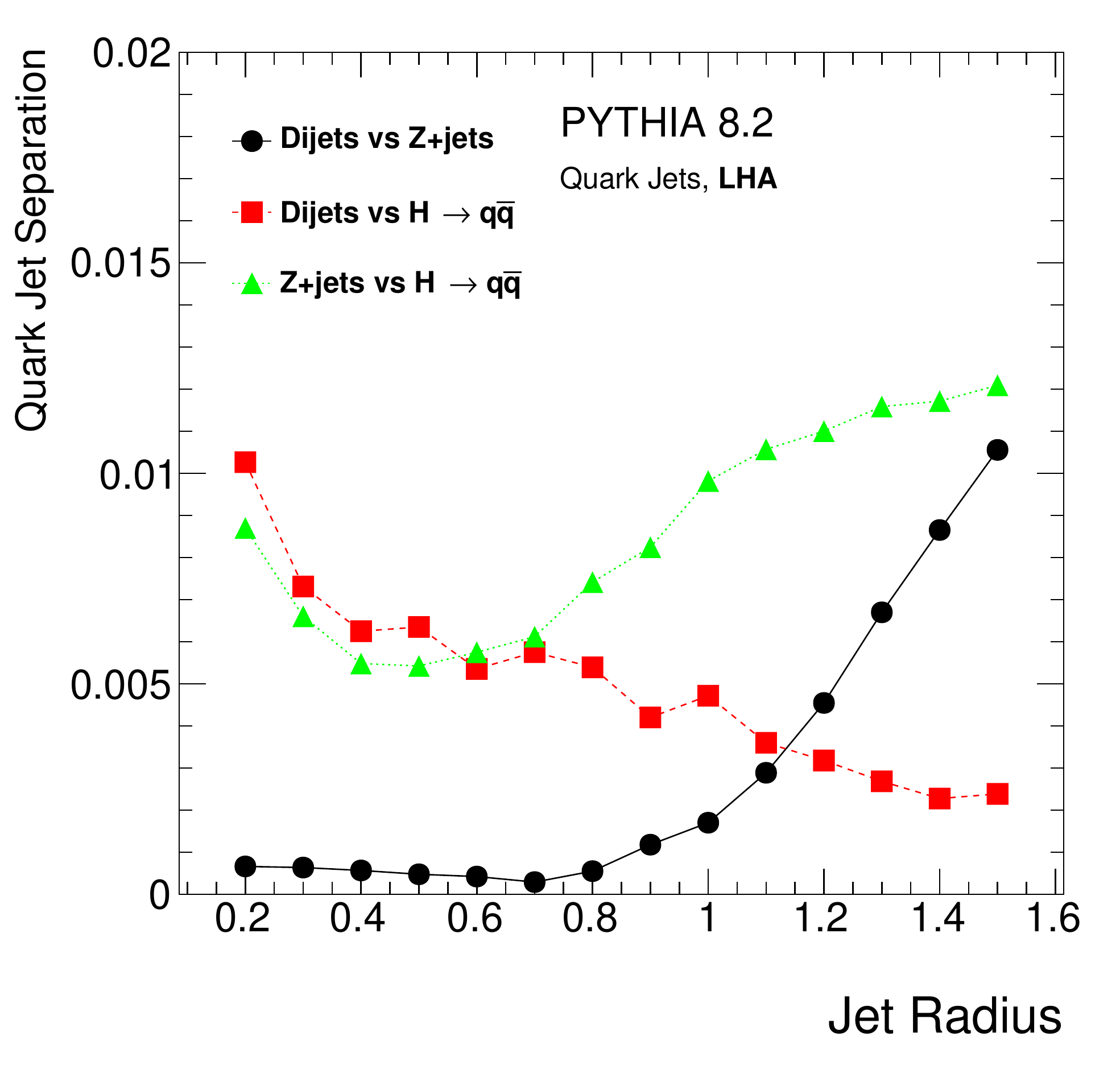}}
\subfloat[]{\label{fig:2d}\includegraphics[width=0.49\textwidth]{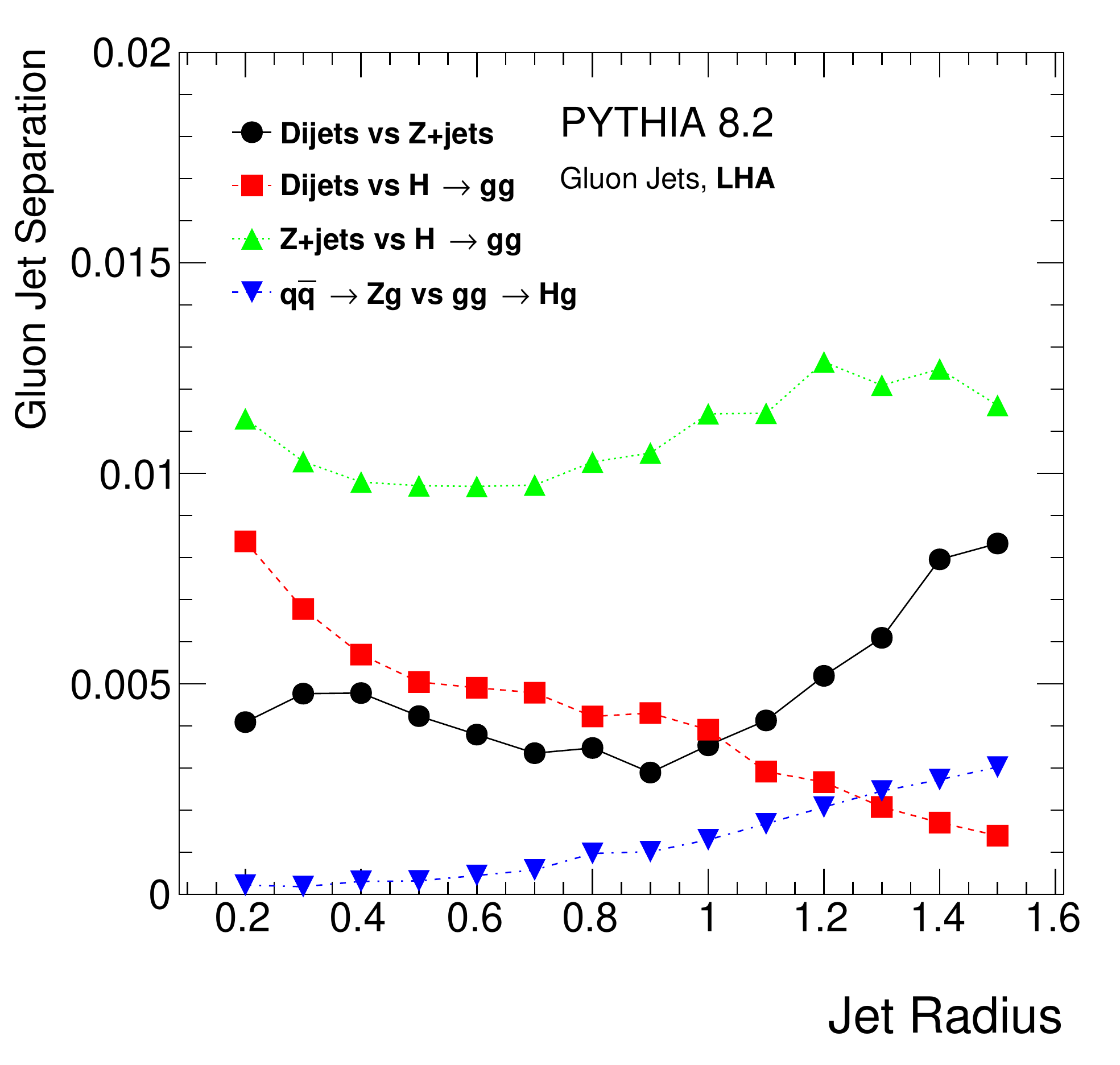}}
\caption{\label{fig:2} Separation power between quark jets (left) and gluon jets (right) from different topologies using LHA. The corresponding plots for mass and width look qualitatively the same.}
\end{figure}

 The radius dependence of classifier separations between topologies for LHA is presented in Fig.~\ref{fig:2}.  There is a strong radius dependence for most of the observables, though the separation does not exceed 1.5\%.  As may be expected from the larger catchment area to event-wide radiation, the quarks in $Z$+jets are more similar to the quark jets in dijets for small jet radii.  A similar trend is observed for gluons down to $R\sim 0.9$, but then the classifier separation becomes independent and even slightly increasing with decreasing jet radius for radii below $R=0.9$.  This increasing trend is observed around the same place for the other comparisons in both quark and gluon jets as well.  The increasing classifier separation with decreasing radius cannot be explained by the size of the catchment area to event-wide radiation.  The counter-intuitive trend could be a feature of the parton labeling scheme, which is also MC-dependent.  Evidence for this is presented in the later sections using an alternative parton labeling scheme (Sec.~\ref{sec:qcdaware}) and an alternative MC setup (Sec.~\ref{sec:her7}) and highlights one of the difficulties in defining quark and gluon jets at the \% level with respect to classifier separation.  Another feature of the radius dependence shown in Fig.~\ref{fig:2}, is that for $R>0.7$, the quark jets display the same splitting between $Z$+jets and dijets that is observed for gluon jets of all the considered radii.

\clearpage

\subsection{Jet grooming}
\label{groomed}

Grooming systematically removes jet constituents in order to reduce contamination from initial-state radiation (ISR), underlying event (UE), and multiple parton-parton/proton-proton collisions (MPI/pileup)~\cite{Butterworth:2008iy,Larkoski:2014wba,Ellis:2009su,Ellis:2009me,Krohn:2009th,Dasgupta:2013ihk}.  By removing radiation that is likely not from a particular parton, grooming may increase the universality of jet parton labels.  For example, groomed observables that are dominated by resummation (and not fixed order) effects are formally process independent when groomed with the soft drop~\cite{Larkoski:2014wba} algorithm~\cite{Frye:2016aiz}.  This also means that groomed jet shapes in $pp$ should be similar to the same observables in jets from $e^+e^-$ (see Sec.~\ref{sec:epem}).  In order to study the impact of grooming on the results presented in the previous section, jets are groomed using the soft drop algorithm with $\beta=0$ and $z_\text{cut}=0.1$ (which is identical to  the modified Mass Drop Tagger (mMDT) \cite{Dasgupta:2013ihk}).

\begin{figure}[h!]
\centering
\subfloat{\label{fig:3a}\includegraphics[width=0.5\textwidth]{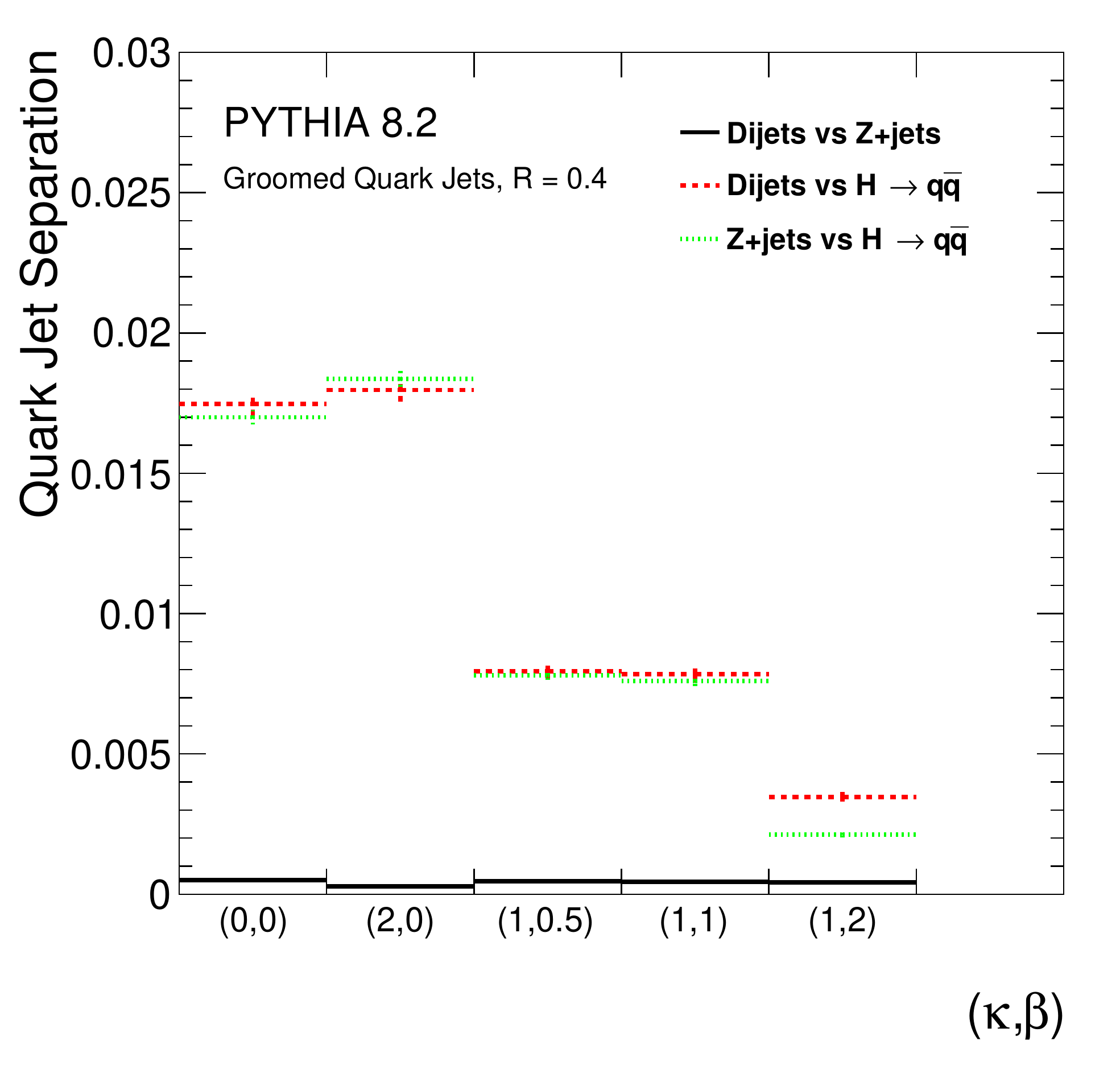}}
\hfill
\subfloat{\label{fig:3b}\includegraphics[width=0.5\textwidth]{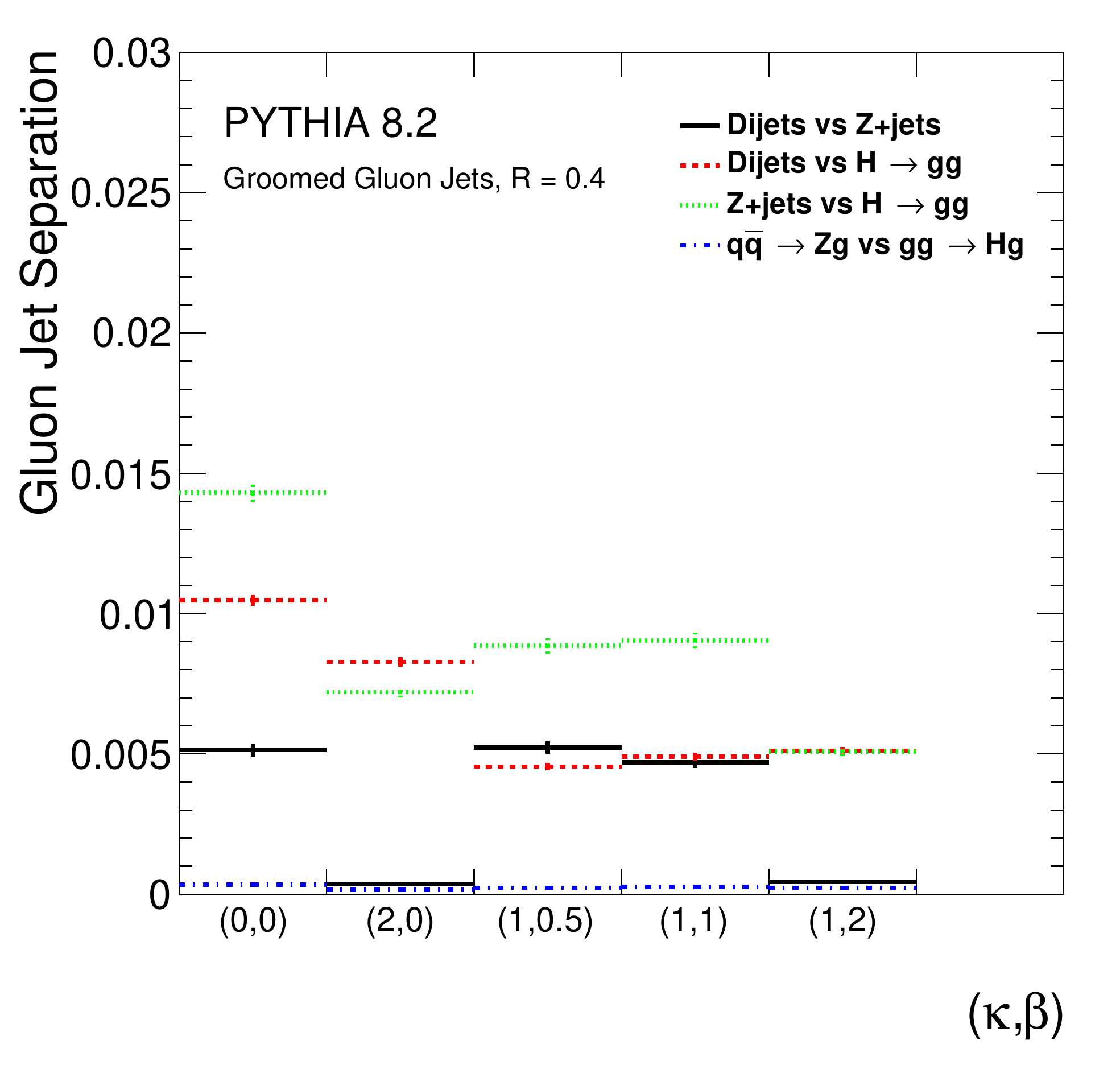}}
\caption{\label{fig:3}A reproduction of Figs.~\ref{fig:1a}-\ref{fig:1b} with soft drop grooming applied to the jets. The plots show classifier separation in the five generalized angularities of our study for (a) leading quark jets from different topologies and (b) leading gluon jets from different topologies.  Error bars represent statistical uncertainty.}
\end{figure}

Figure~\ref{fig:3} is the analog of Figs.~\ref{fig:1a}--\ref{fig:1b} using the same samples, but now with groomed jets.  Grooming reduces the separation power by about 25\% between topologies for the IRC unsafe observables for both quark and gluon jets.  The separation power for the IRC safe observables is about the same, except for mass, where it is reduced by about 50\% from the ungroomed case.  For the groomed jets, the separation power is much more similar across angularities than for ungroomed jets.  

\par
\begin{figure}
\centering
\subfloat[]{\label{fig:4a}\includegraphics[width=0.49\textwidth]{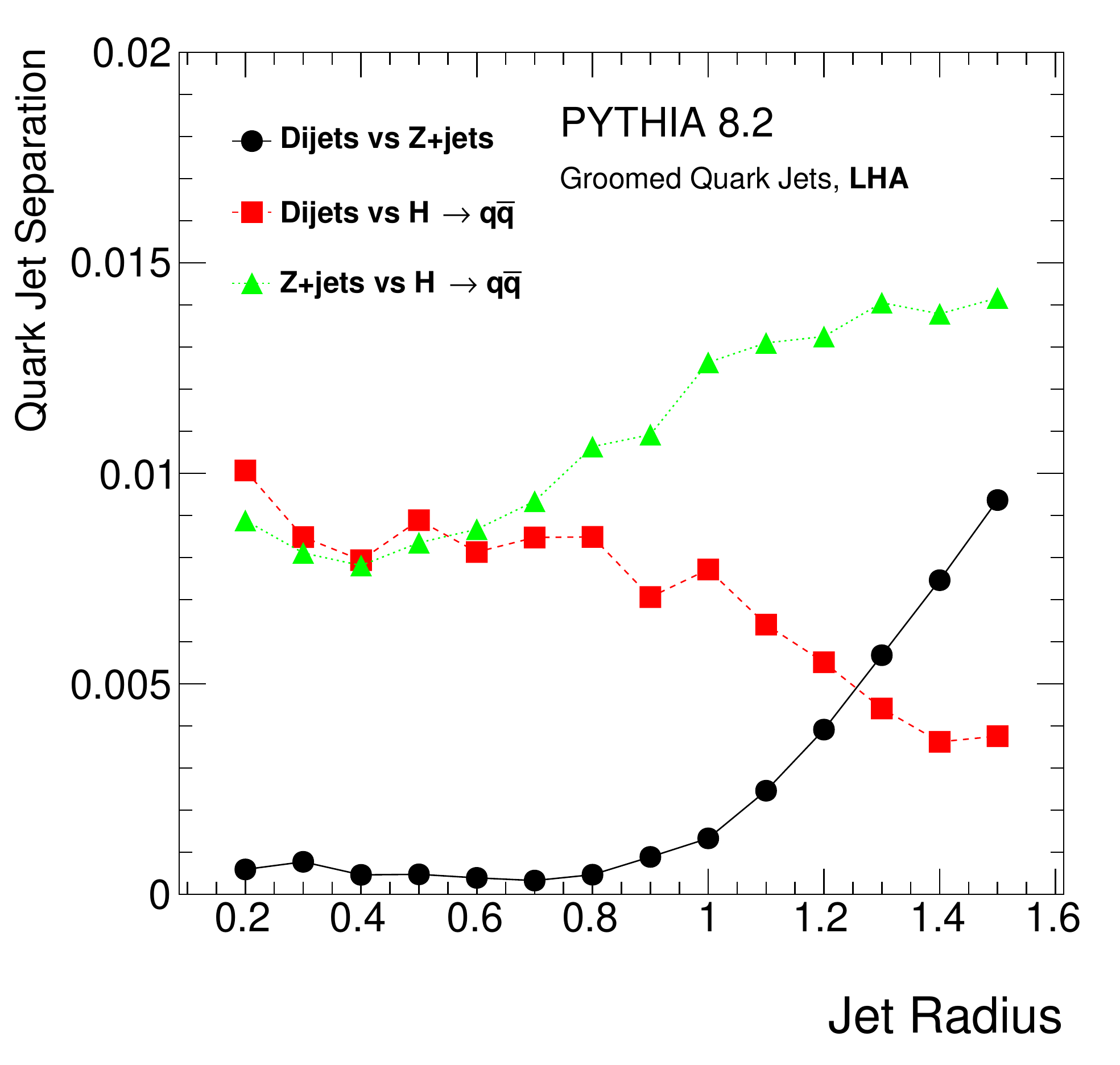}}
\hfill
\subfloat[]{\label{fig:4b}\includegraphics[width=0.49\textwidth]{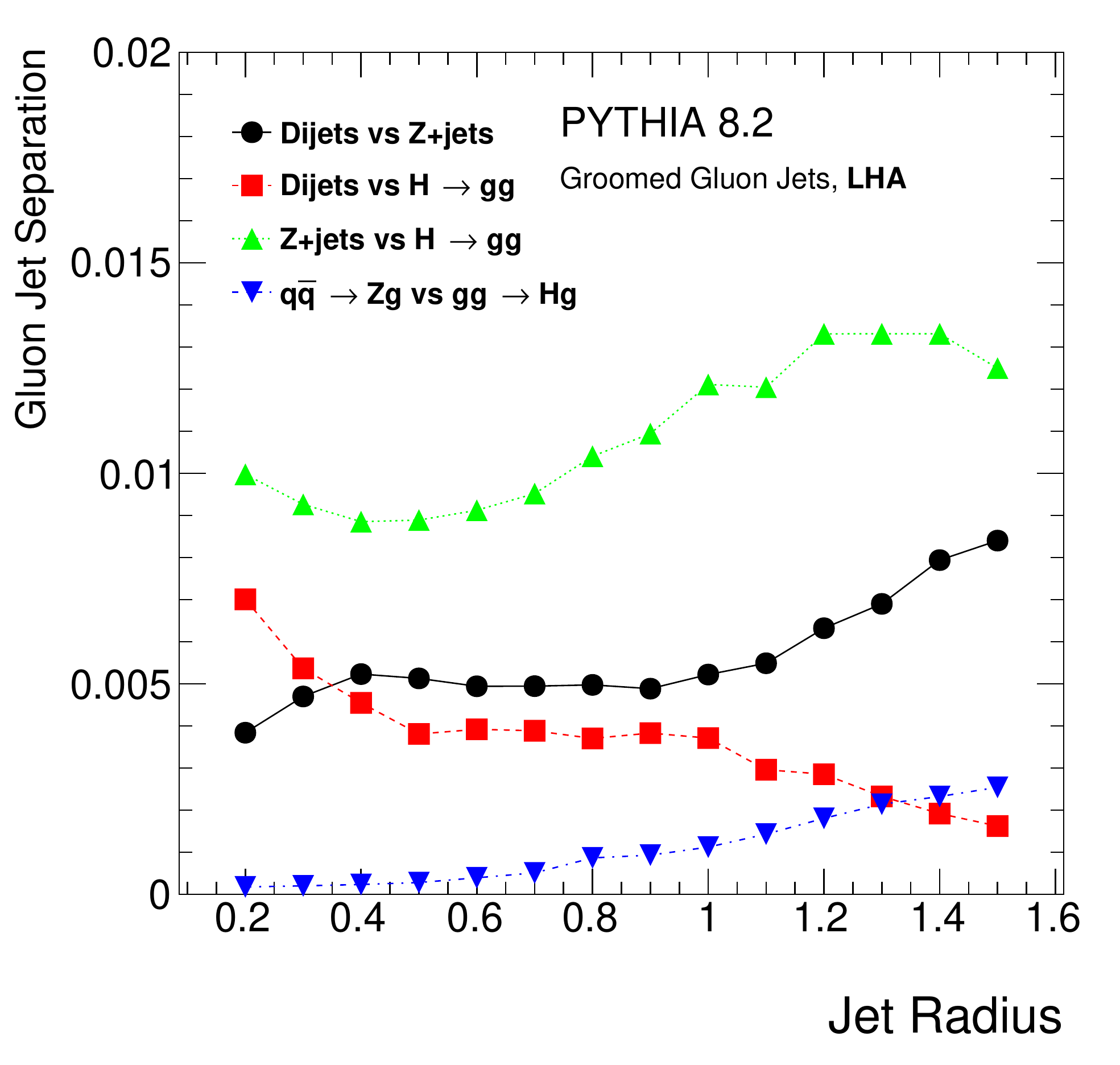}}
\caption{\label{fig:4} 
A reproduction of Fig.~\ref{fig:2} with soft drop grooming applied to the jets.  The plots show the separation power between quark jets (left) and gluon jets (right) from different topologies using LHA. The corresponding plots for mass and width look qualitatively the same.}
\end{figure}

 Figure~\ref{fig:4} is the analog to Fig.~\ref{fig:2}, but with groomed jets.  The jet radius dependence of the classifier separation for the IRC safe angularities is about the same for groomed jets as for ungroomed jets.  A notable exception is that the increasing separation for lower radii below about $R\sim1 $ for gluon jets has been eliminated.  Interestingly, the relative effect of grooming is nearly the same across radii and is not enhanced at the largest radii where the impact of contaminating radiation is largest.

\subsection{$pp$ vs. $e^+e^-$}
\label{sec:epem}

\begin{figure}[h!]
\centering
\subfloat[]{\label{fig:5a}\includegraphics[width=0.49\textwidth]{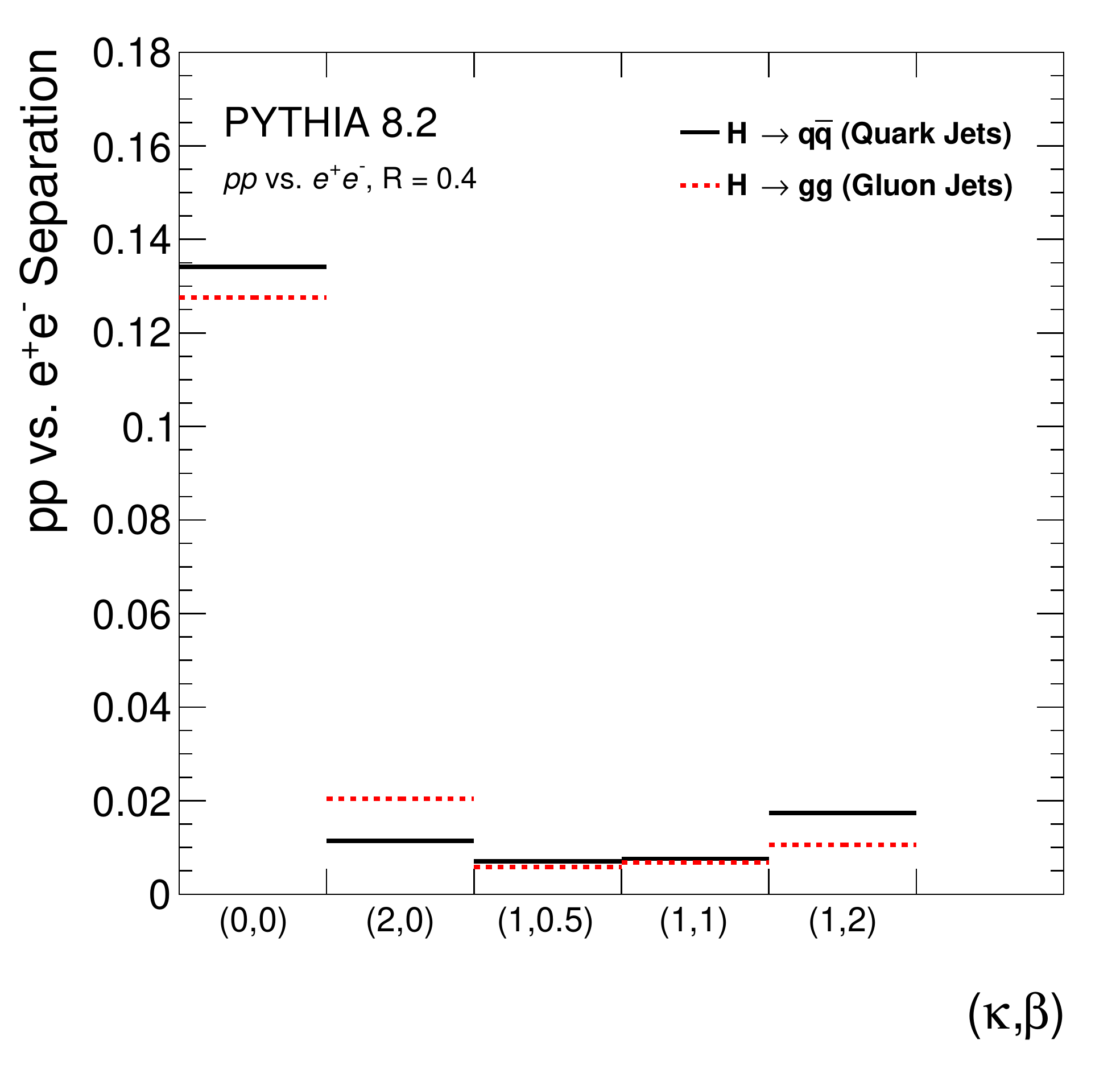}}
\hfill
\subfloat[]{\label{fig:5b}\includegraphics[width=0.49\textwidth]{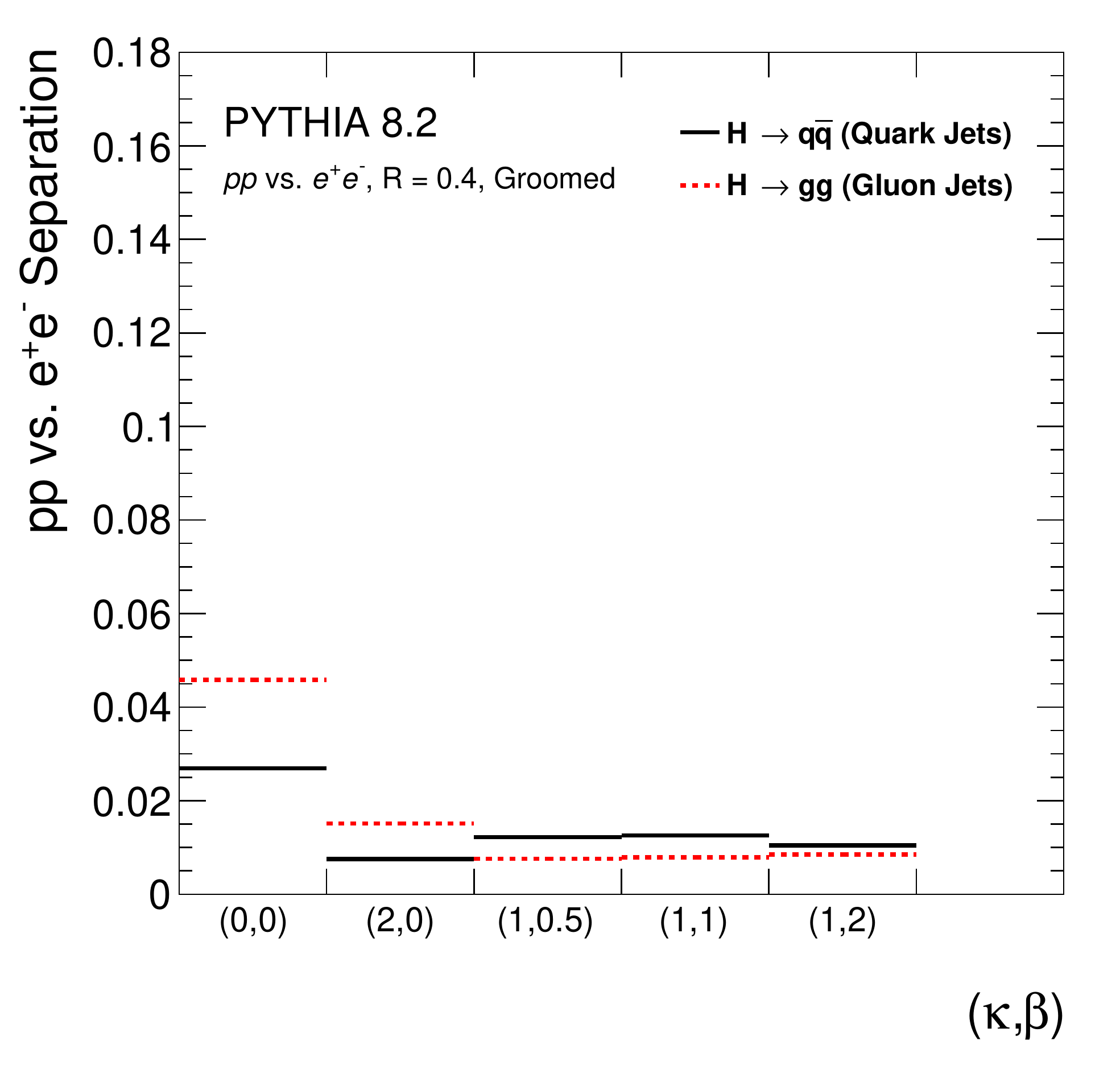}}
\caption{\label{fig:5}Classifier separation $\Delta$ of the five generalized angularities between jets of the same type from $pp$ and $e^+e^-$ collisions. Results are shown using (a) un-groomed jets and (b) groomed jets.}
\end{figure}

Electron-positron collisions lack the initial-state complexity of proton-proton collisions, providing an idealized environment to study jets due to the absence of ISR, UE, and pileup.  Jets produced in this clean environment are expected to be as different as possible than their $pp$ counterparts.   This is demonstrated quantitatively in Fig.~\ref{fig:5}, using $H \rightarrow q\bar{q}$ and $H \rightarrow gg$ in both $e^+e^-$ and $pp$ collisions. The $e^+e^-$ samples were generated with a center-of-mass energy $E_\text{CM} = 200$ GeV (equal to the Higgs mass), but were otherwise treated exactly like the $pp$ samples for jet clustering, $p_T$ and $\eta$ re-weighting, and kinematic cuts.  In contrast to Fig.~\ref{fig:1} and Fig.~\ref{fig:3}, the classifier separations shown in Fig.~\ref{fig:5} are much larger (though still well below the $q/g$ separation from Fig.~\ref{fig:1c}).   For multiplicity, the difference is nearly a factor of six, while it is only about a factor of two for the IRC safe angularities.  While multiplicity and $(p_\text{T}^D)^2$ behaved similarly in $pp$, multiplicity is much more different between $pp$ and $e^+e^-$.  This could be because $(p_\text{T}^D)^2$ is IR safe and so the contaminating soft radiation is suppressed.  Grooming significantly reduces the classifier separation for multiplicity, but has little effect on the IR(C) safe observables, for which $1\%\lesssim \Delta\lesssim 2\%$ for both the groomed and ungroomed jets.

\section{Investigating $p_\text{T}$ dependence}
\label{ptdep}

The low jet $p_\text{T}$ studies in Sec.~\ref{sec:base} showed that differences between same-flavor jets in different topologies were much smaller than differences between opposite-flavor jets.  This section examines the behavior of higher $p_\text{T}$ jets ($200 < p_\text{T}^{\text{jet}} < 350$ GeV).  Since contaminating radiation and other sources of non-universality are expected to be relatively soft, it is expected that higher $p_\text{T}$ jets will be more universal than low $p_\text{T}$ jets.  To test this hypothesis, for all topologies not involving the Higgs, the $\hat{p}_T$ range is changed to $160 \leq \hat{p}_T \leq 400$ GeV (all other settings are as described in Sec.~\ref{sec:base}).  For the topologies involving the Higgs, two configurations are used in order to probe different kinematics.  First, a sample is generated with $m_H=1$ TeV, which is essentially the same as the sample from Sec.~\ref{sec:base}, only it produces harder jets.  A second sample uses $m_H = 100$ GeV and a $\hat{p}_T$ range of $300 < \hat{p}_T < 900$ GeV.  This second sample produces boosted Higgs bosons whose daughter jets are collimated.  The presence of nearby jets originating from color-connected partons is known to distort a jet's substructure~\cite{Gallicchio:2010sw,Aad:2015lxa,Aaboud:2018ibj,Abazov:2011vh} and is thus a source of non-universality that can be probed with this setup.  In order to avoid cases where all of the Higgs decay produces are collected into a single jet (relevant especially for larger jet radii), only those jets with $m_{\text{jet}} < 80$ GeV are considered.  Constraining the jet mass in this manner may have an effect on classifier separation that is independent of the topology. Lower-mass jets tend to be more quark-like, and selecting jets from this subset could alter the fraction of mislabeled jets and change the classifier separation. The boosted Higgs case will be referred to as $H'$ throughout the rest of the section.

\par
\begin{figure}[h!]
\centering
\subfloat[]{\label{fig:6a}\includegraphics[width=0.49\textwidth]{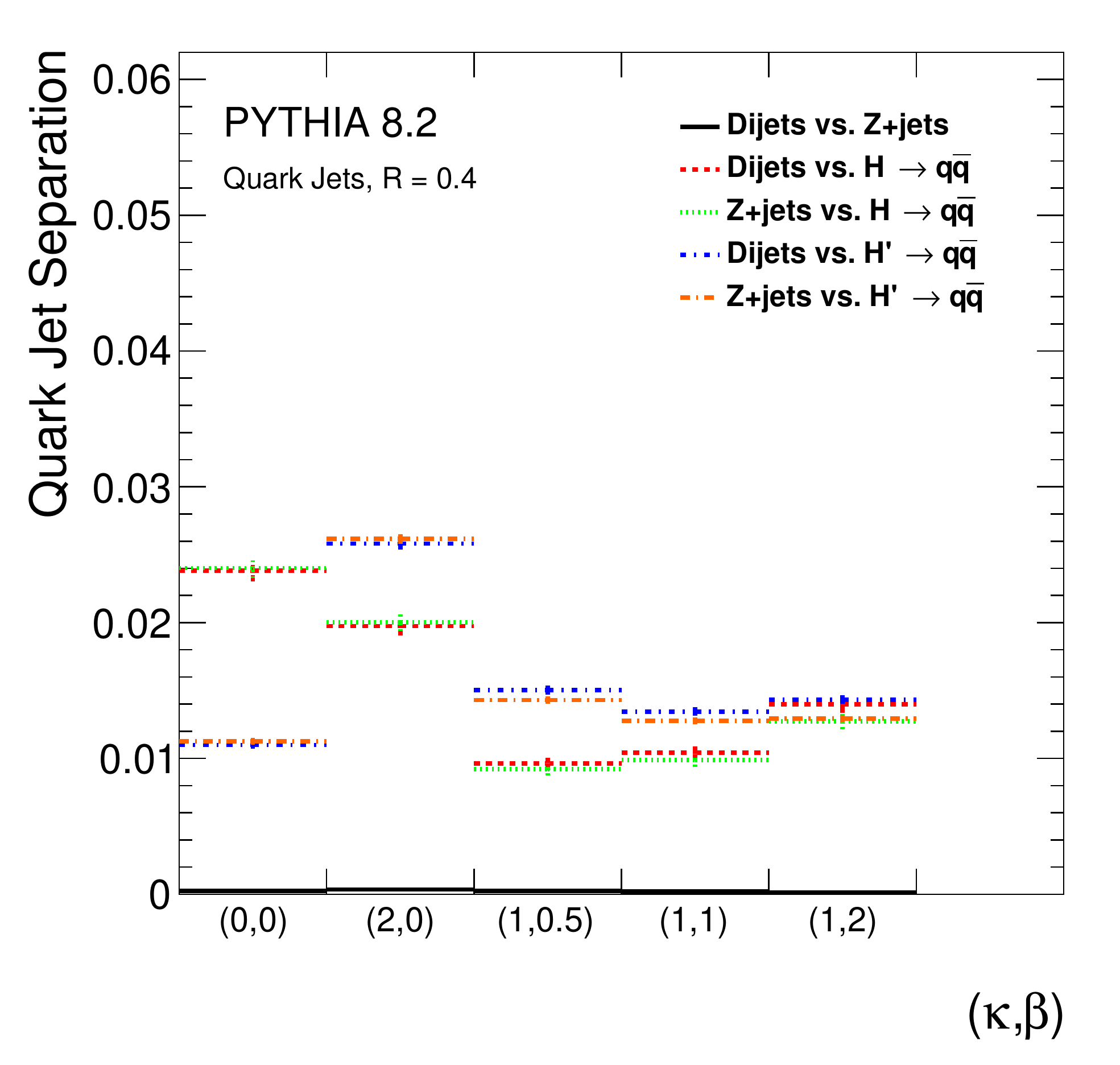}}
\hfill
\subfloat[]{\label{fig:6b}\includegraphics[width=0.49\textwidth]{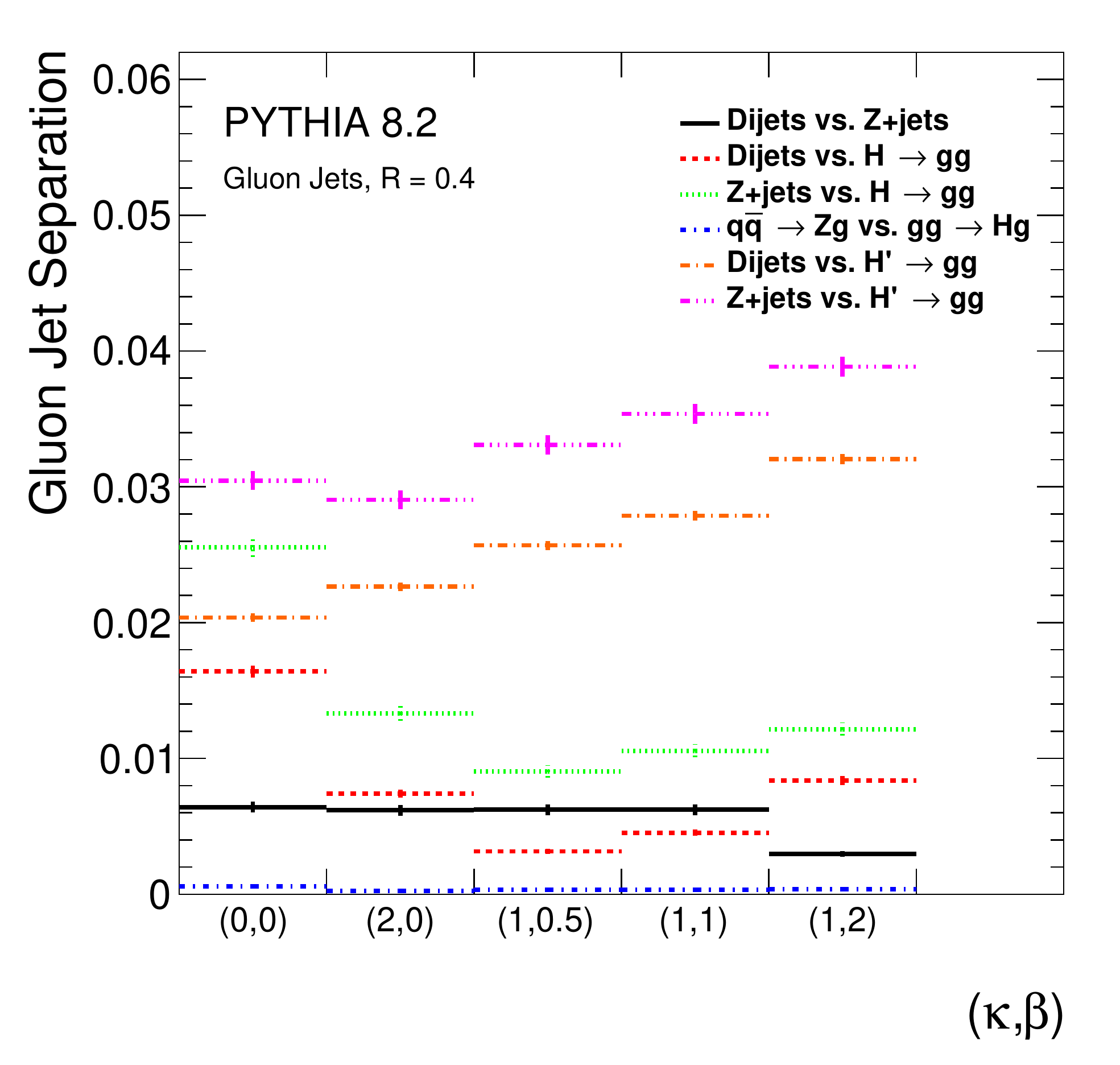}}
\\
\subfloat[]{\label{fig:6b}\includegraphics[width=0.49\textwidth]{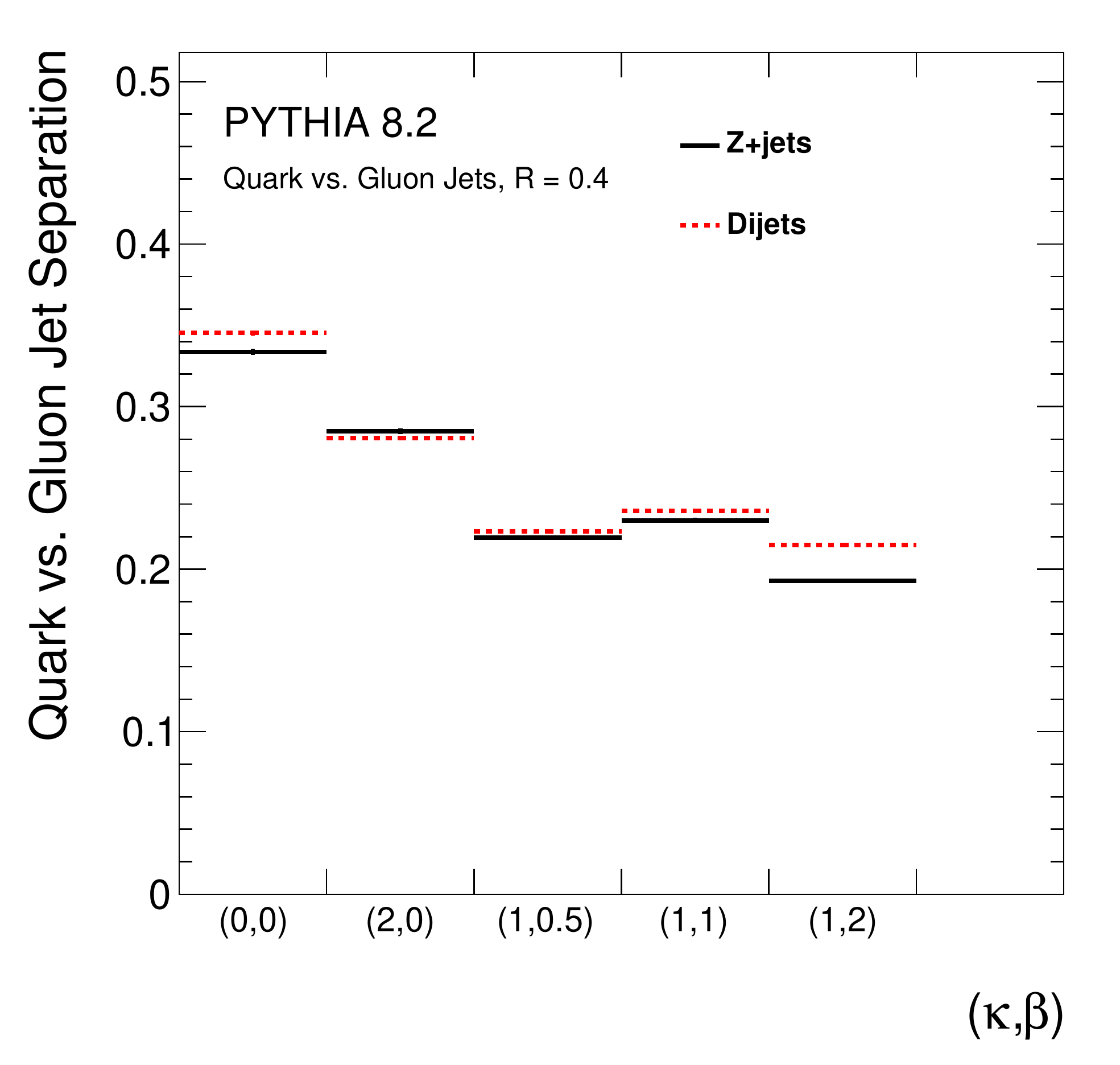}}
\caption{\label{fig:6}Classifier separation  $\Delta$ of the five generalized angularities between leading (a) quark jets and (b) gluon jets from different topologies. Jets have radius $R = 0.4$, and are selected with $200 < p_\text{T}^{\text{jet}} < 400$ GeV and $|\eta| < 2.0$. Higgs samples labeled with $H'$ denote $m_H = 100$ GeV, whereas the label $H$ denotes the default $m_H = 1$ TeV.   Error bars represent statistical uncertainty.}
\end{figure}

 The classifier separation for the various angularities in the high $p_\text{T}$ sample (to be compared with the low $p_\text{T}$ case in Fig.~\ref{fig:1}) are presented in Fig.~\ref{fig:6}.  The trends for high $p_\text{T}$ are nearly the same as for low $p_\text{T}$, with slightly higher classifier separation for the IRC safe observables for quark jets and slightly lower for gluon jets.  Each plot in Fig.~\ref{fig:6} has two new lines with respect to Fig.~\ref{fig:1} from the boosted Higgs topologies.   For the IR(C) safe angularities, the classifier separation between dijets/$Z$+jets and $H \rightarrow q\bar{q}/gg$ is larger for the boosted Higgs compared with the high mass Higgs.  This is also true for multiplicity for gluons but not for quarks. As mentioned above, the jet mass selection applied to the boosted Higgs samples may be the source of this difference. Another difference between quarks and gluons is that the classifier separation is nearly independent of the angular weighting for quarks for the IRC safe angularities while it increases with increasing angular weighting for gluons.  Despite the increased classifier separation for the boosted Higgs case, the overall separation is still below a few percent, well below the q/g separation. 

\graphicspath{ {./highpt_figs/} }

\clearpage

\section{QCD-aware jet flavor tagging}
\label{sec:qcdaware}

Some of the apparent topology effects observed in previous sections may be due in part to artifacts of the parton labeling scheme and generator dependence (Sec.~\ref{sec:her7}).  In order to probe the impact of the parton labeling scheme, this section explores an alternative scheme known as the QCD-aware method~\cite{Buckley:2015gua}.  This alternative scheme has been shown to be relatively robust to variations between and within MC models.  The algorithm acts on parton jets, formed from the final partons produced by a generator before hadronization begins.  These partons have the lowest virtuality and the resulting labels reduce the dependence on many features of the generation.  Partons are clustered with a modified version of the anti-$k_t$ algorithm that incorporates information about the QCD and QED Feynman rules.  The hadron-level jets used in this analysis are assigned a QCD-aware label using the label of the nearest parton-level jet.

\begin{figure}[h!]
\centering
\subfloat[]{\label{fig:8a}\includegraphics[width=0.33\textwidth]{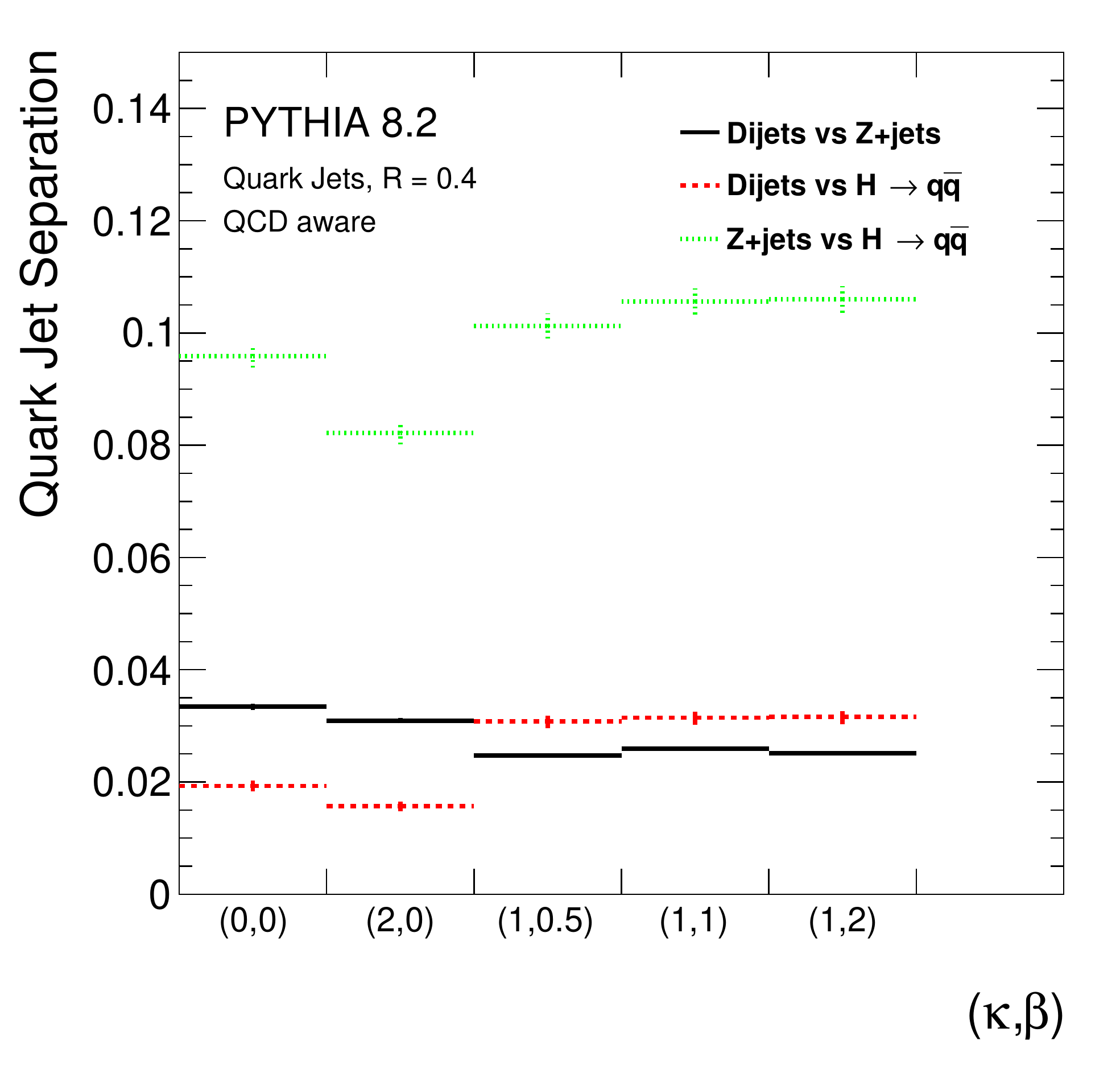}}
\hfill
\subfloat[]{\label{fig:8b}\includegraphics[width=0.33\textwidth]{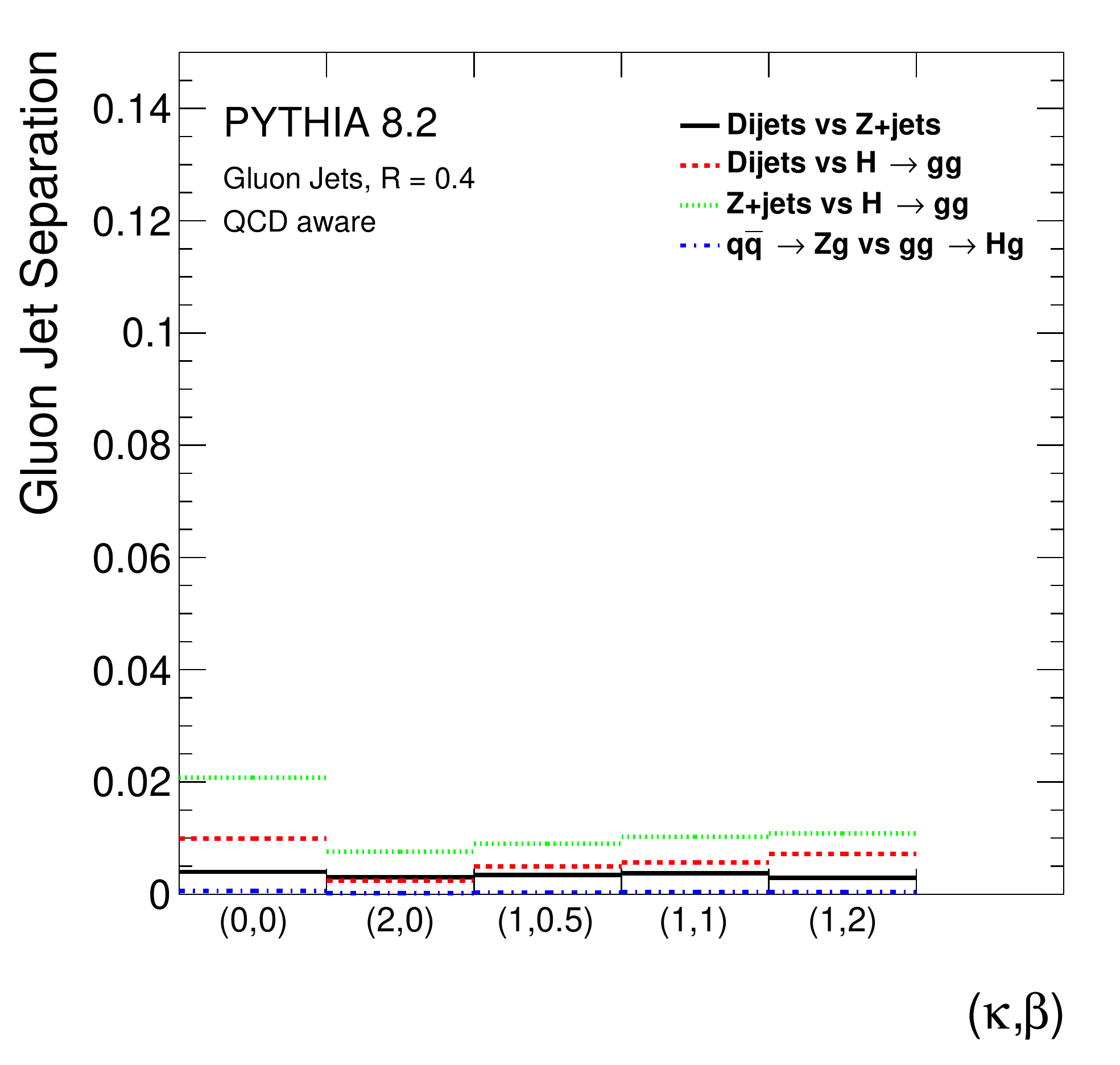}}
\subfloat[]{\label{fig:8c}\includegraphics[width=0.33\textwidth]{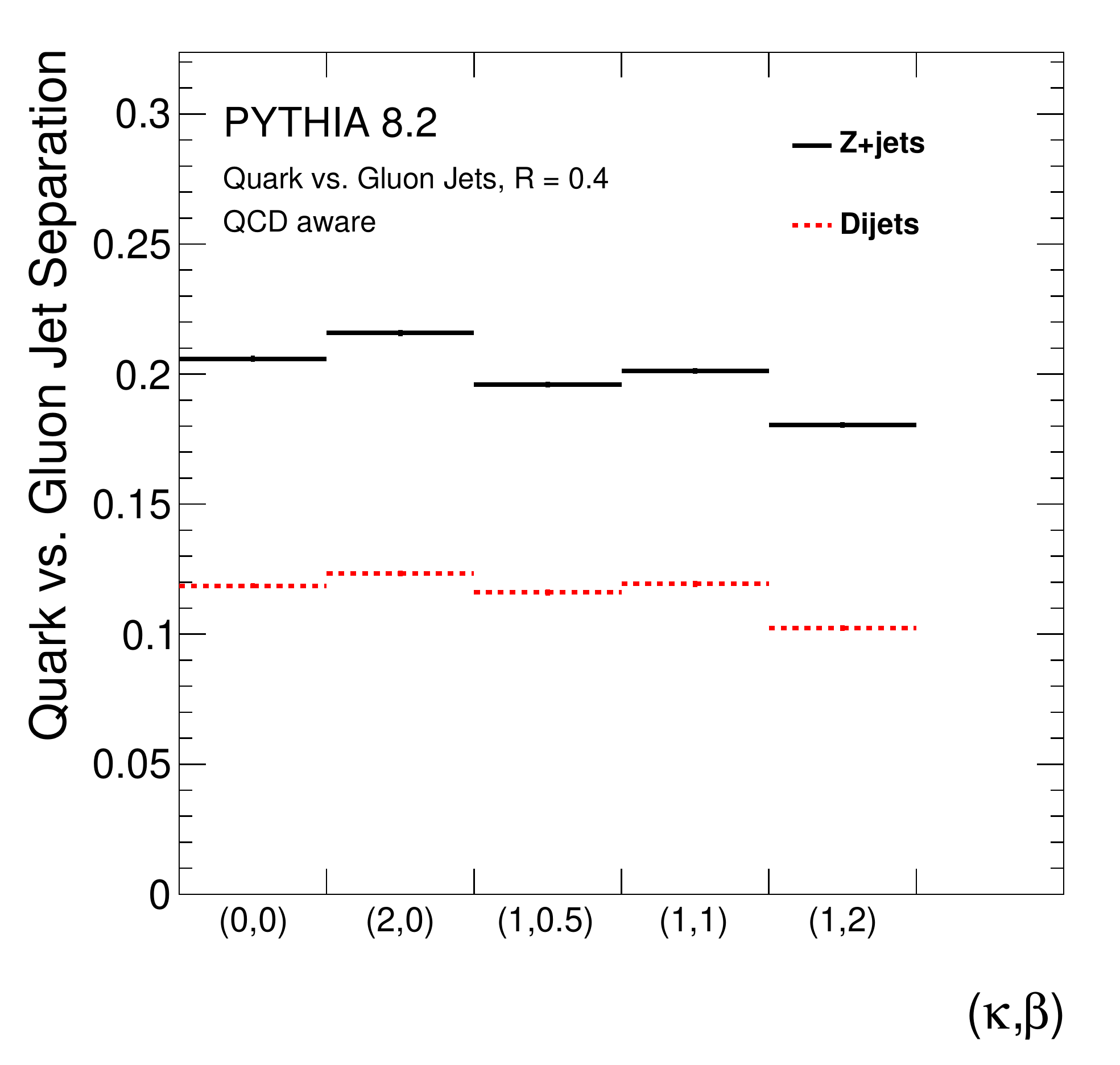}}
\\
\subfloat[]{\label{fig:8d}\includegraphics[width=0.5\textwidth]{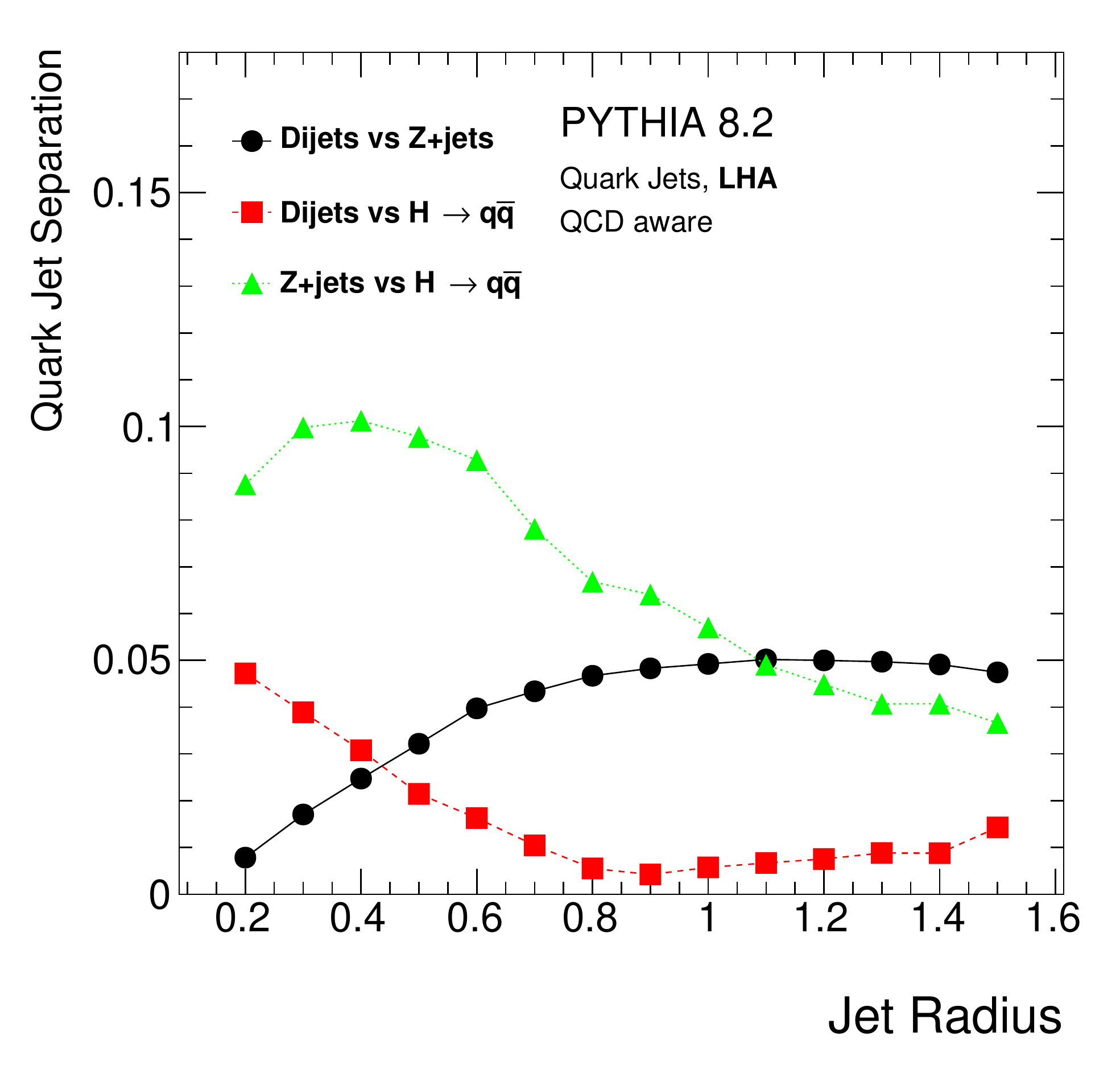}}
\hfill
\subfloat[]{\label{fig:8e}\includegraphics[width=0.5\textwidth]{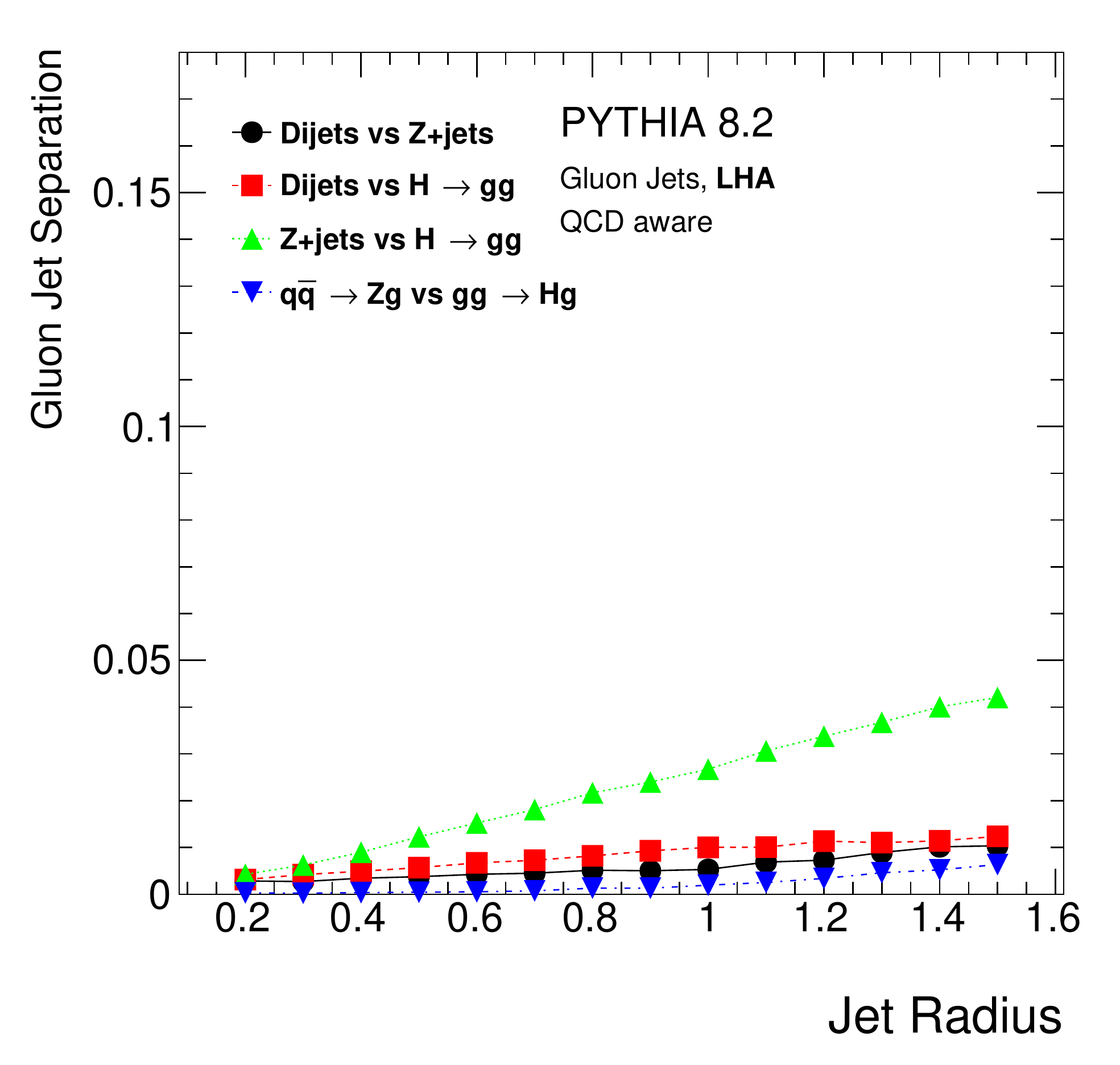}}
\caption{\label{fig:8} Replications of plots from Figs.~\ref{fig:1} and \ref{fig:2} using quark/gluon jets tagged with the QCD aware flavor tagging method. The plots in (a)--(c) replicate Figs.~\ref{fig:1a}--\ref{fig:1c}, showing separation power between various pairs of topologies as a function of angularity at a jet radius of $R = 0.4$. The plots in (c) and (d) replicate \ref{fig:2a} and \ref{fig:2d}, showing how separation power in the LHA angularity varies as a function of jet radius in the quark and gluon jet channels, respectively. Error bars represent statistical uncertainty.}
\end{figure}

The classifier separation for various angularities and a scan in the jet radius is shown in Fig.~\ref{fig:8}.  First, the QCD-aware method predicts a different baseline quark-versus-gluon jet separation compared with the labeling scheme from previous sections (Fig.~\ref{fig:8c}).  The trend as a function of $\kappa$ and $\beta$ is nearly identical, but the overall separation is slightly lower for $Z$+jets and about a factor of two lower for dijets.  While the same-parton comparisons still have classifier separations that are generally much lower than this, the $Z$+jets versus $H\rightarrow q\bar{q}$ for quarks is an exception -- now about 10\% - same as $q$ versus $g$ in dijets and about ten times more than before.  The other comparisons are at or below about 3\%.  In the QCD-aware scheme, dijets and $Z$+jets are generally more different than the scheme used in previous sections.  Furthermore, the difference between the IR(C) safe and unsafe angularities is smaller with the QCD-aware algorithm.

Many of the features shown in the radius dependence (Fig.~\ref{fig:8d} and~\ref{fig:8e}) for the QCD-aware scheme are also qualitatively different than the baseline method.  In particular, the classifier separation for gluon jets now goes to zero for all methods at low radius (closer to what is expected).  However, this is not the case for quark jets and the $Z$+jets/dijets versus $H\rightarrow q\bar{q}$ comparison, which is more like the default parton tagging method.  The ordering of the topologies by classifier separation is the same for QCD-aware algorithm and the default scheme.

\graphicspath{ {./figs_QCDa/} }

\par Overall, the results of the QCD-aware tagging method differ significantly from the results of the default method\footnote{A direct comparison between the default and QCD-aware tagging results are presented in Appendix~\ref{sec:labeling} and indicates that while they usually label jets the same, there is a large fraction of the time where the two do not agree on the label.}, and while many of the qualitative trends are similar between methods, the observed differences underscore the difficulty of determining a robust definition of jet flavor. 
\clearpage

\section{\texttt{MadGraph5} and \texttt{HERWIG 7}}
\label{sec:her7}

As noted earlier, some of the apparent topology effects could be due to the chosen generator (as both the labeling and radiation patterns are model-dependent) so it is important to compare with a different MC.  For this purpose, events are generated with \texttt{MadGraph5 2.6.3.2}~\cite{Alwall:2014hca} for the matrix elements and \texttt{HERWIG 7.1.3} for fragmentation~\cite{Bellm:2015jjp,Reichelt:2017hts}. \texttt{MadGraph5} was run with jet parameters of $45 < p_\text{T}^\text{jet} < 200$ GeV and $\Delta R_\text{jj} > 0.4$. All event generation parameters are the same as described in Sec.~\ref{sec:base}, and jet flavor-tagging is done using the default parton-matching method.

\begin{figure}
\centering
\subfloat[]{\label{fig:9a}\includegraphics[width=0.33\textwidth]{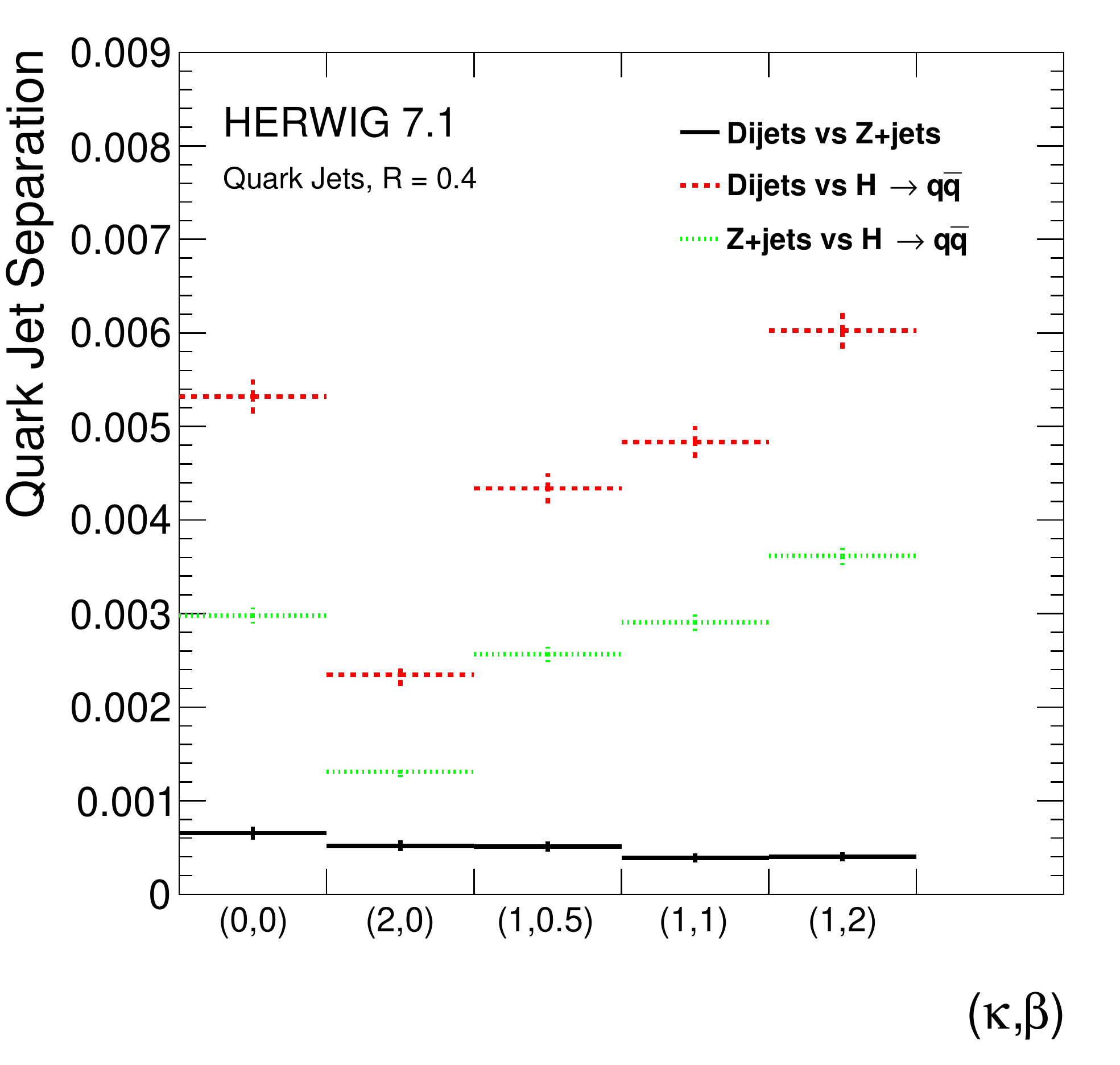}}
\hfill
\subfloat[]{\label{fig:9b}\includegraphics[width=0.33\textwidth]{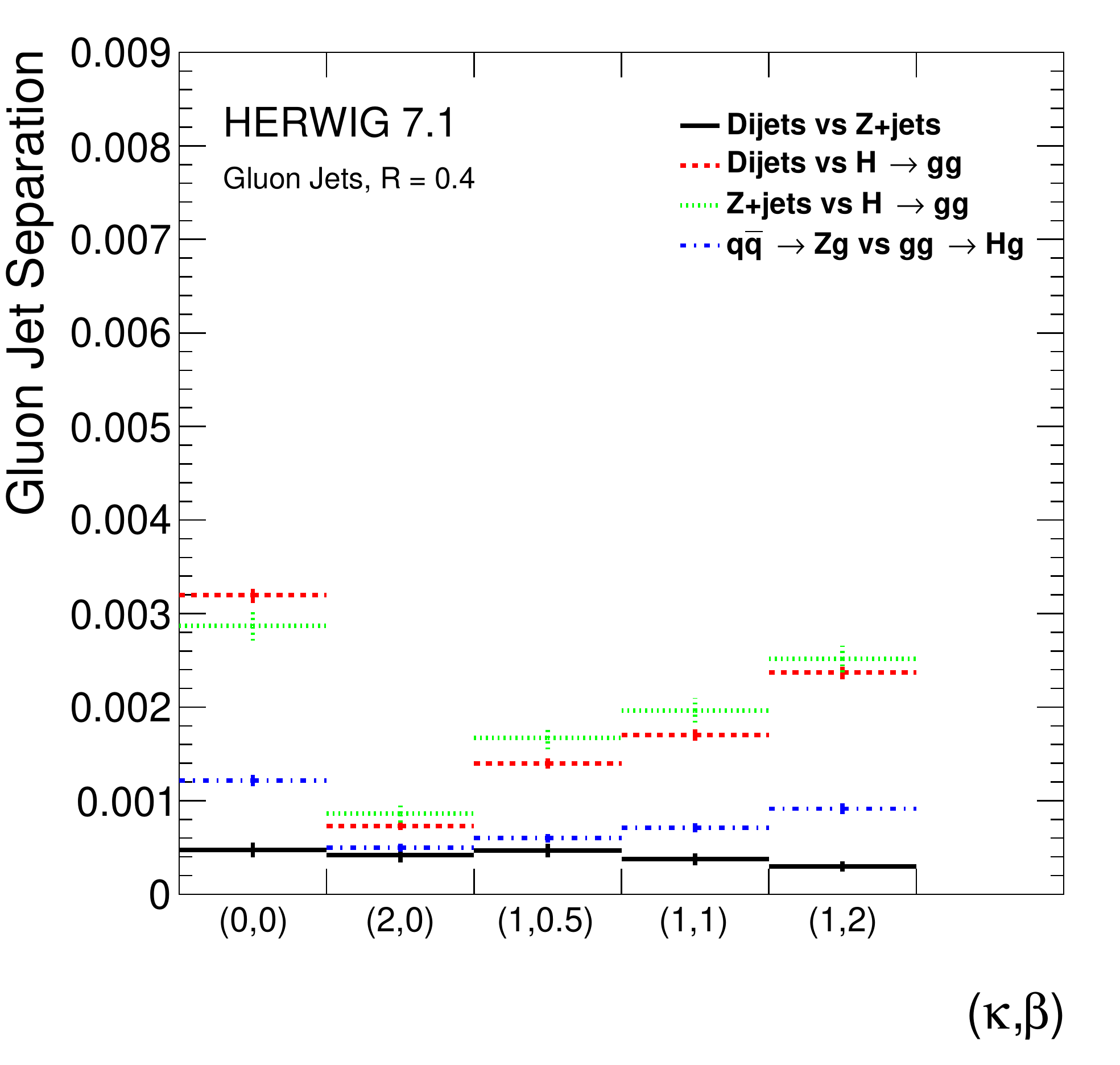}}
\subfloat[]{\label{fig:9c}\includegraphics[width=0.33\textwidth]{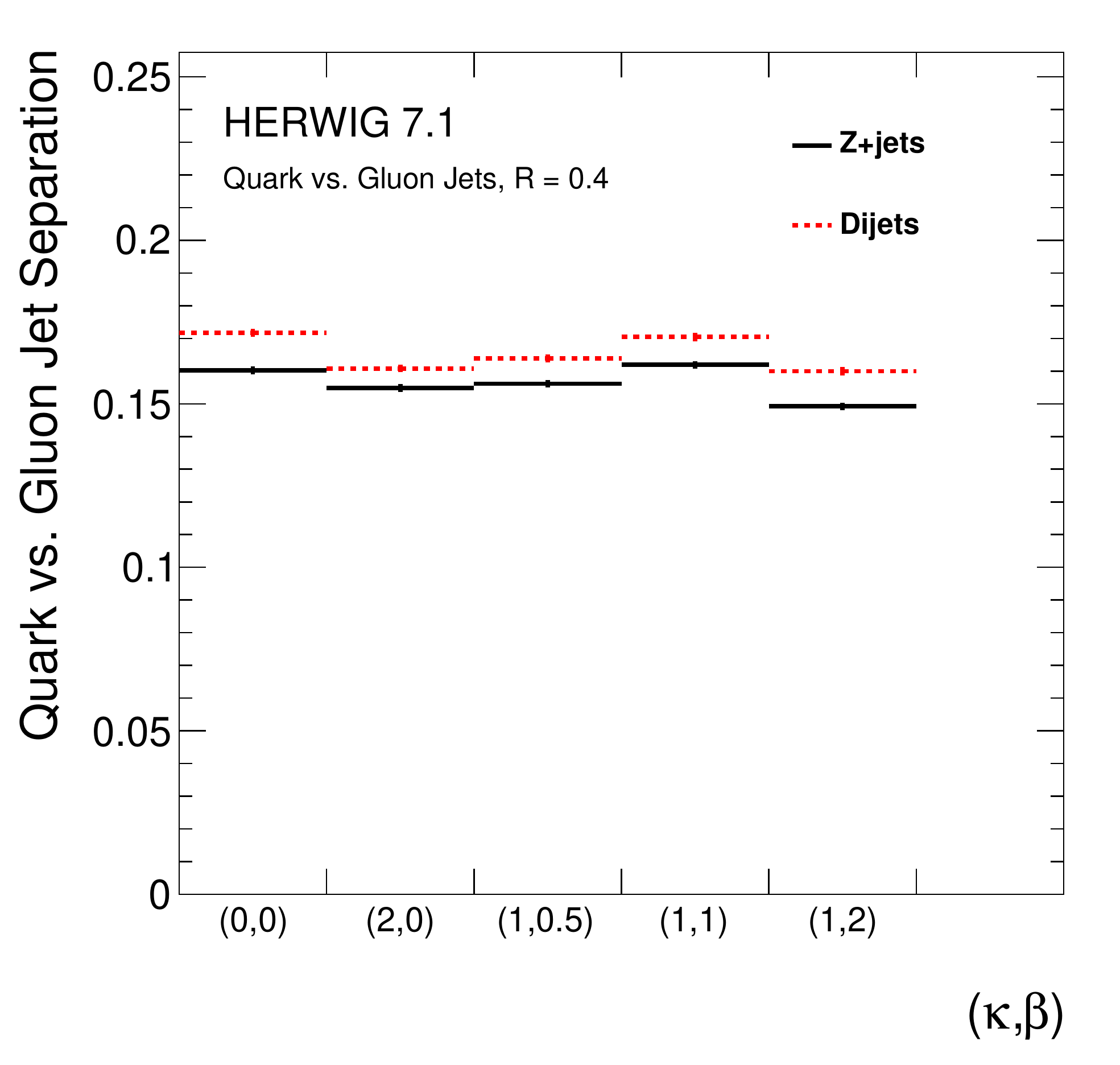}}
\\
\subfloat[]{\label{fig:9d}\includegraphics[width=0.49\textwidth]{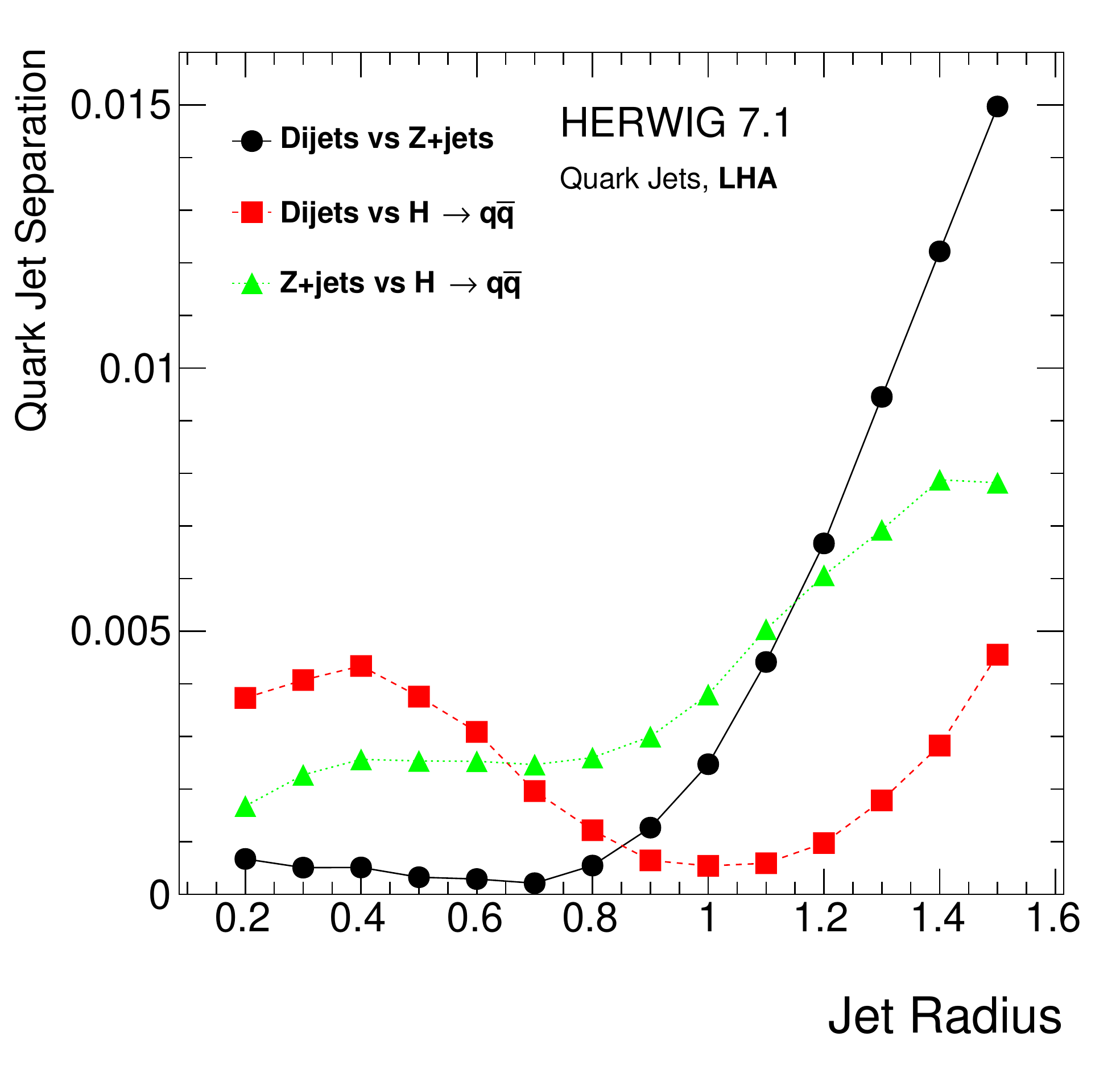}}
\hfill
\subfloat[]{\label{fig:9e}\includegraphics[width=0.49\textwidth]{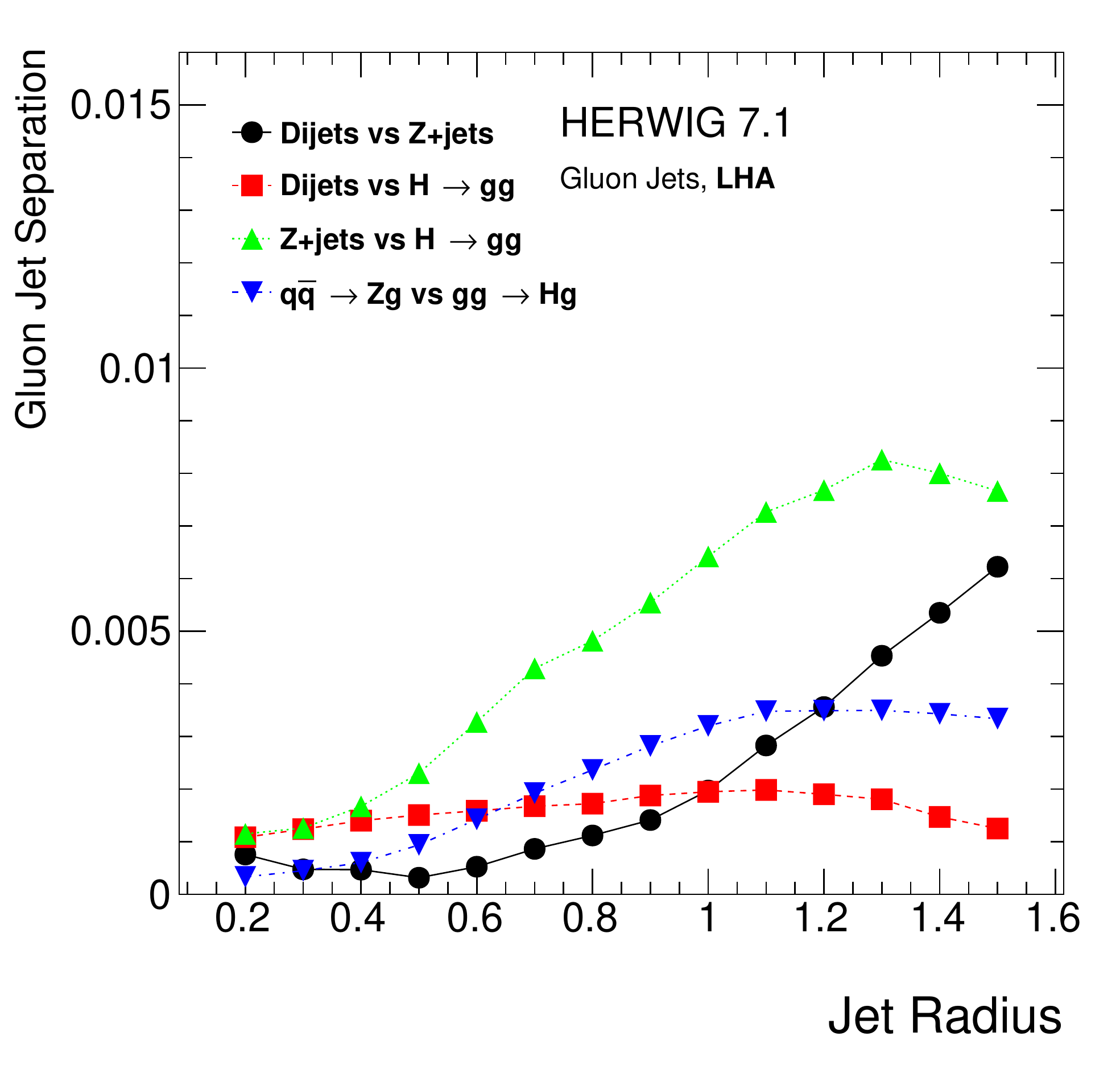}}
\caption{\label{fig:9} Replications of plots from Figs.~\ref{fig:1} and \ref{fig:2} using jets from the \texttt{MadGraph5 + HERWIG 7} samples. Plots (a) and (b) replicate Figs.~\ref{fig:1a} and \ref{fig:1b}, showing separation power between various pairs of topologies as a function of angularity at a jet radius of $R = 0.4$ and plots (c) and (d) replicate Figs.~\ref{fig:2a} and \ref{fig:2d}, showing how separation power in the LHA angularity varies as a function of jet radius in the quark and gluon jet channels, respectively.  Error bars represent statistical uncertainty.}
\end{figure}

Classifier separations for multiple angularities and a scan in the jet radius is presented in Fig.~\ref{fig:9}.   Many of the trends are similar to the ones observed with \texttt{PYTHIA 8.2}, but there are a few key differences.  For example, the same-flavor classifier separation for multiplicity and $(p_\text{T}^D)^2$ is much smaller by factors of about five and ten, respectively, for \texttt{HERWIG 7.1} than for \texttt{PYTHIA 8.2}.  Furthermore, even though the classifier separation scale is about the same, the increasing angular dependence of the IRC safe observables is more pronounced for \texttt{HERWIG 7.1} compared with \texttt{PYTHIA 8.2}.  The level and shape of the classifier separation for dijets versus $Z$+jets is about the same between the two generators, as is the ordering of $Z$+jets versus $H$ and dijets versus $H$ for both quarks and gluons.  \texttt{HERWIG 7.1} also predicts a larger difference between $q\bar{q}\rightarrow Zg$ and $gg\rightarrow Hg$ than \texttt{PYTHIA 8.2}, though in both cases, the separation is 0.1\% or below.

The radius dependence shown in Fig.~\ref{fig:9c} for quark jets is qualitatively the same as for \texttt{PYTHIA 8.2}.  The exact classifier separation that the $Z$+jets versus dijets approaches at the largest radius is higher for \texttt{HERWIG 7.1} than for \texttt{PYTHIA 8.2} by about 50\%.    The low radius behavior of the other comparisons in Figs.~\ref{fig:9c} are also slightly different than for \texttt{PYTHIA 8.2}: the dijets versus $H\rightarrow q\bar{q}$ and $Z$+jets versus $H\rightarrow q\bar{q}$ are more separated at low radius and overall have a lower classifier separation than present in \texttt{PYTHIA 8.2}.  The ordering and numerical values of the separations at high radius are the same between generators for gluons, but the trends toward lower radii are qualitatively different. All of the curves seem to approach zero (as expected) for \texttt{HERWIG 7.1}, while only the $q\bar{q}\rightarrow Zg$ versus $gg\rightarrow Hg$ monotonically approached zero for \texttt{PYTHIA 8.2}.

Overall, the two generators give a similar picture for the topology dependence of quark and gluon jets, though there are some differences in the classifier separation scale and trends with the radii and angular exponents as remarked above.  Some differences may be expected, given the known large differences between generators in describing quark and gluon jets~\cite{Gras:2017jty}.

\graphicspath{ {./figs_Herwig/} }
\clearpage

\section{Conclusions}
\label{sec:conclusions}

Since most measurements and searches at the LHC target topologies with either mostly quark or mostly gluon jets, quark-versus-gluon jet tagging offers a promising set of tools to improve analysis precision and sensitivity.  There is an extensive literature developing observables for distinguishing quark jets from gluon jets and also many studies probing the topology dependence of quark-versus-gluon jets tagging.  This analysis reports the first systematic study in simulation of the topology dependence of quark and gluon jets separately.  Overall, the topology dependence of quark and gluon jets separately is much smaller than for quark versus gluon jets.  This is important for quark-versus-gluon jet tagging at the LHC, where quark and gluon jets are widely treated as universal objects; the study presented here shows that this is true up to $\sim10\%$ corrections ($\sim2\%$ for IRC safe observables in Pythia and for nearly all observables in Herwig; also typically less for smaller jet radii and for IRC unsafe groomed observables).  These corrections have a structure that depends on how the radiation pattern inside jets is probed and what jet radius and jet $p_\text{T}$ are examined.  Many of the qualitative features of the residual topology dependence are robust to changes in kinematics, parton labeling, and MC generator, but there are also some significant differences with these variations as well.  Detailed studies of the residual topology dependence will be a challenging and interesting study for the future\footnote{Appendix~\ref{appUE} provides some evidence that the underlying event may play a key role in these residual differences.  Other differences at small opening angle have an analytic understanding in terms of flavor changing from collinear splittings~\cite{Dasgupta:2014yra}.  This effect can be enhanced with grooming, which increases the sensitivity to small angular scales~\cite{Dasgupta:2013ihk}.}.

Now that jet substructure is reaching a mature level of precision, it may be possible to explain some of the topology-dependent trends observed in the above studies.  This would be helpful to explain the features that are common for all of the studies and would provide critical insight to resolving differences between configurations.  An important first step in this direction for the non-perturbative corrections to the jet mass in Ref.~\cite{stewart2015dissecting,Dasgupta:2007wa} and it would be a significant next step to see such studies applied more broadly, also including observable quantities.

At the same time, there is a plethora of data at the LHC which can be used to probe the trends in situ.  Measurements of jet substructure in complex topologies, such as $t\bar{t}$ events~\cite{Gallicchio:2010sw,Aad:2015lxa,Aaboud:2018ibj,Abazov:2011vh} will continue to provide an important handle on non-universal behavior.  One of the biggest challenges with any study of quark and gluon jets is the assignment and interpretation of parton labels.  New ideas for a pragmatic and generator-independent definition may hold the key to making progress in this area~\cite{Komiske:2018vkc}.  In particular, a definition of quark or gluon jet defined at the level of cross-sections and using pairs of samples in the construction could be used to study the topology dependence of quark and gluon jets separately by combining multiple pairs of samples and multiple observables for extracting the distributions.  Such a study could provide an entirely data-driven probe of quark and gluon jet universality.  

While there are now many new features of jet substructure to investigate, the near universality of quark and gluon jets suggests that the work to develop, calibrate, and deploy powerful taggers to the rich LHC data should continue along the current trajectory.  Investigations of the non-universal behavior will improve our understanding of QCD and may lead to the development of more robust taggers as rarer signals are probed at the LHC and beyond.

\section{Acknowledgments}

We are grateful to Patrick Komiske, Eric Metodiev, and Jesse Thaler for detailed comments on the analysis as well as the manuscript.  This work was supported by the U.S.~Department of Energy, Office of Science under contract DE-AC02-05CH11231.
\FloatBarrier

\appendix
\section{Varying the underlying event}
\label{appUE}

\begin{figure}[h!]
\centering
\subfloat[]{\label{fig:10a}\includegraphics[width=0.49\textwidth]{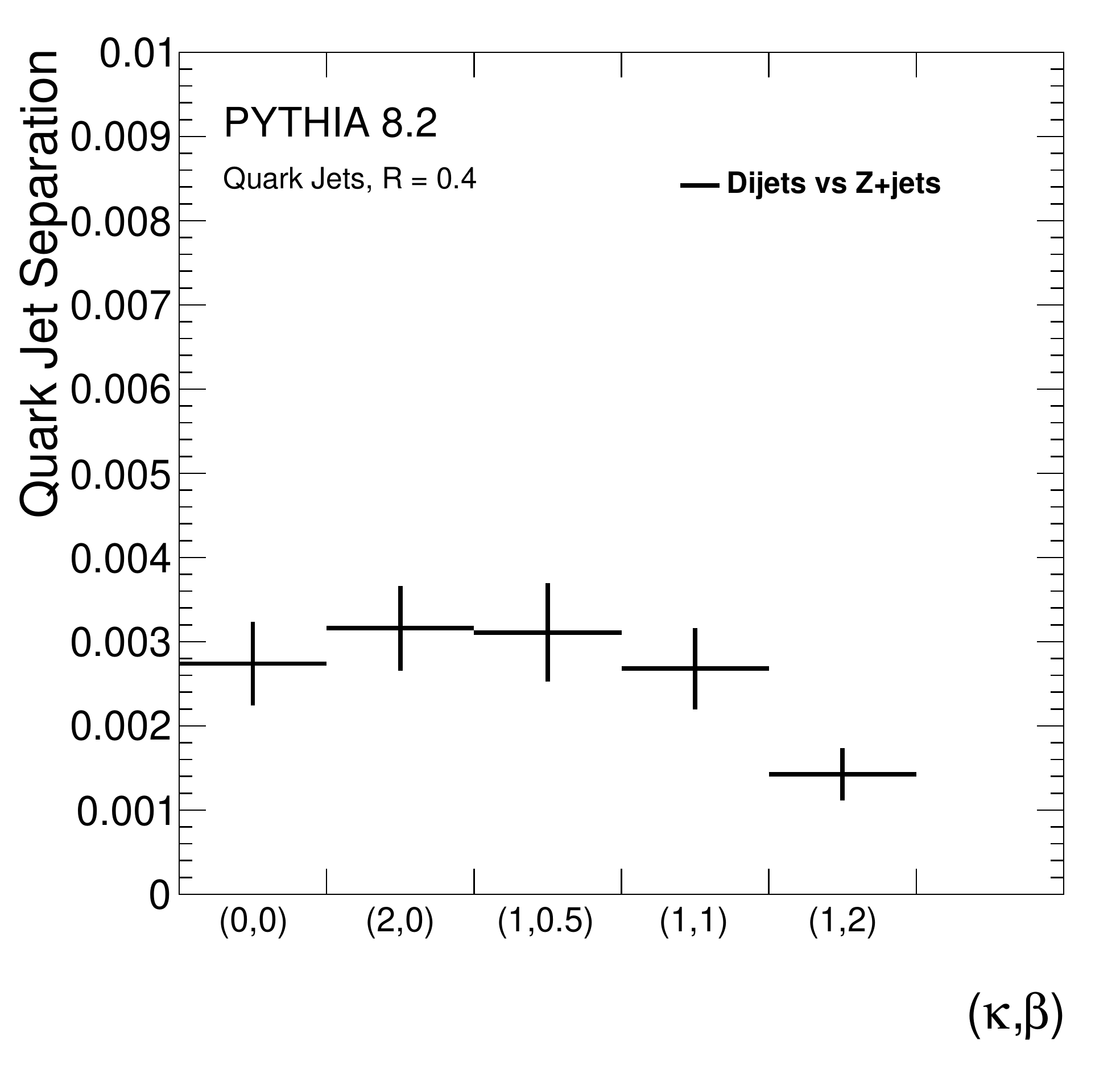}}
\subfloat[]{\label{fig:10b}\includegraphics[width=0.49\textwidth]{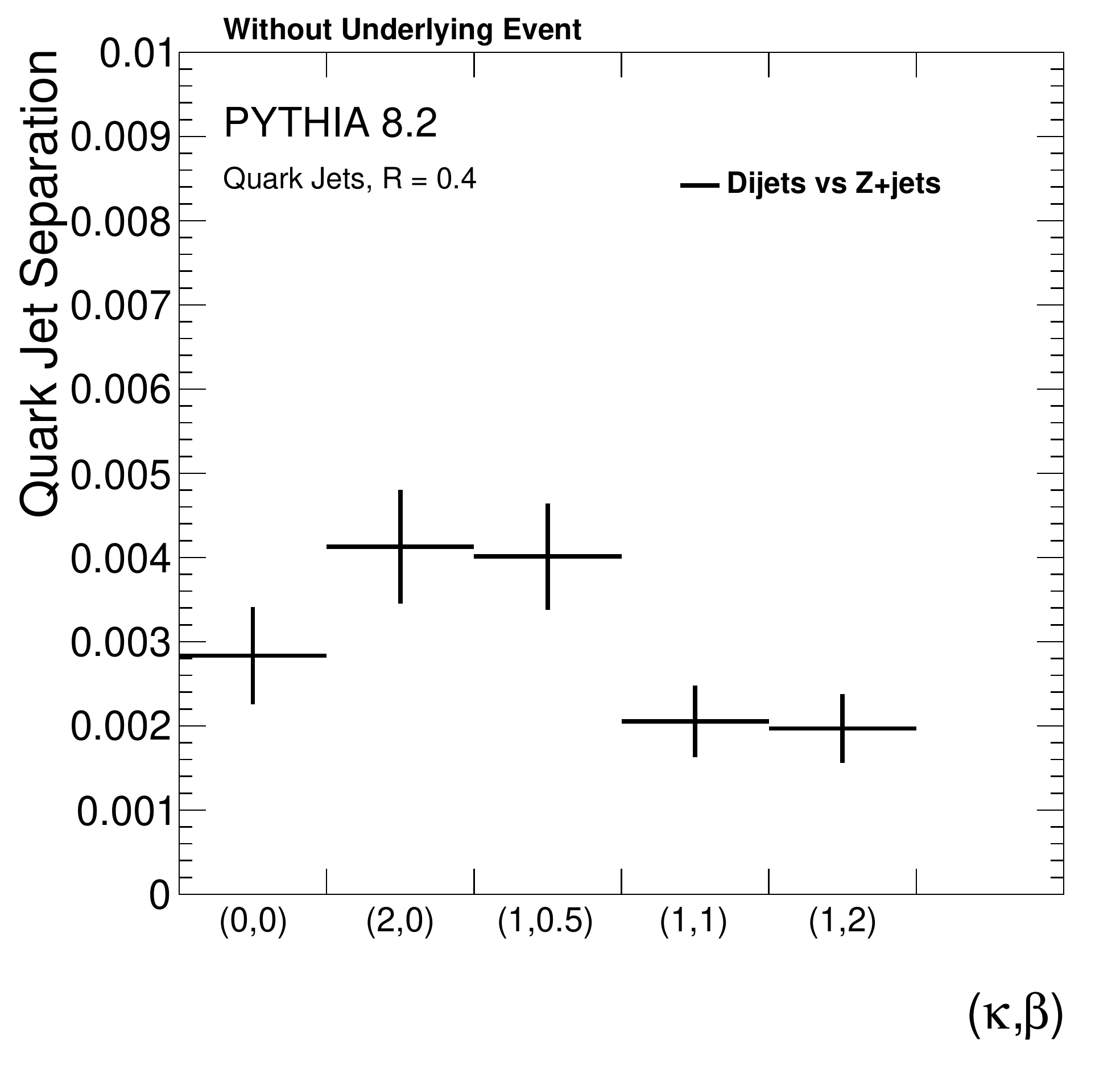}}
\\
\subfloat[]{\label{fig:10a}\includegraphics[width=0.49\textwidth]{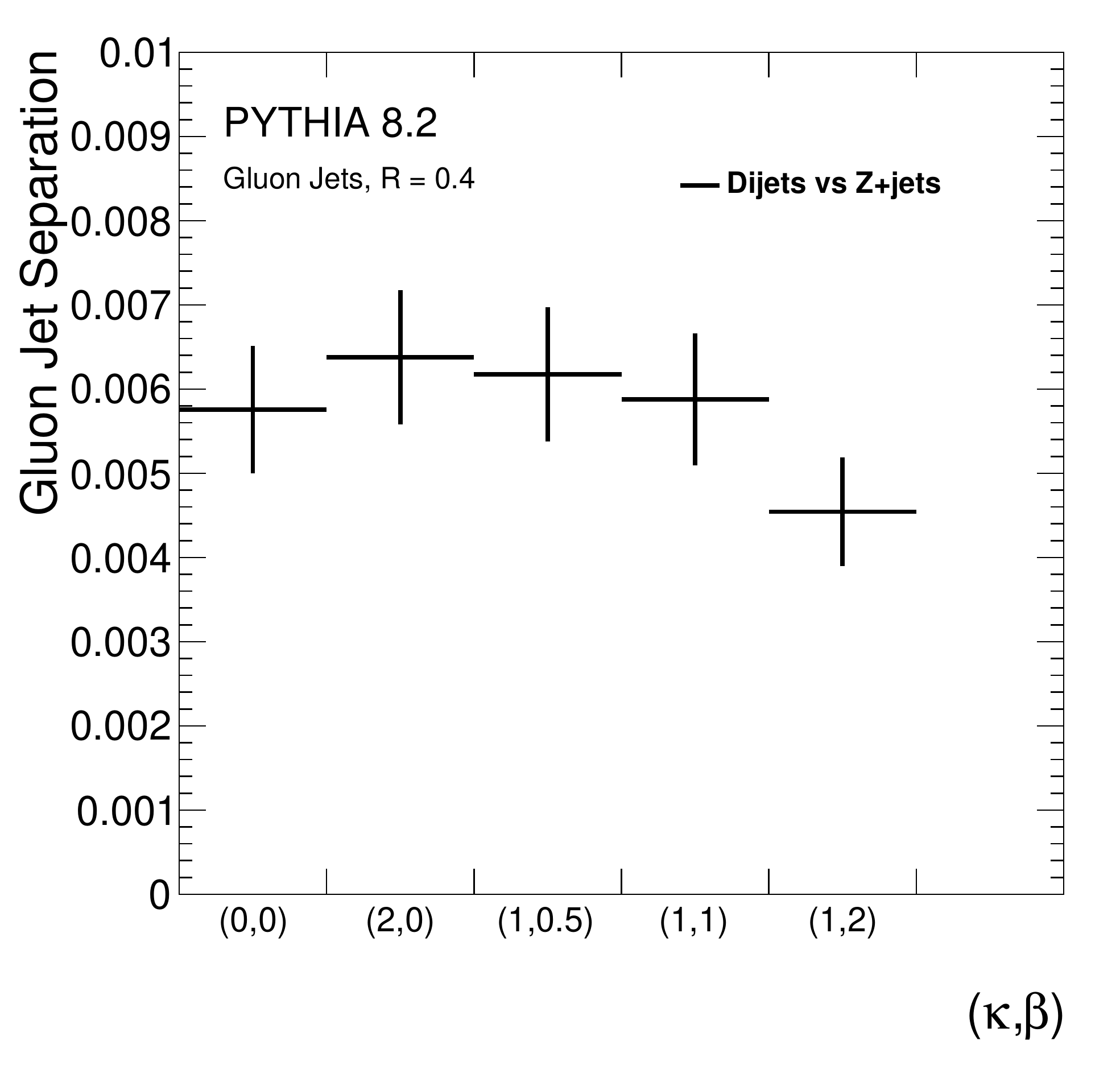}}
\subfloat[]{\label{fig:10b}\includegraphics[width=0.49\textwidth]{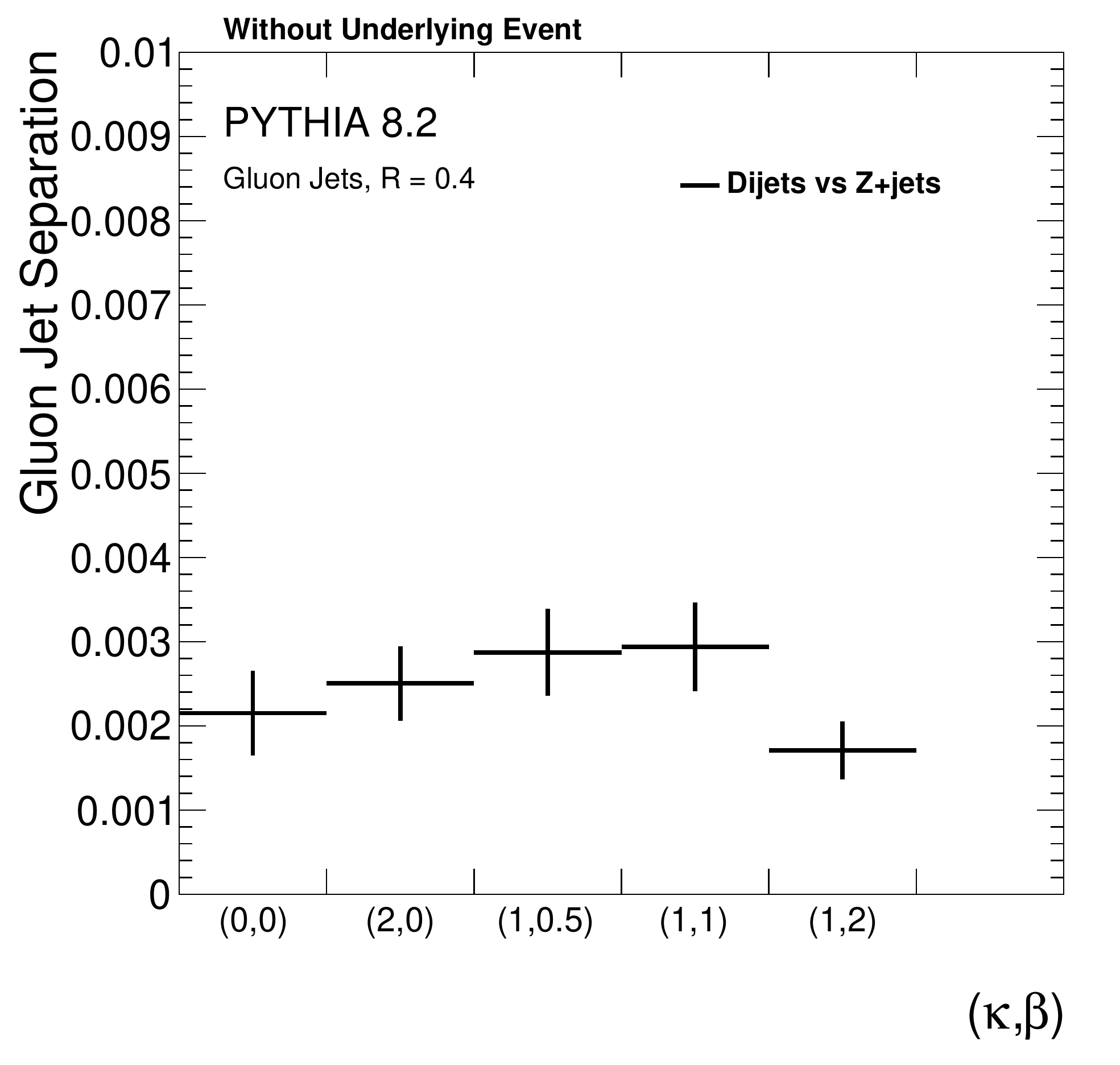}}
\caption{\label{fig:10} Plots showing same-flavor classifier separation in the five angularities for Dijets versus $Z$+jets using samples generated with UE (left) and without UE (right). The top row corresponds to quark jet separation, and the bottom row to gluon jet separation.}
\end{figure}

In order to better understand the origin of some of the observed topology dependence, we conducted a short study comparing samples generated with and without Underlying Event (UE). Two sets of samples of 100,000 events were generated with \texttt{PYTHIA 8.226} in the Dijets, $Z$+jets, and $H \rightarrow q\bar{q}/gg$ topologies, one with and one without UE. Plots showing same-flavor classifier separation for Dijets versus $Z$+jets are presented in Fig.~\ref{fig:10}. In the gluon jet channel, classifier separation in all angularities is reduced when UE is turned off. In the quark jet channel, separation is only reduced for width when UE is deactivated, whereas small increases are seen in the other angularities. The underlying event is expected to be different in Dijets and $Z$+jets, so the results in the gluon jet channel are reasonable. The behavior of the quark jets is less consistent, but the separation power remains small (at or below the 0.4\% level) in both cases. 

Lines for the $H \rightarrow q\bar{q}/gg$ topologies were not included in Fig.~\ref{fig:10}, but an increase in classifier separation was observed for these topologies (when compared to Dijets and $Z$+jets) when UE was deactivated. These results were excluded because we believe they are partially due to the altered labeling scheme used for the Higgs samples, wherein the set of partons used for assigning a jet flavor label was reduced to ancestors of the Higgs. 

Overall, Fig.~\ref{fig:10} suggests that underlying event may indeed be a contributing factor to the small topology dependence observed in the main results of the paper. In the gluon jet channel, classifier separation is reduced by approximately 50\% after UE is turned off, corresponding to a reduction in classifier separation of approximately 0.2--0.3\%. Considering that separation of 0.5--1.5\% was seen in Sec.~\ref{sec:base}, this is a significant effect. However, the inconsistent results of turning off UE in the quark jet channel make it difficult to definitively identify UE as a main source of topology dependence.

\section{Comparing the default and QCD-aware labeling schemes}
\label{sec:labeling}
The difference between the results using the default and QCD-aware labeling schemes is relatively large, and in this section we probe the relative agreement of these two methods on assigning jet labels. In particular, we investigate how often (i.e. for what proportion of jets) the two schemes agree on a jet label, and, given angularity spectra for quark and gluon jets from the two labeling schemes in the same topology, how well these spectra agree. 

In Table~\ref{table:1}, we show statistics on the agreement between the default and QCD-aware methods in the Dijets and $Z$+jets topology. We list how many of each jet type (quark, gluon, or neither; all passing the selection detailed in Sec.~\ref{sec:base}) is found by the default method, then list for what percentage of those jets the QCD-aware method agreed or disagreed on the label. In both topologies, we find that QCD-aware labeled approximately 20\% of default-labeled gluon jets as quark jets. Furthermore, QCD-aware labeled significant fractions of default-labeled quark jets as `neither' (19\% for Dijets, 11\% for $Z$+jets).

\begin{figure}
\centering
\subfloat[]{\label{fig:11a}\includegraphics[width=0.49\textwidth]{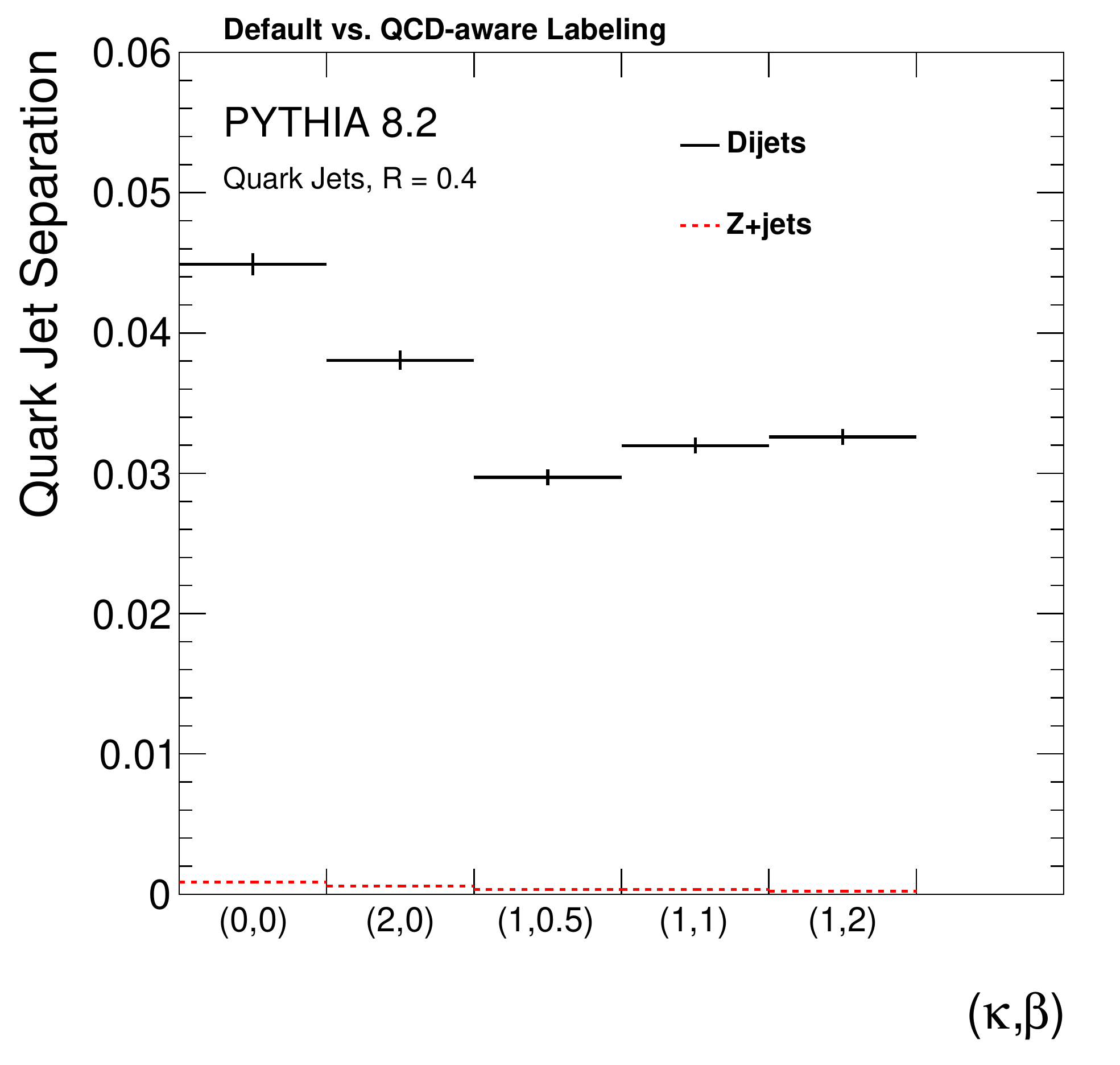}}
\subfloat[]{\label{fig:11b}\includegraphics[width=0.49\textwidth]{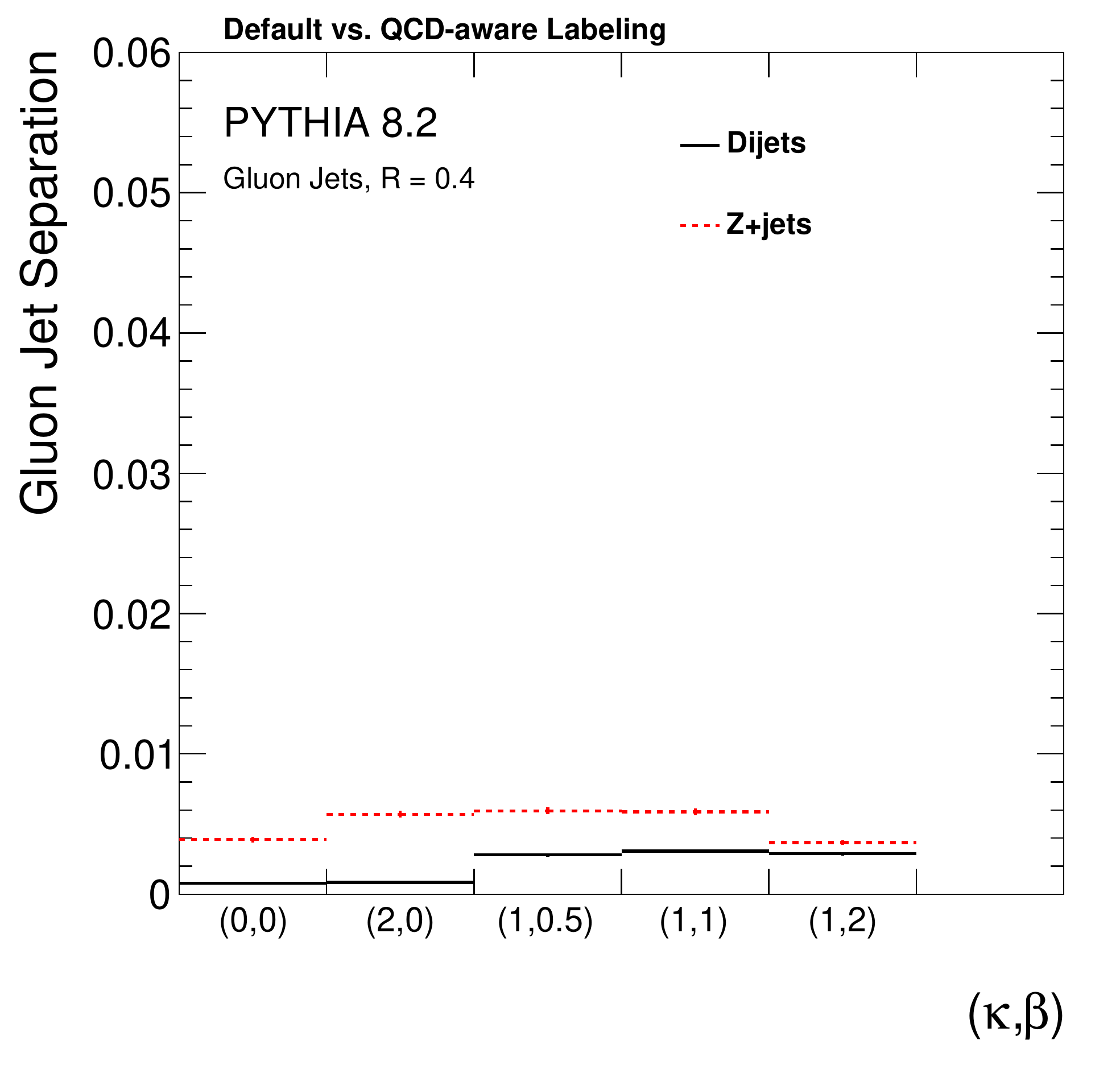}}
\caption{\label{fig:11} Plots showing classifier separation between jets labeled as the same flavor in the same topology by the default and QCD-aware labeling schemes. The left hand plot shows results for quark jets, and the right hand plot shows results for gluon jets.}
\end{figure}

In Figure~\ref{fig:11}, we complement the results of Table~\ref{table:1} by showing same-flavor classifier separation between jets labeled by the default scheme and the QCD-aware scheme in the same topology. Figure~\ref{fig:11a} shows quark jet separation in the five angularities, and Fig.~\ref{fig:11b} shows gluon jet separation. The most striking difference between the two is the Dijets line in the quark jet channel. While default and QCD-aware quark jets are separated by approximately 0.1\% in $Z$+jets, they are separated by about 3\% in the IRC-safe angularities in Dijets. This makes some sense, given that QCD-aware labeled about 20\% of default-tagged quark jets as `neither', but the degree to which the two schemes are separated is nonetheless striking. Considering this difference, the consistently low ($\approx 0.6$\%) classifier separation between gluon jets labeled by the two schemes in both angularities is interesting. While gluon jets in both topologies share a similar 20\% disagreement rate between default and QCD-aware (where QCD-aware labels them as quark jets) as the quark jets in the Dijets sample, the classifier separation remains much lower.

For similar reasons to those stated in Appendix~\ref{appUE}, we exclude results from the $H \rightarrow q\bar{q}/gg$ topologies. Given that the QCD-aware method relies on parton-level information to construct parton-jets that are assigned a flavor label and matched to final-state jets, restricting the set of partons considered for clustering to Higgs ancestors could have a significant impact on the total number and flavor composition of the jets that the QCD-aware scheme labels.

It is difficult to make a definitive statement about the general agreement of the default and QCD-aware labeling schemes, but the statistics in Table~\ref{table:1} and the plots in Fig.~\ref{fig:11} do suggest a significant disagreement about the labeling. In light of this, the difference between results of the two labeling schemes in the main body of the paper makes better sense. This disagreement is not investigated in more detail here, but further analysis of the relative agreement of different quark/gluon jet labeling schemes may be a fruitful topic for further study.

\begin{table}[t]
\centering
\begin{tabular}{|c|c|c|c|c|c|}
\hline
Topology & Jet type & \multicolumn{1}{p{3cm}|}{\centering \# labeled by \\ default tagger} & \multicolumn{1}{p{3cm}|}{\centering \% QCD-aware \\ labeled quark} & \multicolumn{1}{p{3cm}|}{\centering \% QCD-aware \\ labeled gluon} & \multicolumn{1}{p{3cm}|}{\centering \% QCD-aware \\ labeled neither} \\
\hline
\hline
\multirow{3}{4em}{Dijets} & {\color{red}quark} & $131118$ & {\color{red}$77$} & $4$ & $19$ \\
& {\color{blue}gluon} & 403565 & $20$ & {\color{blue}$69$} & $11$ \\
& neither & 42835 & $9$ & $10$ & $81$ \\
\hline
\multirow{3}{4em}{$Z+$jets} & {\color{red}quark} & 343548 & {\color{red}$84$} & $5$ & $11$ \\
& {\color{blue}gluon} & 144898 & $21$ & {\color{blue}$69$} & $10$ \\
& neither & 21379 & $1$ & $3$ & $96$ \\
\hline
\end{tabular}
\caption{\label{table:1}A breakdown of the consistency between the default labeling scheme (simply called quark, gluon, or neither) and the QCD-aware scheme. }
\end{table}
\FloatBarrier

\FloatBarrier

\bibliographystyle{jhep}
\bibliography{myrefs}

\providecommand{\href}[2]{#2}\begingroup\raggedright\begin{thebibliography}{10}

\bibitem{Aaboud:2016cns}
{\bf ATLAS} Collaboration, M.~Aaboud et~al., {\it {Search for the Standard
  Model Higgs boson produced by vector-boson fusion and decaying to bottom
  quarks in $ \sqrt{s}=8 $ TeV pp collisions with the ATLAS detector}},  {\em
  JHEP} {\bf 11} (2016) 112, [\href{http://arxiv.org/abs/1606.02181}{{\tt
  arXiv:1606.02181}}].

\bibitem{ATLAS-CONF-2016-063}
{\textbf{ATLAS} Collaboration}, {\it {Search for Higgs boson production via
  weak boson fusion and decaying to $b \bar b$ in association with a
  high-energy photon in the ATLAS detector}},  {\em ATLAS-CONF-2016-063} (2016)
  [\href{https://cds.cern.ch/record/2206201}{cds.cern.ch:2206201}].

\bibitem{Aad:2016oit}
{\bf ATLAS} Collaboration, G.~Aad et~al., {\it {Measurement of the
  charged-particle multiplicity inside jets from $\sqrt{s}=8$ TeV $pp$
  collisions with the ATLAS detector}},  {\em Eur. Phys. J.} {\bf C76} (2016)
  322, [\href{http://arxiv.org/abs/1602.00988}{{\tt arXiv:1602.00988}}].

\bibitem{ATLAS-CONF-2015-002}
{\bf ATLAS} Collaboration, {\it {Jet global sequential corrections with the
  ATLAS detector in proton-proton collisions at $\sqrt{s} = 8$ TeV}},  {\em
  ATLAS-CONF-2015-002} (2015)
  [\href{https://cds.cern.ch/record/2001682}{cds.cern.ch:2001682}].

\bibitem{Sirunyan:2017jej}
{\bf CMS} Collaboration, A.~M. Sirunyan et~al., {\it {Electroweak production of
  two jets in association with a Z boson in proton-proton collisions at
  $\sqrt{s}= $ 13 $\,\text {TeV}$}},  {\em Eur. Phys. J.} {\bf C78} (2018) 589,
  [\href{http://arxiv.org/abs/1712.09814}{{\tt arXiv:1712.09814}}].

\bibitem{Sirunyan:2018ygk}
{\bf CMS} Collaboration, A.~M. Sirunyan et~al., {\it {Search for
  $t\overline{t}H$ production in the all-jet final state in proton-proton
  collisions at $\sqrt{s}=$ 13 TeV}},  {\em JHEP} {\bf 06} (2018) 101,
  [\href{http://arxiv.org/abs/1803.06986}{{\tt arXiv:1803.06986}}].

\bibitem{Khachatryan:2015bnx}
{\bf CMS} Collaboration, V.~Khachatryan et~al., {\it {Search for the standard
  model Higgs boson produced through vector boson fusion and decaying to $b
  \overline{b}$}},  {\em Phys. Rev.} {\bf D92} (2015) 032008,
  [\href{http://arxiv.org/abs/1506.01010}{{\tt arXiv:1506.01010}}].

\bibitem{Khachatryan:2014dea}
{\bf CMS} Collaboration, V.~Khachatryan et~al., {\it {Measurement of
  electroweak production of two jets in association with a Z boson in
  proton-proton collisions at $\sqrt{s}=8\,\text {TeV}$}},  {\em Eur. Phys. J.}
  {\bf C75} (2015) 66, [\href{http://arxiv.org/abs/1410.3153}{{\tt
  arXiv:1410.3153}}].

\bibitem{Aaboud:2017eta}
{\bf ATLAS} Collaboration, M.~Aaboud et~al., {\it {Search for diboson
  resonances with boson-tagged jets in $pp$ collisions at $\sqrt{s}=13$ TeV
  with the ATLAS detector}},  {\em Phys. Lett.} {\bf B777} (2018) 91--113,
  [\href{http://arxiv.org/abs/1708.04445}{{\tt arXiv:1708.04445}}].

\bibitem{Sirunyan:2017pjw}
{\bf CMS} Collaboration, A.~M. Sirunyan et~al., {\it {Search for supersymmetry
  in proton-proton collisions at 13 TeV using identified top quarks}},  {\em
  Phys. Rev.} {\bf D97} (2018) 012007,
  [\href{http://arxiv.org/abs/1710.11188}{{\tt arXiv:1710.11188}}].

\bibitem{Sirunyan:2017lzl}
{\bf CMS} Collaboration, A.~M. Sirunyan et~al., {\it {Search for vector-like
  light-flavor quark partners in proton-proton collisions at $\sqrt{s}=8$
  TeV}},  {\em Phys. Rev.} {\bf D97} (2017) 072008,
  [\href{http://arxiv.org/abs/1708.02510}{{\tt arXiv:1708.02510}}].

\bibitem{Sirunyan:2017wif}
{\bf CMS} Collaboration, A.~M. Sirunyan et~al., {\it {Search for direct
  production of supersymmetric partners of the top quark in the all-jets final
  state in proton-proton collisions at $ \sqrt{s}=13 $ TeV}},  {\em JHEP} {\bf
  10} (2017) 005, [\href{http://arxiv.org/abs/1707.03316}{{\tt
  arXiv:1707.03316}}].

\bibitem{Altarelli:1977zs}
G.~Altarelli and G.~Parisi, {\it {Asymptotic Freedom in Parton Language}},
  {\em Nucl. Phys.} {\bf B126} (1977) 298.

\bibitem{Larkoski:2014pca}
A.~J. Larkoski, J.~Thaler, and W.~J. Waalewijn, {\it {Gaining (Mutual)
  Information about Quark/Gluon Discrimination}},  {\em JHEP} {\bf 11} (2014)
  129, [\href{http://arxiv.org/abs/1408.3122}{{\tt arXiv:1408.3122}}].

\bibitem{Gras:2017jty}
P.~Gras, S.~H{\"o}che, D.~Kar, A.~Larkoski, L.~L{\"o}nnblad, S.~Pl{\"a}tzer,
  A.~Si{\'o}dmok, P.~Skands, G.~Soyez, and J.~Thaler, {\it {Systematics of
  quark/gluon tagging}},  {\em JHEP} {\bf 07} (2017) 091,
  [\href{http://arxiv.org/abs/1704.03878}{{\tt arXiv:1704.03878}}].

\bibitem{Frye:2017yrw}
C.~Frye, A.~J. Larkoski, J.~Thaler, and K.~Zhou, {\it {Casimir Meets Poisson:
  Improved Quark/Gluon Discrimination with Counting Observables}},  {\em JHEP}
  {\bf 09} (2017) 083, [\href{http://arxiv.org/abs/1704.06266}{{\tt
  arXiv:1704.06266}}].

\bibitem{Gallicchio:2011xq}
J.~Gallicchio and M.~D. Schwartz, {\it {Quark and Gluon Tagging at the LHC}},
  {\em Phys. Rev. Lett.} {\bf 107} (2011) 172001,
  [\href{http://arxiv.org/abs/1106.3076}{{\tt arXiv:1106.3076}}].

\bibitem{Gallicchio:2012ez}
J.~Gallicchio and M.~D. Schwartz, {\it {Quark and Gluon Jet Substructure}},
  {\em JHEP} {\bf 04} (2013) 090, [\href{http://arxiv.org/abs/1211.7038}{{\tt
  arXiv:1211.7038}}].

\bibitem{Aad:2014gea}
{\textbf{ATLAS} Collaboration}, {\it {Light-quark and gluon jet discrimination
  in $pp$ collisions at $\sqrt{s}=7$ TeV with the ATLAS detector}},  {\em Eur.
  Phys. J.} {\bf C74} (2014) 3023, [\href{http://arxiv.org/abs/1405.6583}{{\tt
  arXiv:1405.6583}}].

\bibitem{ATLAS-CONF-2016-034}
{\textbf{ATLAS} Collaboration}, {\it {Discrimination of Light Quark and Gluon
  Jets in $pp$ collisions at $\sqrt{s} = 8$ TeV with the ATLAS Detector}},
  {\em ATLAS-CONF-2016-034} (2016)
  [\href{https://cds.cern.ch/record/2200202}{cds.cern.ch:2200202}].

\bibitem{ATL-PHYS-PUB-2017-009}
{\textbf{ATLAS} Collaboration}, {\it {Quark versus Gluon Jet Tagging Using
  Charged Particle Multiplicity with the ATLAS Detector}},  {\em
  ATL-PHYS-PUB-2017-009} (2017)
  [\href{https://cds.cern.ch/record/2263679}{cds.cern.ch:2263679}].

\bibitem{CMS-PAS-JME-13-002}
{\textbf{CMS} Collaboration}, {\it {Performance of quark/gluon discrimination
  in 8 TeV pp data}},  {\em CMS-PAS-JME-13-002} (2013)
  [\href{https://cds.cern.ch/record/1599732}{cds.cern.ch:1599732}].

\bibitem{CMS-DP-2016-070}
{\textbf{CMS} Collaboration}, {\it {Performance of quark/gluon discrimination
  in 13 TeV data}},  {\em CMS-DP-2016-070} (2016)
  [\href{https://cds.cern.ch/record/2234117}{cds.cern.ch:2234117}].

\bibitem{ATL-PHYS-PUB-2017-017}
{\bf {ATLAS}} Collaboration, {\it {Quark versus Gluon Jet Tagging Using Jet
  Images with the ATLAS Detector}},  {\em ATL-PHYS-PUB-2017-017} (2017)
  [\href{https://cds.cern.ch/record/2275641}{cds.cern.ch:2275641}].

\bibitem{CMS-DP-2017-027}
{\bf {CMS}} Collaboration, {\it {New Developments for Jet Substructure
  Reconstruction in CMS}},  {\em CMS-DP-2017-027} (2017)
  [\href{https://cds.cern.ch/record/2275226}{cds.cern.ch:2275226}].

\bibitem{Komiske:2016rsd}
P.~T. Komiske, E.~M. Metodiev, and M.~D. Schwartz, {\it {Deep learning in
  color: towards automated quark/gluon jet discrimination}},  {\em JHEP} {\bf
  01} (2017) 110, [\href{http://arxiv.org/abs/1612.01551}{{\tt
  arXiv:1612.01551}}].

\bibitem{Dery:2017fap}
L.~M. Dery, B.~Nachman, F.~Rubbo, and A.~Schwartzman, {\it {Weakly Supervised
  Classification in High Energy Physics}},  {\em JHEP} {\bf 05} (2017) 145,
  [\href{http://arxiv.org/abs/1702.00414}{{\tt arXiv:1702.00414}}].

\bibitem{Metodiev:2017vrx}
E.~M. Metodiev, B.~Nachman, and J.~Thaler, {\it {Classification without labels:
  Learning from mixed samples in high energy physics}},  {\em JHEP} {\bf 10}
  (2017) 174, [\href{http://arxiv.org/abs/1708.02949}{{\tt arXiv:1708.02949}}].

\bibitem{Luo:2017ncs}
H.~Luo, M.-x. Luo, K.~Wang, T.~Xu, and G.~Zhu, {\it {Quark jet versus gluon
  jet: deep neural networks with high-level features}},
  \href{http://arxiv.org/abs/1712.03634}{{\tt arXiv:1712.03634}}.

\bibitem{Komiske:2018oaa}
P.~T. Komiske, E.~M. Metodiev, B.~Nachman, and M.~D. Schwartz, {\it {Learning
  to classify from impure samples with high-dimensional data}},  {\em Phys.
  Rev.} {\bf D98} (2018) 011502, [\href{http://arxiv.org/abs/1801.10158}{{\tt
  arXiv:1801.10158}}].

\bibitem{Cheng:2017rdo}
T.~Cheng, {\it {Recursive Neural Networks in Quark/Gluon Tagging}},  {\em
  Comput. Softw. Big Sci.} {\bf 2} (2018) 3,
  [\href{http://arxiv.org/abs/1711.02633}{{\tt arXiv:1711.02633}}].

\bibitem{Gallicchio:2010sw}
J.~Gallicchio and M.~D. Schwartz, {\it {Seeing in Color: Jet Superstructure}},
  {\em Phys. Rev. Lett.} {\bf 105} (2010) 022001,
  [\href{http://arxiv.org/abs/1001.5027}{{\tt 1001.5027}}].

\bibitem{Aad:2015lxa}
{\bf ATLAS} Collaboration, G.~Aad et~al., {\it {Measurement of colour flow with
  the jet pull angle in $t\bar{t}$ events using the ATLAS detector at
  $\sqrt{s}=8$ TeV}},  {\em Phys. Lett.} {\bf B750} (2015) 475,
  [\href{http://arxiv.org/abs/1506.05629}{{\tt arXiv:1506.05629}}].

\bibitem{Aaboud:2018ibj}
{\bf ATLAS} Collaboration, M.~Aaboud et~al., {\it {Measurement of colour flow
  using jet-pull observables in $t\bar{t}$ events with the ATLAS experiment at
  $\sqrt{s} = 13$ TeV}},  \href{http://arxiv.org/abs/1805.02935}{{\tt
  arXiv:1805.02935}}.

\bibitem{Abazov:2011vh}
{\bf D0} Collaboration, V.~M. Abazov et~al., {\it {Measurement of color flow in
  $\mathbf{t\bar{t}}$ events from $\mathbf{p\bar{p}}$ collisions at
  $\mathbf{\sqrt{s}=1.96}$ TeV}},  {\em Phys. Rev.} {\bf D83} (2011) 092002,
  [\href{http://arxiv.org/abs/1101.0648}{{\tt arXiv:1101.0648}}].

\bibitem{stewart2015dissecting}
I.~W. Stewart, F.~J. Tackmann, and W.~J. Waalewijn, {\it {Dissecting Soft
  Radiation with Factorization}},  {\em Phys. Rev. Lett.} {\bf 114} (2015)
  092001, [\href{http://arxiv.org/abs/1405.6722}{{\tt arXiv:1405.6722}}].

\bibitem{Metodiev:2018ftz}
E.~M. Metodiev and J.~Thaler, {\it {Jet Topics: Disentangling Quarks and Gluons
  at Colliders}},  {\em Phys. Rev. Lett.} {\bf 120} (2018) 241602,
  [\href{http://arxiv.org/abs/1802.00008}{{\tt arXiv:1802.00008}}].

\bibitem{Komiske:2018vkc}
P.~T. Komiske, E.~M. Metodiev, and J.~Thaler, {\it {An operational definition
  of quark and gluon jets}},  \href{http://arxiv.org/abs/1809.01140}{{\tt
  arXiv:1809.01140}}.

\bibitem{Larkoski:2017jix}
A.~J. Larkoski, I.~Moult, and B.~Nachman, {\it {Jet Substructure at the Large
  Hadron Collider: A Review of Recent Advances in Theory and Machine
  Learning}},  \href{http://arxiv.org/abs/1709.04464}{{\tt arXiv:1709.04464}}.

\bibitem{Asquith:2018igt}
L.~Asquith et~al., {\it {Jet Substructure at the Large Hadron Collider :
  Experimental Review}},  \href{http://arxiv.org/abs/1803.06991}{{\tt
  arXiv:1803.06991}}.

\bibitem{Chatrchyan:2012sn}
{\bf CMS} Collaboration, S.~Chatrchyan et~al., {\it {Search for a Higgs boson
  in the decay channel $H$ to ZZ(*) to $q\bar{q}$ $\ell^-\ell^+$ in $pp$
  collisions at $\sqrt{s}=7$ TeV}},  {\em JHEP} {\bf 04} (2012) 036,
  [\href{http://arxiv.org/abs/1202.1416}{{\tt arXiv:1202.1416}}].

\bibitem{catani1992jet}
S.~Catani, G.~Turnock, and B.~R. Webber, {\it {Jet broadening measures in
  $e^{+} e^{-}$ annihilation}},  {\em Phys. Lett.} {\bf B295} (1992) 269.

\bibitem{rakow1981transverse}
P.~E.~L. Rakow and B.~R. Webber, {\it {Transverse Momentum Moments of Hadron
  Distributions in {QCD} Jets}},  {\em Nucl. Phys.} {\bf B191} (1981) 63--74.

\bibitem{Ellis:1986ig}
R.~K. Ellis and B.~R. Webber, {\it {QCD Jet Broadening in Hadron Hadron
  Collisions}},  {\em Conf. Proc.} {\bf C860623} (1986) 74.

\bibitem{Farhi:1977sg}
E.~Farhi, {\it {A QCD Test for Jets}},  {\em Phys. Rev. Lett.} {\bf 39} (1977)
  1587.

\bibitem{Larkoski:2015lea}
A.~J. Larkoski, S.~Marzani, and J.~Thaler, {\it {Sudakov Safety in Perturbative
  QCD}},  {\em Phys. Rev.} {\bf D91} (2015) 111501,
  [\href{http://arxiv.org/abs/1502.01719}{{\tt arXiv:1502.01719}}].

\bibitem{Harrison:1998yr}
{\bf BaBar} Collaboration, D.~Boutigny et~al., {\it {The BABAR physics book:
  Physics at an asymmetric $B$ factory}},  in {\em {Workshop on Physics at an
  Asymmetric $B$ Factory}}, 1998.

\bibitem{Hocker:2007ht}
A.~Hocker et~al., {\it {TMVA - Toolkit for Multivariate Data Analysis}},
  \href{http://arxiv.org/abs/physics/0703039}{{\tt physics/0703039}}.

\bibitem{csiszar63}
I.~Csisz\'ar, {\it Eine informationstheoretische ungleichung und ihre anwendung
  auf den beweis der ergodizitat von markoffschen ketten},  {\em Magyar. Tud.
  Akad. Mat. Kutat\'o Int. K\"ozl} {\bf 8} (1963) 85.

\bibitem{citeulike:11934748}
S.~M. Ali and S.~D. Silvey, {\it {A General Class of Coefficients of Divergence
  of One Distribution from Another}},  {\em Journal of the Royal Statistical
  Society. Series B} {\bf 28} (1966) 131.

\bibitem{doi:10.1143/JPSJ.18.328}
T.~Morimoto, {\it Markov processes and the h-theorem},  {\em Journal of the
  Physical Society of Japan} {\bf 18} (1963) 328.

\bibitem{Nachman:2016qyc}
B.~Nachman, {\em {Investigating the Quantum Properties of Jets and the Search
  for a Supersymmetric Top Quark Partner with the ATLAS Detector}}.
\newblock PhD thesis, Stanford U., Phys. Dept., 2016.
\newblock \href{http://arxiv.org/abs/1609.03242}{{\tt arXiv:1609.03242}}.

\bibitem{8507032}
G.~Lu and B.~Li, {\it A class of new metrics based on triangular
  discrimination},  {\em Information} {\bf 6} (2015) 361.

\bibitem{850703}
F.~Topsoe, {\it Some inequalities for information divergence and related
  measures of discrimination},  {\em IEEE Transactions on Information Theory}
  {\bf 46} (2000) 1602.

\bibitem{Sjostrand:2007gs}
T.~Sjostrand, S.~Mrenna, and P.~Z. Skands, {\it {A Brief Introduction to PYTHIA
  8.1}},  {\em Comput. Phys. Commun.} {\bf 178} (2008) 852,
  [\href{http://arxiv.org/abs/0710.3820}{{\tt arXiv:0710.3820}}].

\bibitem{skands2014tuning}
P.~Skands, S.~Carrazza, and J.~Rojo, {\it {Tuning PYTHIA 8.1: the Monash 2013
  Tune}},  {\em Eur. Phys. J.} {\bf C74} (2014) 3024,
  [\href{http://arxiv.org/abs/1404.5630}{{\tt 1404.5630}}].

\bibitem{Cacciari:2011ma}
M.~Cacciari, G.~P. Salam, and G.~Soyez, {\it {FastJet User Manual}},  {\em Eur.
  Phys. J.} {\bf C72} (2012) 1896, [\href{http://arxiv.org/abs/1111.6097}{{\tt
  arXiv:1111.6097}}].

\bibitem{Cacciari:2008gp}
M.~Cacciari, G.~P. Salam, and G.~Soyez, {\it {The anti-$k_t$ jet clustering
  algorithm}},  {\em JHEP} {\bf 04} (2008) 063,
  [\href{http://arxiv.org/abs/0802.1189}{{\tt arXiv:0802.1189}}].

\bibitem{Butterworth:2008iy}
J.~M. Butterworth, A.~R. Davison, M.~Rubin, and G.~P. Salam, {\it {Jet
  substructure as a new Higgs search channel at the LHC}},  {\em Phys. Rev.
  Lett.} {\bf 100} (2008) 242001, [\href{http://arxiv.org/abs/0802.2470}{{\tt
  arXiv:0802.2470}}].

\bibitem{Larkoski:2014wba}
A.~J. Larkoski, S.~Marzani, G.~Soyez, and J.~Thaler, {\it {Soft Drop}},  {\em
  JHEP} {\bf 05} (2014) 146, [\href{http://arxiv.org/abs/1402.2657}{{\tt
  arXiv:1402.2657}}].

\bibitem{Ellis:2009su}
S.~D. Ellis, C.~K. Vermilion, and J.~R. Walsh, {\it {Techniques for improved
  heavy particle searches with jet substructure}},  {\em Phys. Rev.} {\bf D80}
  (2009) 051501, [\href{http://arxiv.org/abs/0903.5081}{{\tt
  arXiv:0903.5081}}].

\bibitem{Ellis:2009me}
S.~D. Ellis, C.~K. Vermilion, and J.~R. Walsh, {\it {Recombination Algorithms
  and Jet Substructure: Pruning as a Tool for Heavy Particle Searches}},  {\em
  Phys. Rev.} {\bf D81} (2010) 094023,
  [\href{http://arxiv.org/abs/0912.0033}{{\tt arXiv:0912.0033}}].

\bibitem{Krohn:2009th}
D.~Krohn, J.~Thaler, and L.-T. Wang, {\it {Jet Trimming}},  {\em JHEP} {\bf 02}
  (2010) 084, [\href{http://arxiv.org/abs/0912.1342}{{\tt arXiv:0912.1342}}].

\bibitem{Dasgupta:2013ihk}
M.~Dasgupta, A.~Fregoso, S.~Marzani, and G.~P. Salam, {\it {Towards an
  understanding of jet substructure}},  {\em JHEP} {\bf 09} (2013) 029,
  [\href{http://arxiv.org/abs/1307.0007}{{\tt arXiv:1307.0007}}].

\bibitem{Frye:2016aiz}
C.~Frye, A.~J. Larkoski, M.~D. Schwartz, and K.~Yan, {\it {Factorization for
  groomed jet substructure beyond the next-to-leading logarithm}},  {\em JHEP}
  {\bf 07} (2016) 064, [\href{http://arxiv.org/abs/1603.09338}{{\tt
  arXiv:1603.09338}}].

\bibitem{Buckley:2015gua}
A.~Buckley and C.~Pollard, {\it {QCD-aware partonic jet clustering for
  truth-jet flavour labelling}},  {\em Eur. Phys. J.} {\bf C76} (2016) 71,
  [\href{http://arxiv.org/abs/1507.00508}{{\tt arXiv:1507.00508}}].

\bibitem{Alwall:2014hca}
J.~Alwall, R.~Frederix, S.~Frixione, V.~Hirschi, F.~Maltoni, O.~Mattelaer,
  H.~S. Shao, T.~Stelzer, P.~Torrielli, and M.~Zaro, {\it {The automated
  computation of tree-level and next-to-leading order differential cross
  sections, and their matching to parton shower simulations}},  {\em JHEP} {\bf
  07} (2014) 079, [\href{http://arxiv.org/abs/1405.0301}{{\tt
  arXiv:1405.0301}}].

\bibitem{Bellm:2015jjp}
J.~Bellm et~al., {\it {Herwig 7.0/Herwig++ 3.0 release note}},  {\em Eur. Phys.
  J.} {\bf C76} (2016) 196, [\href{http://arxiv.org/abs/1512.01178}{{\tt
  arXiv:1512.01178}}].

\bibitem{Reichelt:2017hts}
D.~Reichelt, P.~Richardson, and A.~Siodmok, {\it {Improving the Simulation of
  Quark and Gluon Jets with Herwig 7}},  {\em Eur. Phys. J.} {\bf C77} (2017),
  no.~12 876, [\href{http://arxiv.org/abs/1708.01491}{{\tt arXiv:1708.01491}}].

\bibitem{Dasgupta:2014yra}
M.~Dasgupta, F.~Dreyer, G.~P. Salam, and G.~Soyez, {\it {Small-radius jets to
  all orders in QCD}},  {\em JHEP} {\bf 04} (2015) 039,
  [\href{http://arxiv.org/abs/1411.5182}{{\tt arXiv:1411.5182}}].

\bibitem{Dasgupta:2007wa}
M.~Dasgupta, L.~Magnea, and G.~P. Salam, {\it {Non-perturbative QCD effects in
  jets at hadron colliders}},  {\em JHEP} {\bf 02} (2008) 055,
  [\href{http://arxiv.org/abs/0712.3014}{{\tt 0712.3014}}].

\end{thebibliography}\endgroup

\end{document}